\documentclass[aps,jcp,reprint,twocolumn,citeautoscript,longbibliography,preprintnumbers]{revtex4}
\usepackage{graphicx}
\usepackage[font={small}]{caption}
\usepackage{subcaption}
\usepackage{amsmath}
\usepackage{amssymb}
\usepackage{tabularx}
\usepackage{placeins}
\usepackage{units}
\usepackage{color}
\usepackage{hyperref}
\usepackage[normalem]{ulem}
\usepackage{dcolumn}% Align table columns on decimal point
\usepackage{breakcites}
\usepackage{bm}% bold math
\usepackage{multirow}
\usepackage{adjustbox}
\usepackage{mathtools}
\newcommand{\fref}[1]{Figure~\ref{#1}}
\newcommand{\tref}[1]{Table~\ref{#1}}

\begin{document}

\renewcommand{\arraystretch}{1.2}

\title{A new generation of effective core potentials from correlated and spin-orbit calculations: selected heavy elements}

\thanks{Notice: This manuscript has been authored by UT-Battelle, LLC, under
contract DE-AC05-00OR22725 with the US Department of Energy (DOE).
The US government retains and the publisher, by accepting
the article for publication, acknowledges that the US
government retains a nonexclusive, paid-up, irrevocable, worldwide
license to publish or reproduce the published form of this manuscript,
or allow others to do so, for US government purposes.
DOE will provide public access to these results
of federally sponsored research in accordance with the DOE Public
Access Plan (\url{http://energy.gov/downloads/doe-public-access-plan}).
}

\author{Guangming Wang$^{1}$}
\email{gwang18@ncsu.edu}
\author{Benjamin Kincaid$^{1}$, Haihan Zhou$^1$, Abdulgani Annaberdiyev$^{1}$, M. Chandler Bennett$^{2}$, Jaron T. Krogel$^{2}$, and Lubos Mitas$^1$}

\affiliation{
1) Department of Physics, North Carolina State University, Raleigh, North Carolina 27695-8202, USA \\
2) Materials Science and Technology Division, Oak Ridge National Laboratory, Oak Ridge, Tennessee, 37831, USA
}

\begin{abstract}
We introduce new correlation consistent effective core potentials (ccECPs) for the elements I, Te, Bi, Ag, Au, Pd, Ir, Mo, and W with $4d$, $5d$, $6s$ and $6p$ valence spaces. 
These ccECPs are given as a sum of spin-orbit averaged relativistic effective potential (AREP) and effective spin-orbit (SO) terms. 
The construction involves several steps with increasing refinements from more simple to fully correlated methods.
The optimizations are carried out with objective functions that include weighted many-body atomic spectra, norm-conservation criteria, and spin-orbit splittings.
Transferability tests involve molecular binding curves of corresponding hydride and oxide dimers. 
The constructed ccECPs are systematically better and in a few cases on par with previous effective core potential (ECP) tables on all tested criteria and provide a significant increase in accuracy for valence-only calculations with these elements. 
Our study confirms the importance of the AREP part in determining the overall quality of the ECP even in the presence of sizable spin-orbit effects. The subsequent quantum Monte Carlo (QMC) calculations point out the importance of accurate trial wave functions which in some cases (mid series transition elements) require treatment well beyond single-reference.  
\end{abstract}

\maketitle

\section{Introduction}
\label{sec:intro}

Most electronic structure calculations aim at valence properties
such as bonding, ground and excited states, and related properties. These characteristics are determined by the
valence electronic states with energy scale of eVs and spatial ranges from covalent bonds to delocalized conduction states. 
On the other hand, the atomic cores of heavier elements are very strongly bonded to nuclei and spatially very localized. Therefore, the cores appear as  almost rigid repulsive charge barriers around the nuclei so that in many quantum chemical and condensed matter electronic structure calculations the cores are  routinely kept static and frozen. Building upon this understanding, it has been realized that the atomic cores can be alternatively represented by properly adjusted effective core 
potentials (ECPs). The effective potentials mimic the action of the core on valence states and allow for valence-only calculations with resulting gains in efficiency.  On a quantitative level, this partitioning is based on significant differences between  spatial and energy domains that are occupied by core vs valence states. ECPs and closely related pseudopotentials in the condensed matter context have been developed over several decades and involve a number of complete tables  \cite{fernandez_pacios_ab_1985, hurley_1986, lajohn_1987, ross_1990, stevens_relativistic_1992, troullier_efficient_1991, burkatzki_energy-consistent_2007-2, burkatzki_energy-consistent_2008, bergner_ab_1993,dolg_energyadjusted_1987, bachelet_pseudopotentials_1982, trail_shape_2017, vanderbilt_soft_1990, dolg_relativistic_2012}
as well as computational tools \cite{fuchs_ab_1999, opium}.

Basic construction of ECPs involves reproducing  valence one-particle eigenvalues and closely related one-particle orbital norm conservation, i.e., the amount of valence charge outside an appropriate 
effective ion radius \cite{bachelet_pseudopotentials_1982}.
Since the number of core states and their spatial properties vary, each angular momentum symmetry channel requires a different effective potential resulting 
in semilocal ECPs with corresponding projectors.
This simplest construction can be further generalized in several directions \cite{kleinman_efficacious_1982, blochl_projector_1994, shirley_many-body_1993, shirley_core_1997}.
Further improvement has been introduced through the so-called energy consistency that requires
reproducing energy differences such as atomic excitations within a given theory, e.g., Hartree-Fock \cite{bachelet_pseudopotentials_1982, burkatzki_energy-consistent_2007-2, burkatzki_energy-consistent_2008, dolg_relativistic_2012}. 
Other directions for improvement has been explored  within Density Functional theory (DFT) using many-body perturbation theory \cite{shirley_extended_1989, kresse_optimized_1992}

Very recently, we have introduced correlation consistent ECPs (ccECPs) 
\cite{bennett_new_2017, bennett_new_2018, annaberdiyev_new_2018, annaberdiyev_accurate_2020}
that build upon previous constructions.
Our overarching goal was to provide ccECPs that would offer the accuracy needed for many-body and highly-accurate calculations. For this purpose our construction has been based on many-body wave functions.  These have enabled us to use nearly-exact results to reach higher levels of accuracy and also to ascertain the robustness and transferability of the constructed potentials. The construction of ccECPs for heavier atoms involves several steps. The initial construction involves reproducing atomic excitations across a range of energies and different states including also a number of highly ionized states. We included scalar relativistic effects from the outset, and as explained below,
spin-orbit effects were added as they can influence the valence accuracy for heavier elements. 

We have taken into account further criteria having in mind the ultimate goal of describing with high accuracy the valence properties in real systems such as molecules or condensed matter systems. Therefore, the transferability of the ECPs 
has been carefully tested by probing the (effective) ion in bonded environments.  In this setting, reproducing molecular bonding curves have become another important criterion both for construction
and for validation. Therefore, validation tests of the ccECP construction have included hydride and oxide dimers for each element. This provides insights into the simplest bonds with covalent and ionic character.  
Molecular calculations involved essentially the full binding curve from the stretched bond lengths to the dissociation limit at short interatomic distances so as to test the restructuring of 
the valence charge at high pressure bonding environments. 
The typical discrepancies that were observed have been within chemical 
accuracy ($\approx$ 0.043 eV).  
There were a few exceptions where discrepancies were larger due to small number of valence electrons, especially at highly compressed bond length, we find the inclusion of norm-conservation are essential to alleviate such discrepancies. Therefore, additional cautions are needed to achieve optimal balance in several properties including spectrum and molecular bindings.

The tests and comparisons with previously generated 
tables have shown that ccECPs represent a significant step forward in achieved accuracy and fidelity within the ECP effective Hamiltonian model. 
In addition, we have found that correlated construction also provides welcome and important gains when compared with frozen core, all-electron treatments. In particular, ccECPs 
capture
core-valence correlations that are  
missed in most frozen core calculations. We have found that using ccECPs provides higher valence 
accuracy than uncorrelated (self-consistent) cores (UC). 

In this work, we extend our generation of ccECPs beyond the $3d$ and $4s4p$ elements  where explicit spin-orbit interaction is an important ingredient for accurate correlated treatments.
The constructed ccECPs are selected heavy elements that include $4d$ and $5d$ as well as
main group atoms, namely, I, Te, Bi, Ag, Au, Pd, Ir, Mo, and W. This choice has been motivated by several considerations. One is the development of an efficient methodology for $4d$ and $5d$ transition elements that require an accurate representation of atomic spin-orbit effects.  Another goal was to include elements that are prominent in a number of technologically important 2D materials \cite{margulis_1993, handy_structural_1952, wang_electronic_2011, li_crxte_2014}.

We mostly employ our previously developed
methodology with several updates and adjustments
needed due to increasing demands on correlated treatment of large atomic cores.  
Intuitively, explicit spin-orbit effects are treated as further refinements and advanced features relying on accurate spin-orbit averaged relativistic effective potentials (AREP) to form the spin-orbit relativistic effective potentials (SOREP).
The detailed methods for construction of these heavy element ccECPs are described in the following section. 

The composition of the paper is as follows:
Section \ref{sec:methods} describes the form parameterizations of ccECPs. In what follows, we discuss the objective function and optimization procedure for AREP and SOREP.
In the subsequent Section \ref{sec:results}, the results are presented including the atomic properties and selected tests on molecular systems using both correlated methods based on basis set expansions such as coupled cluster with singles, doubles, and perturbative triples (CCSD(T)) method and fixed-phase spin-orbit diffusion Monte Carlo (FPSODMC) approaches \cite{melton_spin-orbit_2016}. 
Each element is presented in greater detail and this is followed by summaries of ccECP properties.
The results are further elaborated in discussion and conclusions in Section \ref{sec:conclusions}.

\section{Methods}
\label{sec:methods}

\subsection{ECP Form and Parameterization}
\label{sec:ecp_form}

The aim of this study is to accurately reproduce the properties of the relativistic all-electron (AE) Hamiltonians with a much smaller valence effective Hamiltonians $H_{val}$. Following the Born-Oppenheimer approximation, the valence Hamiltonian $H_{val}$ in atomic units (a.u) is expressed as:
\begin{equation}
   {H}_{val} = \sum_i[T^{\rm kin}_i + V^{\rm {SOREP}}_i] +\sum_{i<j} 1/r_{ij}.
\end{equation}
The full spin-orbit relativistic effective potential (SOREP) ECPs are of the form proposed by Lee \cite{lee_ab_1977}:
\begin{equation}
\label{eqn:sorep_form}
\begin{split}
& V^{SOREP} = V^{SOREP}_{LJ}(r) + \\ 
& + \sum ^{L-1}_{l=0} \sum ^{l+1/2}_{j=|l-1/2|} \sum ^{j}_{m=-j} [V_{lj}^{SOREP}(r) - V^{SOREP}_{LJ}(r)] |ljm \rangle \langle ljm |
\end{split}
\end{equation}
%\begin{equation}
%\label{eqn:sorep_form}
%\begin{split}
%& V^{SOREP} = V^{SOREP}_{LJ}(r) + \\ 
%& + \sum ^{L-1}_{l=0} \sum ^{l+1/2}_{j=|l-1/2|} \sum ^{j}_{m=-j} V_{lj}^{SOREP}(r) |ljm \rangle \langle ljm |
%\end{split}
%\end{equation}
where $r$ is electron-ion distance and $V^{SOREP}_{LJ}(r)$ is a local potential.
This form can be split into the averaged relativistic effective core potential (AREP) and spin-orbit potential (SO) terms \cite{ermler_ab_1981}:
\begin{equation}
\label{eqn:split_form}
V^{SOREP} = V^{AREP} + V^{SO}
\end{equation}
where $V^{AREP}_{\ell}$ is the weighted $J-$average of $V^{SOREP}$:
\begin{equation}
    V^{AREP}_{\ell}(r) = \frac{1}{2 \ell + 1} \left[\ell \cdot V^{SOREP}_{\ell,\ell - \frac{1}{2}}(r) + (\ell + 1) \cdot V^{SOREP}_{\ell,\ell + \frac{1}{2}}(r) \right]
\end{equation}
The set of SO potentials $V^{SO}$ are defined using $V^{SOREP}$ as:
\begin{equation}
\begin{split}
    V^{SO} = s \cdot & \sum^{L}_{\ell = 1} \frac{2}{2\ell+1} \Delta V^{SOREP}_{\ell}(r) \cdot \\
    &\sum^{\ell}_{m=-\ell}\sum^{\ell}_{m'=-\ell} 
    |\ell m\rangle\langle \ell m| \ell |\ell m'\rangle\langle \ell m'|
\end{split}
\end{equation}
where
\begin{equation}
\Delta V^{SOREP}_{\ell}(r) = V^{SOREP}_{\ell,\ell+\frac{1}{2}}(r) - V^{SOREP}_{\ell,\ell-\frac{1}{2}}(r).
\end{equation}
$V^{AREP}$ is defined as follows similar to our previous works:
\begin{equation}
\label{eqn:arep_form}
\begin{split}
& V^{AREP}(r) = V^{AREP}_{L}(r) + \\ 
& + \sum_{\ell=0}^{\ell_{max} = L - 1} [V^{AREP}_\ell(r) - V^{AREP}_{L}(r)] \sum_{m}|\ell m\rangle\langle \ell m|.
\end{split}
\end{equation}

The latter part of Eqn. \eqref{eqn:arep_form} involves the non-local $|\ell m\rangle\langle \ell m|$ spherical harmonics projectors.
$V^{AREP}_{L}(r_i)$ is again a local channel that acts on all valence electrons and parametrized as: 
\begin{equation}
    V^{AREP}_{L}(r) = -\frac{Z_{\rm eff}}{r}(1 -e^{-\alpha r^2}) + \alpha Z_{\rm eff} re^{-\beta r^2} + \sum_{k=1} \gamma_{k} e^{-\delta_{k} r^2},
\end{equation} 
where the $Z_{eff} = Z - Z_{core}$ is the effective core charge, $\alpha$, $\beta$, $\gamma_{k}$, and $\delta_{k}$ are optimization coefficients. 
With the given format of local potential, the Coulomb singularity is explicitly cancelled out and first derivative at the origin vanishes.
The non-local potential is expressed as:
\begin{equation}
    V^{AREP}_\ell(r) - V^{AREP}_{L}(r) = \sum_{p=1}\beta_{\ell p} r^{n^{\ell p} - 2} e^{-\alpha_{\ell p} r^2}
\end{equation}
where $\beta_{\ell p}$ and $\alpha_{\ell p}$ are parameters to be optimized.
In most cases, $n_{\ell p}$ were set to be $n_{\ell p}$ = 2, however, we included $n_{\ell p}$ = 4 terms in some cases to achieve the desired accuracy.

The ccECP SO parameters are provided as:
\begin{equation}
    V_{\ell}^{SO, ccECP} = \frac{2}{2 \ell + 1}\Delta V_{\ell}^{SOREP}
\end{equation}
which is also the convention adopted by codes such as \textsc{Dirac}, \textsc{Molpro}, and \textsc{NWChem}.
$V_{l}^{SO, ccECP}$ are parametrized similarly to AREP non-local potentials:
\begin{equation}
    V^{SO, ccECP}_{\ell}(r) = \sum_{p'=1}\beta_{\ell p'} r^{n^{\ell p'} - 2} e^{-\alpha_{\ell p'} r^2}.
\end{equation}
Together with parameters described in AREP part, $n_{\ell p'}$, $\beta_{\ell p'}$, and $\alpha_{\ell p'}$, are the full sets of variables to be determined and treated by the optimizer in minimizing the chosen objective functions, separately at AREP and SO level in sequence. 
The optimization is in such a way that SO parameters are pursued after we have ensured the desirable AREP accuracy.
AREP parameters are kept fixed during SO optimizations. 

We aimed to keep the parameterization of ECP in a simple and compact form.
For the $4^{th}$ row elements, I, Te, Ag, Pd, and Mo, $\ell_{max}=2$ is used as there are $3d$ electrons being included in the core.
For the $5^{th}$ row elements, Bi, Au, Ir, and W, $\ell_{max}=3$ is employed because the $4f$ electrons are present in the core.
For each channel, the chosen number of Gaussian terms vary from 2 to 4 due to the different scenarios in the ECP construction of each element.
Note that we did not include any core polarization potential (CPP) terms or considered these in our calculations.

\subsection{Objective Function and Optimization - AREP}
\label{sec:arep_obj}

The objective functions used to optimize the AREP portion of each ccECP depended on the particular element in question.
The general recipe follows our previous works \cite{bennett_new_2017, bennett_new_2018, annaberdiyev_new_2018, wang_new_2019}.
We include the many-body energy-consistency and single-particle eigenvalues in the definition of the objective function:
\begin{equation}
\label{eqn:msf_obj_fun}
\begin{split}
    \mathcal{O}^2_{AREP}= &\sum_{s\in S} w_s(\Delta E_{AE}^{(s)} - \Delta E_{ECP}^{(s)})^2 + \\
    &\sum_{i \in L} w_i (\epsilon_{AE}^{(i)} - \epsilon_{ECP}^{(i)})^2
\end{split}
\end{equation}
where the $\Delta E_{X}^{(s)}$, $X \in \{ ECP, AE \}$, denotes the atomic energy gaps referenced to the ground state for given Hamiltonians. 
The subset of states are chosen by picking a representative set of states for the pertinent valence space.
In particular, the electron affinity (EA), neutral excitations, and various ionization levels (IP$n$) are included.
$w_s$ are weights corresponding to spectral states, decreasing as the energy gap increases due to the deep ionizations.
$\epsilon_{X}^{(i)}$, $X \in \{ ECP, AE \}$, labels the one-particle eigenvalue in ground state for $i^{th}$ valence eigenvalue with a weight $w_i$.
The ECP parameters were initialized from either MDFSTU \cite{STU-MDF} or BFD \cite{burkatzki_energy-consistent_2007-2, burkatzki_energy-consistent_2008} ECP parameters.

All energies in AREP case are calculated using CCSD(T) method with large uncontracted \mbox{aug-cc-p(wC)V$n$Z} ($n \in {T, Q, 5}$) basis sets while adjustments are made either to add some diffuse primitives or to remove primitives which
cause near-linear dependencies, as necessary.
Here AE calculations are \textit{fully} correlated which includes core-core, core-valence, and valence-valence correlations and use a scalar relativistic $10^{th}$ order DKH Hamiltonian \cite{reiherExactDecouplingDirac2004}.
All ECP calculations correlate the full valence space as well (including the semi-cores if present).
\textsc{Molpro} code \cite{werner_molpro_2012} was used for all AREP calculations.

For the elements Ag, I, Mo, and W, the simple objective function given in Eqn. \ref{eqn:msf_obj_fun} produced accurate ECPs. 
Some of the other elements in this series did not yield as accurate ECPs when this method was employed leading to two other ECP optimization schemes being used.
The first of these was a modification of the objective function (Eqn. \ref{eqn:msf_obj_fun}) by adding the norm-conservation criteria as used in previous work \cite{bennett_new_2018}. 
This norm-conserving method was only used for Bi.

Lastly, a method similar to the single-electron fit (SEFIT), and multi-electron fit (MEFIT) optimizations developed by the Stuttgart group \cite{dolg_energyadjusted_1987} has been employed.  
In this method, the ECP parameters are first optimized using single-particle energies only, and then the exponents are kept fixed in the MEFIT optimization.
This method was used for the elements Au, Ir, and Pd with objective function given in Eqn. \ref{eqn:msf_obj_fun} being used in the MEFIT step.

\subsection{Objective Function and Optimization - SOREP}
\label{sec:sorep_obj}

The spin-orbit coupling terms were optimized using the DIRAC {\cite{DIRAC21}} code and the Complete Open-Shell Configuration Interaction (COSCI) method.
The reference AE atomic states were calculated using the exact two-component (X2C) Hamiltonian as implemented in DIRAC.
The SO parameters were initialized from MDFSTU ECP values.
In general, we keep the energy-consistency scheme in spin-orbit terms optimization:

\begin{equation}
\label{eqn:so_obj_func}
\begin{split}
\mathcal{O}^2_{SO} = & \sum_{s\in S'} w_s(\Delta E_{AE}^{(s)} - \Delta E_{ECP}^{(s)})^2 + \\ 
& \sum_{m\in M} w_m(\Delta E_{AE}^{(m)} - \Delta E_{ECP}^{(m)})^2
\end{split}
\end{equation}
where the $\Delta E_{X}^{(Y)}$, $X \in \{ ECP, AE \}$ and $Y \in \{ s, m \}$ denotes the atomic gaps using COSCI method.
Here $Y$ labels the different kind of states included in the SO optimization.
For $Y=s$, states with different charges referenced to ground state were included such as EA, IP, IP2 similar to the AREP case.
For $Y=m$, states with the same charges were included, in particular, the $^{2S+1}L_{J}$ multiplet splittings with various $S$, $L$, and $J$ values were included by referencing the lowest energy for the given electronic charge.
This separation of $Y \in \{ s, m \}$ was motivated by the goal of capturing the SO effect better for small gaps. 
Additionally, since COSCI is generally less accurate than CCSD(T), our expectation is that referencing the lowest energy within a given charge will result in a better error cancellation producing accurate multiplet gaps.
This will be apparent in the Results section where COSCI gaps are directly compared to experimental gaps.

\section{Results}
\label{sec:results}

%\subsection{AREP accuracy}

We use three different metrics to assess the errors of the pseudoatom spectrum at AREP level. 
One is mean absolute deviation (MAD) of all considered $N$ atomic gaps:
\begin{equation}
\label{eqn:mad}
    \mathrm{MAD} = \frac{1}{N} \sum_{i}^{N} \left| \Delta E_{i}^{\textrm{ECP}} - \Delta E_{i}^{\textrm{AE}} \right|.
\end{equation}
Another metric is the MAD of selected low-lying $n$ gaps (LMAD):
\begin{equation}
    \mathrm{LMAD} = \frac{1}{n} \sum_{i}^{n} \left| \Delta E_{i}^{\textrm{ECP}} - \Delta E_{i}^{\textrm{AE}} \right|.
\end{equation}
For LMAD states, we chose electron affinity (EA), first ionization potential (IP), and second ionization potential (IP2) states only.
Finally, we also consider a weighted-MAD (WMAD) of all considered $N$ gaps as follows:
\begin{equation}
    \mathrm{WMAD} = \frac{1}{N} \sum_{i}^{N} \frac{100 \%}{\sqrt{|\Delta E_{i}^{\textrm{AE}}|}} \left| \Delta E_{i}^{\textrm{ECP}} - \Delta E_{i}^{\textrm{AE}} \right|.
\end{equation}

%The above quantities of energy and atomic energy gaps are provided for uncontracted aug-cc-pwCV$n$Z basis sets $n = \{ T, Q, 5 \}$,
%%which is largest available basis, 
%while adjustments are made to either add some diffuse primitives or to remove primitives which
%cause near-linear dependences, as necessary.
The pseudoatom spectrum errors are evaluated also for various other tabulated ECPs so as to assess the quality of our constructions. 
The ECPs are compared with MWBSTU\cite{STU-MWB}, MDFSTU\cite{STU-MDF}, BFD\cite{burkatzki_energy-consistent_2007-2,burkatzki_energy-consistent_2008}, LANL2\cite{LANL2}, CRENB(S/L)\cite{lajohn_ab_1987, ross__1990}, and SBKJC\cite{SBKJC-345} ECPs.
Additionally, to further demonstrate the improvement of ccECPs, we include uncorrelated-core (UC) calculations which are self-consistent AE calculations but with the only valence electrons correlated, and the cores are frozen with the same size of ECPs.

\begin{figure}[!htbp]
\centering
\includegraphics[width=1.00\columnwidth]{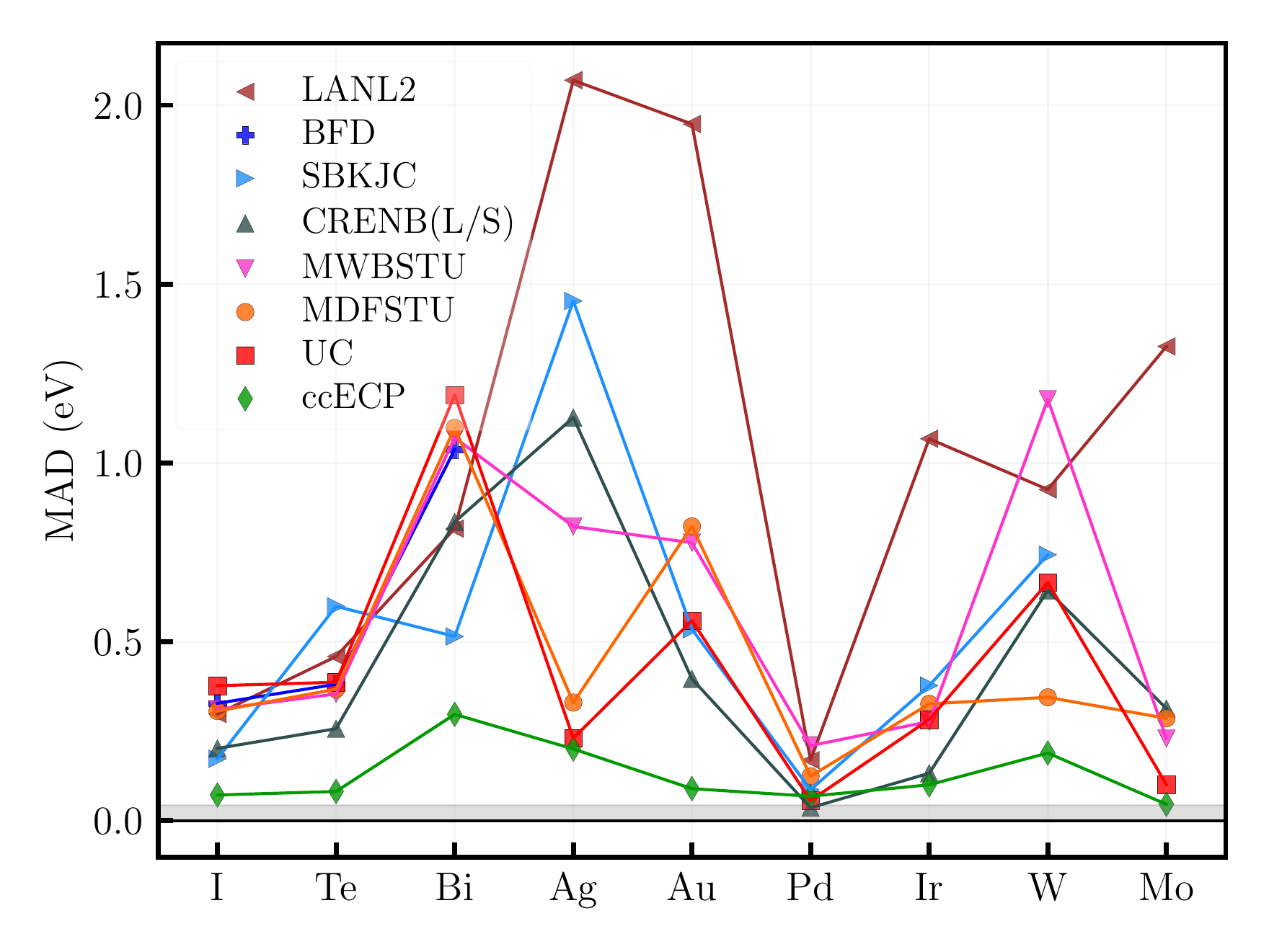}
\caption{
Scalar relativistic AE gap MADs for various core approximations using RCCSD(T) method.
}
\label{fig:MAD_in_elements}
\end{figure}

\begin{figure}[!htbp]
\centering
\includegraphics[width=1.00\columnwidth]{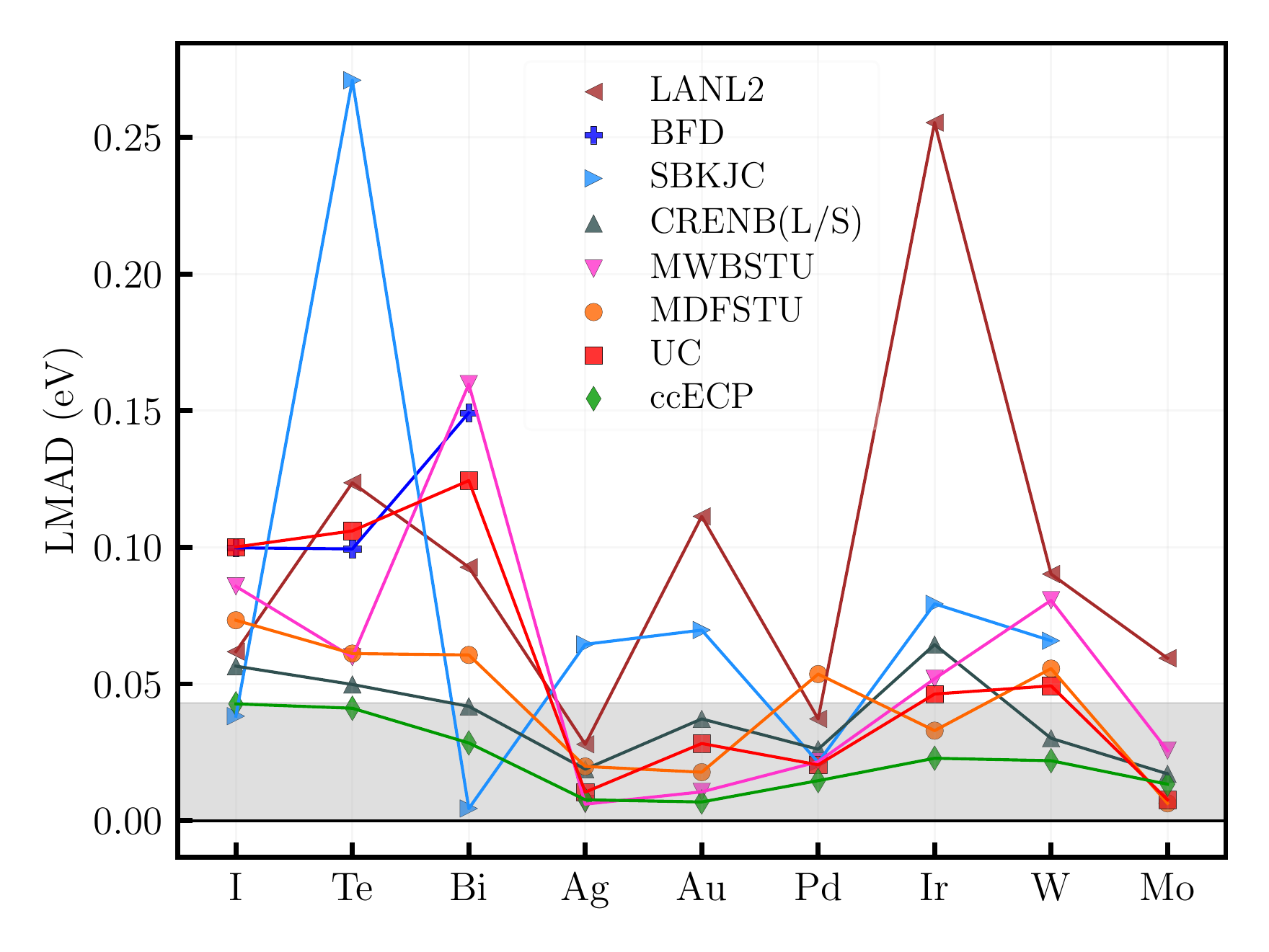}
\caption{
Scalar relativistic AE gap LMADs for various core approximations using RCCSD(T) method.
}
\label{fig:LMAD_in_elements}
\end{figure}

\begin{figure}[!htbp]
\centering
\includegraphics[width=1.00\columnwidth]{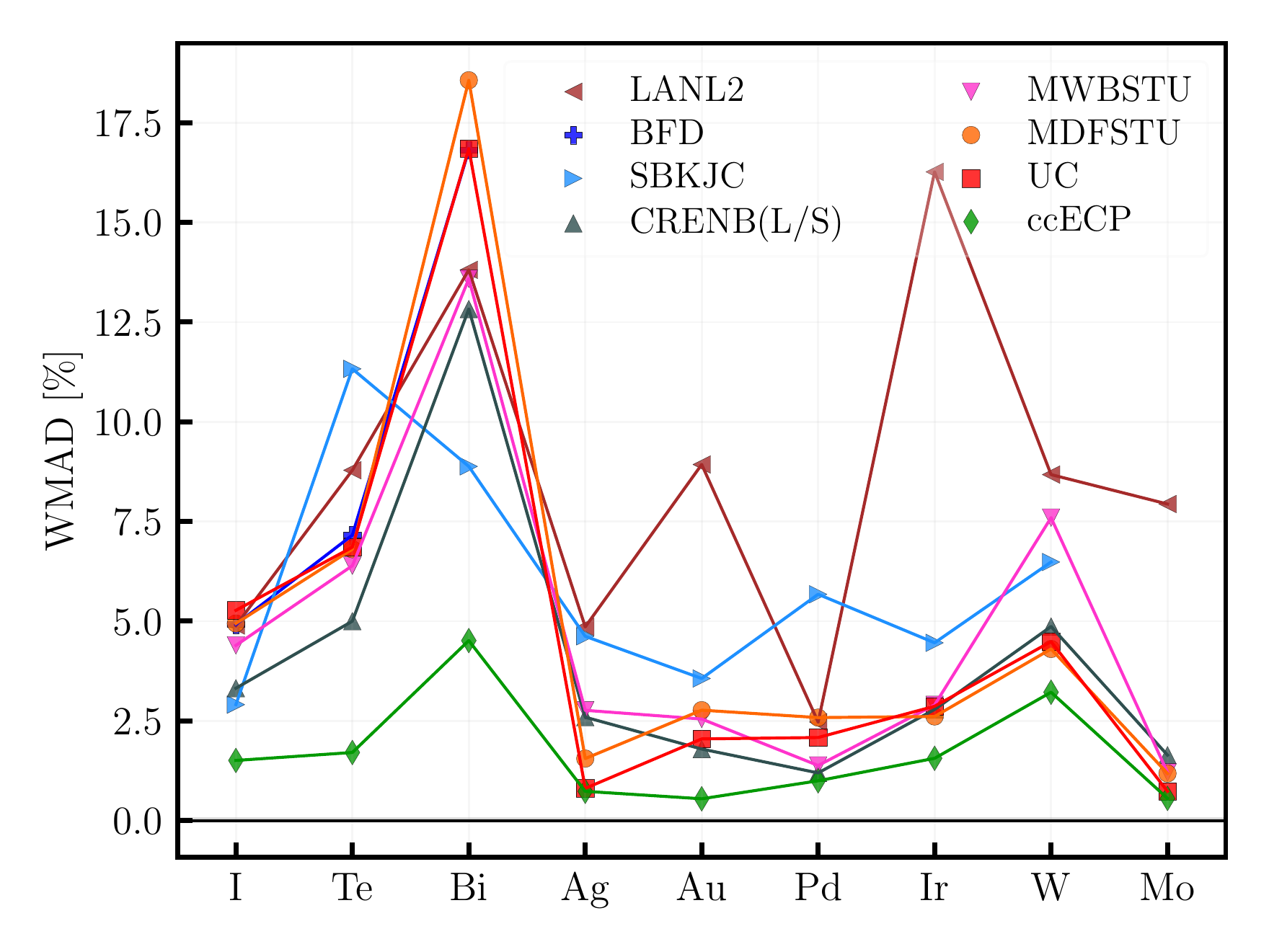}
\caption{
Scalar relativistic AE gap WMADs for various core approximations using RCCSD(T) method.
}
\label{fig:WMAD_in_elements}
\end{figure}

The summary of defined quantities, MAD, LMAD and WMAD, for selected elements are provided in \fref{fig:MAD_in_elements}, \fref{fig:LMAD_in_elements}, and \fref{fig:WMAD_in_elements} respectively.
Clearly, ccECP achieves consistent improvements overall.
Detailed discussions for each element will follow in the parts below.

Additionally, for each element, we show the transferability tests of all core approximations using molecular binding energies compared to the fully correlated AE case.
We use the AE discrepancy of dimer binding energies as one criterion of ECP quality:
\begin{equation}
    \Delta(r) = D(r)^{ECP} - D(r)^{AE}
\end{equation}
where the $D(r)$ is the binding energy at given separation $r$ for dimer atoms.

For SOREP pseudoatom spectrum, we adopt MAD definition defined above in Eqn. \eqref{eqn:mad} where the atomic gaps are chosen for the pertinent states from spin-orbit calculations.
The MADs from COSCI calculations for MDFSTU and ccECPs are compared with AE values to show the optimization gains.
Additionally, we provide error analysis by comparing both ECP gaps to experimental values using FPSODMC method which also provides the spin-orbit splittings.

The SOREP transferability tests are done for several mono-atomic dimers for both MDFSTU and ccECPs using AREP CCSD(T) and SOREP FPSODMC calculations which are compared to experimental data.
The binding curve is fitted to Morse potential:
\begin{equation}
\label{morse_pot}
    V(r)=D_{e}\left(e^{-2 a\left(r-r_{e}\right)}-2 e^{-a\left(r-r_{e}\right)}\right),
\end{equation}
where $D_{e}$ labels the dissociation energy, $r_e$ is the equilibrium bond length, and $a$ is a fitting parameter related to vibrations:
\begin{equation}
    \omega_{e}=\sqrt{\frac{2 a^{2} D_{e}}{\mu}},
\end{equation}
where $\mu$ is the reduced mass of the molecule.
Note that we use Morse potential for fitting all molecules both for AREP and SOREP treatments.

All optimized parameters for ccECPs are shown in \tref{tab:4th_ecp_params} and \tref{tab:5th_ecp_params}, respectively for the $4^{th}-$row elements and the $5^{th}-$row elements.
Results for each element are discussed separately in the following sections.

\begin{table*}%[htbp!]
\small
\centering
\caption{Parameter values for the 4th row selected heavy-element ccECPs. For all ECPs, the highest $\ell$ value corresponds to the local channel $L$.
Note that the highest non-local angular momentum channel $\ell_{max}$ is related to it as $\ell_{max}=L-1$.
}

\label{tab:4th_ecp_params}
\begin{adjustbox}{width=0.8\textwidth, center}
\begin{tabular}{ccrrrrrccrrrrrrr}
\hline\hline
\multicolumn{1}{c}{Atom} & \multicolumn{1}{c}{$Z_{\rm eff}$} & \multicolumn{1}{c}{Hamiltonian} & \multicolumn{1}{c}{$\ell$} & \multicolumn{1}{c}{$n_{\ell k}$} & \multicolumn{1}{c}{$\alpha_{\ell k}$} & \multicolumn{1}{c}{$\beta_{\ell k}$} & & \multicolumn{1}{c}{Atom} & \multicolumn{1}{c}{$Z_{\rm eff}$} & \multicolumn{1}{c}{Hamiltonian} & \multicolumn{1}{c}{$\ell$} & \multicolumn{1}{c}{$n_{\ell k}$} & \multicolumn{1}{c}{$\alpha_{\ell k}$} & \multicolumn{1}{c}{$\beta_{\ell k}$} \\
\hline
I  &  7 & AREP &  0 & 2 &    3.150276 &  114.023762  && Pd & 14 & AREP &  0 & 2 &   12.451647 &  222.004870   \\
   &    &      &  0 & 2 &    1.008080 &    1.974146  &&    &    &      &  0 & 2 &    9.656958 &    4.624354   \\
   &    &      &  1 & 2 &    2.954190 &  111.466589  &&    &    &      &  0 & 2 &    6.301508 &   55.838586   \\
   &    &      &  1 & 2 &    0.730822 &    1.710240  &&    &    &      &  1 & 2 &   12.376014 &  166.644960   \\
   &    &      &  2 & 2 &    2.604979 &   45.300844  &&    &    &      &  1 & 2 &    9.850179 &   28.721347   \\
   &    &      &  2 & 2 &    0.899593 &    8.704290  &&    &    &      &  1 & 2 &    5.706187 &   42.681167   \\
   &    &      &  3 & 1 &   12.001337 &    7.000000  &&    &    &      &  2 & 2 &    9.195344 &   57.976811   \\
   &    &      &  3 & 3 &   12.035882 &   84.009359  &&    &    &      &  2 & 2 &    7.923467 &   53.576592   \\
   &    &      &  3 & 2 &    3.022296 &   -3.998559  &&    &    &      &  2 & 2 &    3.358692 &    4.674921   \\
   &    &      &  3 & 2 &    0.977906 &   -3.050310  &&    &    &      &  3 & 1 &   15.997664 &   18.000000   \\
   &    &      &    &   &             &              &&    &    &      &  3 & 3 &   15.964009 &  287.957945   \\
   &    &  SO  &  1 & 2 &    3.086882 &   55.714747  &&    &    &      &  3 & 2 &   16.128878 & -176.382552   \\
   &    &      &  1 & 2 &    2.867556 &  -54.814595  &&    &    &      &  3 & 2 &    9.592539 &  -44.253623   \\
   &    &      &  1 & 2 &    1.921809 &   -1.544859  &&    &    &      &    &   &             &               \\
   &    &      &  1 & 2 &    1.061605 &    1.770002  &&    &    &  SO  &  1 & 2 &   17.775389 &   -4.392610   \\
   &    &      &  2 & 2 &    2.000739 &   -8.269917  &&    &    &      &  1 & 2 &    6.010086 &    4.167684   \\
   &    &      &  2 & 2 &    1.969122 &    8.345637  &&    &    &      &  2 & 2 &    8.310412 &  -28.617555   \\
   &    &      &  2 & 2 &    1.002536 &   -2.185466  &&    &    &      &  2 & 2 &    8.113178 &   28.944376   \\
   &    &      &  2 & 2 &    0.976089 &    2.118147  &&    &    &      &  2 & 2 &    3.313498 &   -1.901878   \\
   &    &      &    &   &             &              &&    &    &      &  2 & 2 &    2.152512 &    0.867558   \\
Te &  6 & AREP &  0 & 2 &    2.656483 &   48.280562  &&    &    &      &    &   &             &               \\
   &    &      &  0 & 2 &    2.281974 &   -1.021525  && Mo & 14 & AREP &  0 & 2 &   10.079375 &  180.089303   \\
   &    &      &  1 & 2 &    2.946988 &   39.656024  &&    &    &      &  0 & 2 &    6.321328 &   16.961392   \\
   &    &      &  1 & 2 &    2.790001 &   79.544830  &&    &    &      &  0 & 2 &    4.386730 &   24.723630   \\
   &    &      &  1 & 2 &    1.909579 &   -2.728455  &&    &    &      &  1 & 2 &    9.027487 &  123.680940   \\
   &    &      &  1 & 2 &    1.750168 &   -2.600491  &&    &    &      &  1 & 2 &    6.340373 &   16.933336   \\
   &    &      &  2 & 2 &    1.107233 &    6.846462  &&    &    &      &  1 & 2 &    3.910532 &   18.784314   \\
   &    &      &  2 & 2 &    1.084059 &    9.411814  &&    &    &      &  2 & 2 &    7.378964 &   48.289191   \\
   &    &      &  3 & 1 &   12.000000 &    6.000000  &&    &    &      &  2 & 2 &    6.390417 &   16.932154   \\
   &    &      &  3 & 3 &   12.000000 &   72.000000  &&    &    &      &  2 & 2 &    2.774830 &    7.940102   \\
   &    &      &  3 & 2 &   12.000000 &  -50.505126  &&    &    &      &  3 & 1 &   10.993527 &   14.000000   \\
   &    &      &    &   &             &              &&    &    &      &  3 & 3 &   11.015078 &  153.909378   \\
   &    &  SO  &  1 & 2 &    2.826025 &  -79.930555  &&    &    &      &  3 & 2 &   10.365343 &  -91.434311   \\
   &    &      &  1 & 2 &    2.772908 &   79.814350  &&    &    &      &  3 & 2 &    6.281056 &  -16.945728   \\
   &    &      &  1 & 2 &    2.087577 &   -1.100622  &&    &    &      &    &   &             &               \\
   &    &      &  1 & 2 &    1.603342 &    1.969873  &&    &    &  SO  &  1 & 2 &    9.121564 &  -82.455357   \\
   &    &      &  2 & 2 &    1.107233 &   -5.059096  &&    &    &      &  1 & 2 &    8.863223 &   82.452670   \\
   &    &      &  2 & 2 &    1.084059 &    4.999134  &&    &    &      &  1 & 2 &    4.044948 &  -12.690183   \\
   &    &      &    &   &             &              &&    &    &      &  1 & 2 &    3.866657 &   12.458423   \\
Ag & 19 & AREP &  0 & 2 &   12.570677 &  281.004418  &&    &    &      &  2 & 2 &    7.535754 &  -19.308744   \\
   &    &      &  0 & 2 &    7.075228 &   40.246565  &&    &    &      &  2 & 2 &    7.278976 &   19.318449   \\
   &    &      &  1 & 2 &   11.402841 &  210.992439  &&    &    &      &  2 & 2 &    2.772085 &    3.133446   \\
   &    &      &  1 & 2 &    6.504915 &   30.829872  &&    &    &      &  2 & 2 &    2.763205 &   -3.189516   \\
   &    &      &  2 & 2 &   10.792675 &  101.033610  &&    &    &      &    &   &             &               \\
   &    &      &  2 & 2 &    4.485396 &   14.813640  &&    &    &      &    &   &             &               \\
   &    &      &  3 & 1 &   11.116996 &   19.000000  &&    &    &      &    &   &             &               \\
   &    &      &  3 & 3 &   11.307067 &  211.222924  &&    &    &      &    &   &             &               \\
   &    &      &  3 & 2 &   10.887465 & -111.385674  &&    &    &      &    &   &             &               \\
   &    &      &  3 & 2 &    6.050896 &  -10.049234  &&    &    &      &    &   &             &               \\
   &    &      &    &   &             &              &&    &    &      &    &   &             &               \\
   &    &  SO  &  1 & 2 &   14.533732 &   -7.088207  &&    &    &      &    &   &             &               \\
   &    &      &  1 & 2 &    7.620883 &    6.990650  &&    &    &      &    &   &             &               \\
   &    &      &  2 & 2 &   11.057856 &   28.649549  &&    &    &      &    &   &             &               \\
   &    &      &  2 & 2 &    9.109402 &  -28.772257  &&    &    &      &    &   &             &               \\
   &    &      &  2 & 2 &    5.540539 &   -5.426715  &&    &    &      &    &   &             &               \\
   &    &      &  2 & 2 &    4.274636 &    4.873157  &&    &    &      &    &   &             &               \\
   &    &      &    &   &             &              &&    &    &      &    &   &             &               \\
\hline\hline
\end{tabular}
\end{adjustbox}
\end{table*}

\begin{table*}[htbp!]
\setlength{\tabcolsep}{4pt} %% default is 6pt
\small
\centering
\caption{ Parameter values for the 5th row selected heavy-element ccECPs. For all ECPs, the highest $\ell$ value corresponds to the local channel $L$.
Note that the highest non-local angular momentum channel $\ell_{max}$ is related to it as $\ell_{max}=L-1$.
}
\label{tab:5th_ecp_params}
\begin{tabular}{ccrrrrrccrrrrrrr}
\hline\hline
\multicolumn{1}{c}{Atom} & \multicolumn{1}{c}{$Z_{\rm eff}$} & \multicolumn{1}{c}{Hamiltonian} & \multicolumn{1}{c}{$\ell$} & \multicolumn{1}{c}{$n_{\ell k}$} & \multicolumn{1}{c}{$\alpha_{\ell k}$} & \multicolumn{1}{c}{$\beta_{\ell k}$} & & \multicolumn{1}{c}{Atom} & \multicolumn{1}{c}{$Z_{\rm eff}$} & \multicolumn{1}{c}{Hamiltonian} & \multicolumn{1}{c}{$\ell$} & \multicolumn{1}{c}{$n_{\ell k}$} & \multicolumn{1}{c}{$\alpha_{\ell k}$} & \multicolumn{1}{c}{$\beta_{\ell k}$} \\
\hline
Bi &  5 & AREP &  0 & 2 &    3.398563 &  137.072914  && Au & 19 & AREP &  0 & 2 &   13.809175 &  423.151508   \\ 
   &    &      &  0 & 2 &    2.236391 &   34.391642  &&    &    &      &  0 & 2 &    6.576682 &   23.880430   \\
   &    &      &  0 & 4 &    1.063682 &   -1.801931  &&    &    &      &  0 & 2 &    4.315266 &   12.729619   \\
   &    &      &  1 & 2 &    1.245620 &   39.826968  &&    &    &      &  1 & 2 &   11.806942 &  260.255986   \\
   &    &      &  1 & 2 &    0.366762 &    0.844734  &&    &    &      &  1 & 2 &    5.843816 &   32.797931   \\
   &    &      &  1 & 4 &    0.823760 &   -4.391855  &&    &    &      &  1 & 2 &    4.462109 &   14.975886   \\
   &    &      &  2 & 2 &    0.924628 &   -3.787918  &&    &    &      &  2 & 2 &    8.633228 &  135.277395   \\
   &    &      &  2 & 2 &    0.835034 &   24.096314  &&    &    &      &  2 & 2 &    4.081605 &   13.920475   \\
   &    &      &  2 & 4 &    1.239855 &    0.456375  &&    &    &      &  2 & 2 &    3.946753 &   12.464658   \\
   &    &      &  3 & 2 &    1.847446 &   -0.382962  &&    &    &      &  3 & 2 &    3.613935 &   15.143598   \\
   &    &      &  3 & 2 &    1.069842 &   -0.546444  &&    &    &      &  3 & 2 &    3.593635 &   14.877722   \\
   &    &      &  3 & 4 &    0.862193 &   -0.236708  &&    &    &      &  3 & 2 &    3.467163 &   12.317560   \\
   &    &      &  4 & 1 &    8.029160 &    5.000000  &&    &    &      &  4 & 1 &    9.985262 &   19.000000   \\
   &    &      &  4 & 3 &    7.481551 &   40.145802  &&    &    &      &  4 & 3 &    9.536259 &  189.719980   \\
   &    &      &  4 & 2 &    1.775683 &   -6.251215  &&    &    &      &  4 & 2 &   10.882390 & -140.528643   \\
   &    &      &  4 & 2 &    0.698085 &   -0.286410  &&    &    &      &  4 & 2 &    3.792290 &   -8.417696   \\
   &    &      &    &   &             &              &&    &    &      &    &   &             &               \\
   &    &  SO  &  1 & 2 &    3.624196 &   -0.056803  &&    &    &  SO  &  1 & 2 &   15.759013 &  -22.092336   \\
   &    &      &  1 & 2 &    2.084017 &    3.659507  &&    &    &      &  1 & 2 &    6.041694 &   22.320517   \\
   &    &      &  1 & 4 &    1.279962 &    3.069035  &&    &    &      &  2 & 2 &    8.904735 &   20.778790   \\
   &    &      &  2 & 2 &    1.877353 &    3.942473  &&    &    &      &  2 & 2 &    6.570928 &  -24.832959   \\
   &    &      &  2 & 2 &    1.118436 &   -4.025036  &&    &    &      &  2 & 2 &    3.389961 &    3.968015   \\
   &    &      &  2 & 4 &    1.002374 &   -0.276729  &&    &    &      &  3 & 2 &    3.387468 &   -5.901882   \\
   &    &      &  3 & 2 &    1.013377 &    1.745433  &&    &    &      &  3 & 2 &    3.083291 &    5.872893   \\
   &    &      &  3 & 2 &    0.959437 &   -1.645441  &&    &    &      &    &   &             &               \\
   &    &      &  3 & 4 &    1.004895 &    0.007556  &&    &    &      &    &   &             &               \\
   &    &      &    &   &             &              &&    &    &      &    &   &             &               \\
Ir & 17 & AREP &  0 & 2 &   13.826287 &  438.873879  &&  W & 14 & AREP &  0 & 2 &   11.273956 &  420.479803   \\ 
   &    &      &  0 & 2 &    6.284095 &   76.599511  &&    &    &      &  0 & 2 &    8.410924 &   39.584943   \\
   &    &      &  1 & 2 &   11.264188 &  262.750135  &&    &    &      &  1 & 2 &    8.517824 &  320.877235   \\
   &    &      &  1 & 2 &    5.346383 &   61.991944  &&    &    &      &  1 & 2 &    1.608091 &   -0.472821   \\
   &    &      &  2 & 2 &    6.815210 &  123.955993  &&    &    &      &  1 & 4 &   10.073490 &    0.112929   \\
   &    &      &  2 & 2 &    4.766746 &   31.123654  &&    &    &      &  2 & 2 &    6.059286 &  158.077959   \\
   &    &      &  3 & 2 &    2.786350 &   10.329276  &&    &    &      &  2 & 2 &    1.104853 &   -0.283299   \\
   &    &      &  3 & 2 &    2.592946 &   11.885698  &&    &    &      &  2 & 4 &    6.262477 &   -0.459490   \\
   &    &      &  4 & 1 &   12.171070 &   17.000000  &&    &    &      &  3 & 2 &    2.075342 &    7.745405   \\
   &    &      &  4 & 3 &   12.170139 &  206.908194  &&    &    &      &  3 & 2 &    1.845947 &    6.916005   \\
   &    &      &  4 & 2 &   12.320343 & -143.208379  &&    &    &      &  4 & 1 &   10.188960 &   14.000000   \\
   &    &      &  4 & 2 &    5.602984 &  -23.157042  &&    &    &      &  4 & 3 &    9.478240 &  142.645440   \\
   &    &      &    &   &             &              &&    &    &      &  4 & 2 &   10.146880 & -100.078727   \\
   &    &  SO  &  1 & 2 &   13.722145 &  -15.366434  &&    &    &      &  4 & 2 &    5.491651 &   -0.941578   \\
   &    &      &  1 & 2 &    4.201427 &    9.843941  &&    &    &      &    &   &             &               \\
   &    &      &  2 & 2 &   10.062658 &   -2.838512  &&    &    &  SO  &  1 & 2 &   11.964154 &   -0.551147   \\
   &    &      &  2 & 2 &    5.177629 &    3.880360  &&    &    &      &  1 & 2 &    4.897325 &    1.427345   \\
   &    &      &  3 & 2 &    2.299318 &    4.142597  &&    &    &      &  1 & 4 &    8.302365 &  194.404774   \\
   &    &      &  3 & 2 &    2.295619 &   -4.159394  &&    &    &      &  2 & 2 &    5.772034 &    7.045184   \\
   &    &      &    &   &             &              &&    &    &      &  2 & 2 &    3.175200 &   -0.115276   \\
   &    &      &    &   &             &              &&    &    &      &  2 & 4 &    6.155484 &    3.907520   \\
   &    &      &    &   &             &              &&    &    &      &  3 & 2 &    2.473295 &    4.123819   \\
   &    &      &    &   &             &              &&    &    &      &  3 & 2 &    2.405627 &   -4.032609   \\
\\
\hline\hline
\end{tabular}
\end{table*}

%\subsection{SOREP}

%%% Include the selected elements here:
\subsection{Iodine (I)}
\subsubsection{AREP: I}

\begin{figure}[!htbp]
\centering
\includegraphics[width=1.00\columnwidth]{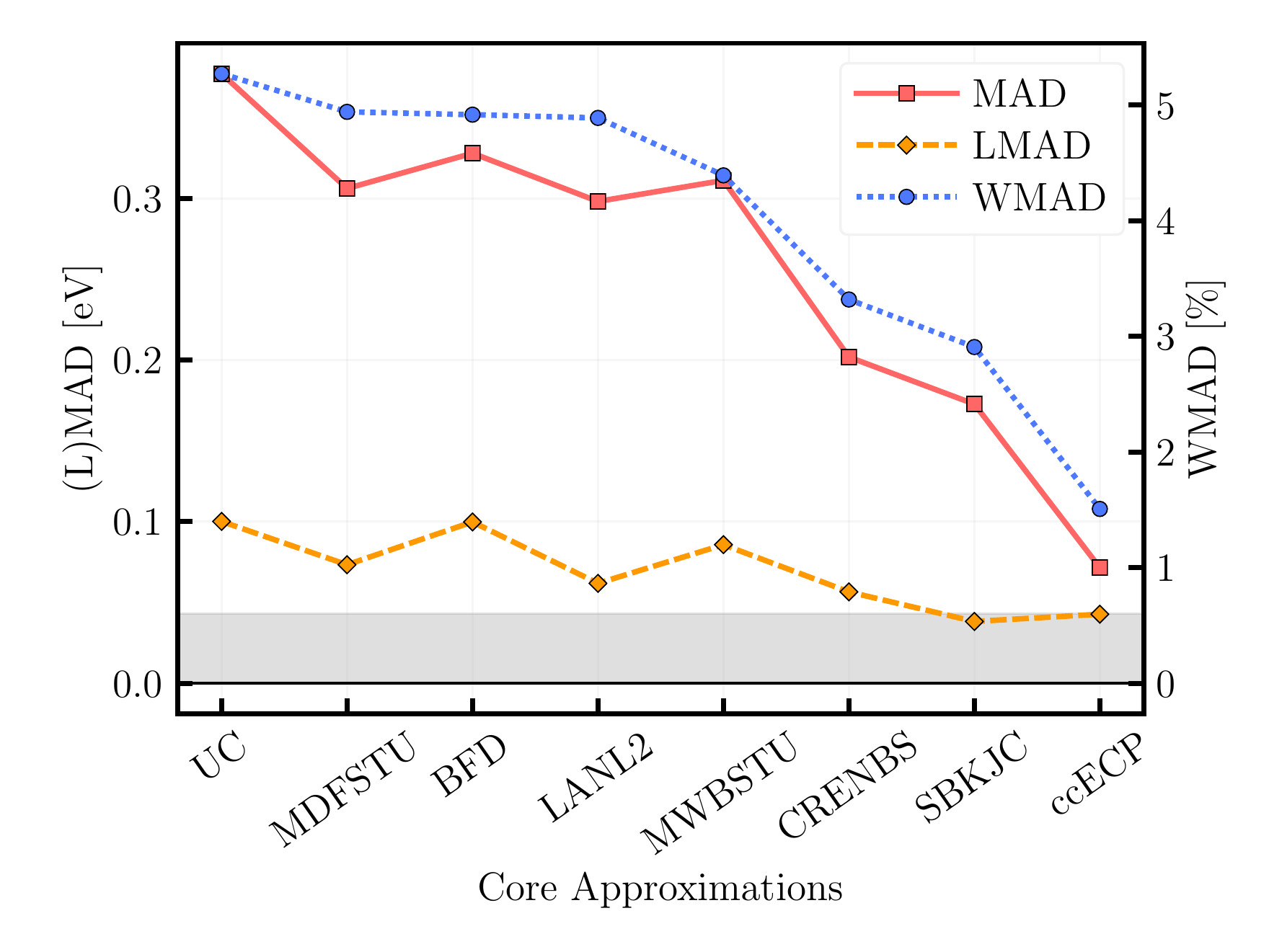}
\caption{
I scalar relativistic AE gap (W/L)MADs for various core approximations. 
\mbox{[core] = [Kr]$4d^{10}$} (46 electrons).
RCCSD(T) method with unc-aug-cc-pwCVQZ basis set was used.
}
\label{fig:I_spectrum}
\end{figure}

\begin{figure*}[!htbp]
\centering
\begin{subfigure}{0.5\textwidth}
\includegraphics[width=\textwidth]{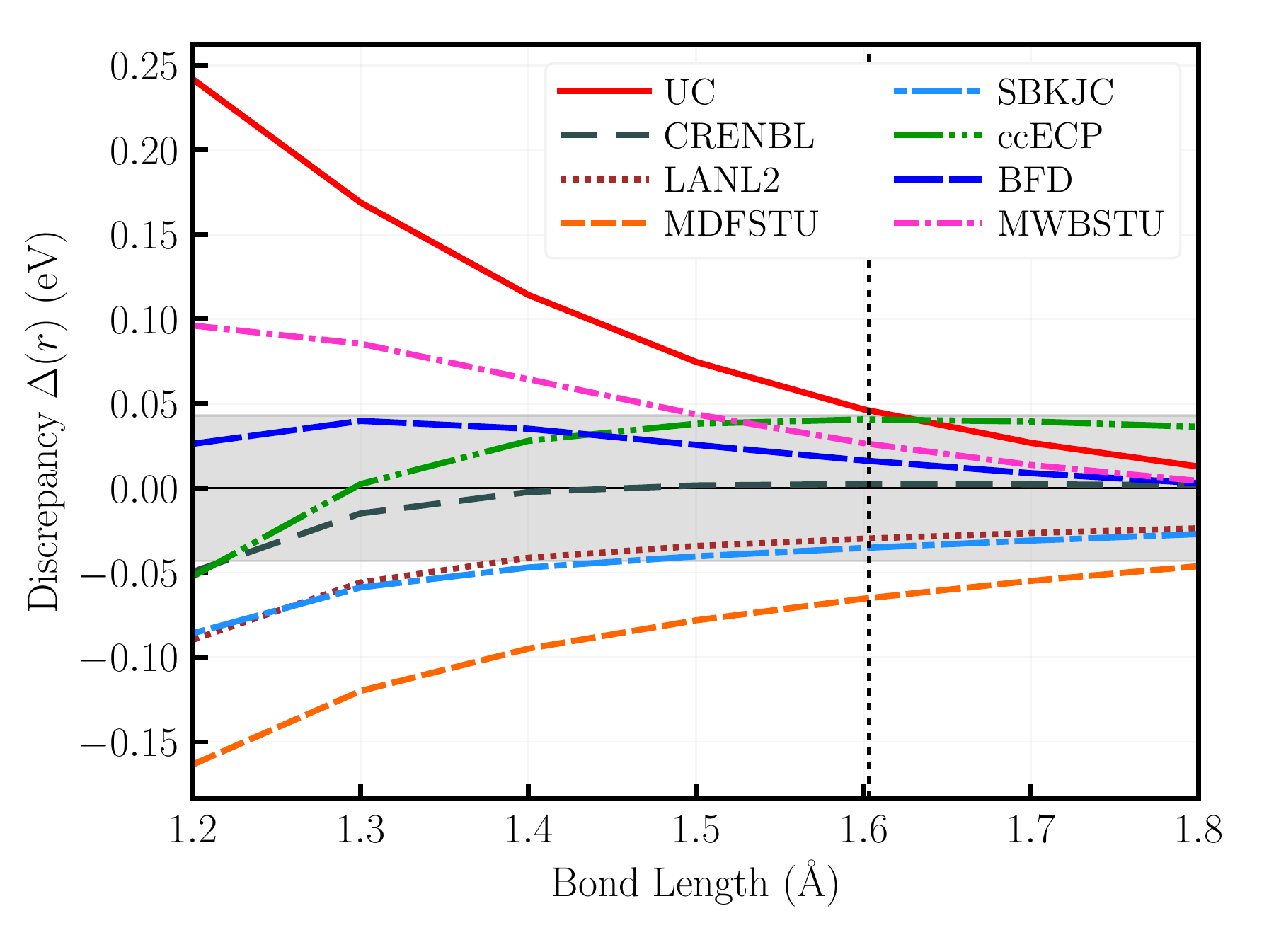}
\caption{IH ($^1\Sigma$) binding curve discrepancies}
%\label{fig:}
\end{subfigure}%
\begin{subfigure}{0.5\textwidth}
\includegraphics[width=\textwidth]{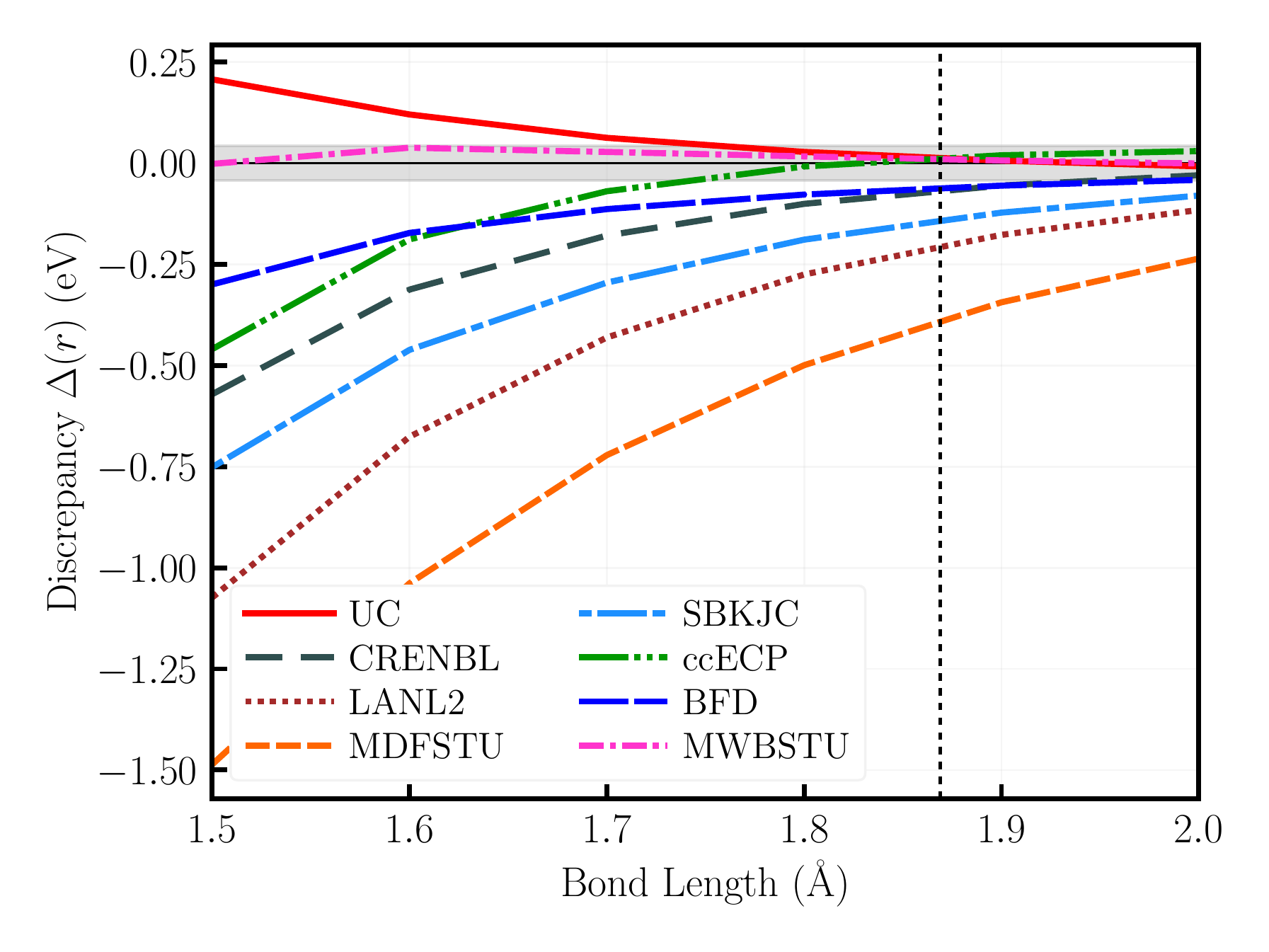}
\caption{IO ($^2\Pi$)  binding curve discrepancies}
%\label{fig:}
\end{subfigure}
\caption{
Binding energy discrepancies for (a) IH and (b) IO molecules relative to scalar relativistic AE CCSD(T).
The shaded region indicates the band of chemical accuracy. The dashed vertical line represents the equilibrium geometry.
}
\label{fig:I_mols}
\end{figure*}

For Iodine, the AREP atomic and molecular results are shared in \fref{fig:I_spectrum} and \fref{fig:I_mols}.
The construction of ccECP for iodine followed some of the adjustments used for $3s,3p$ elements which are in the same group of elements.
The atomic spectral properties are significantly improved when compared with previous constructions especially in MAD and WMAD.
The low-lying excitations 
are on par with the best available construction SBKJC
while our oxide molecule shows much improved binding curve (Figure \ref{fig:I_mols}).
MWBSTU shows less errors in IO dimers, however it underbinds in short bond length IH and has larger errors in atomic spectrum.
At low IO distances, we see similar inaccuracies 
that we observed for 
phosphorus \cite{bennett_new_2018}
that is manifested by 
overbinding at significantly compressed bond lengths.
Within the presented data, our construction represents the most consistent option with
clear advantages for broad range of valence-only  calculations.

\subsubsection{SOREP: I}

\tref{tab:I_sorep} shows SOREP iodine atomic excitation errors. We see 
that the spin-orbit splittings are reproduced very well, on par with MDFSTU results (that are very accurate to begin with) or marginally better.
However, FPSODMC MAD shows an improvement over MDFSTU.

\begin{figure}[!htbp]
\centering
\includegraphics[width=0.925\columnwidth]{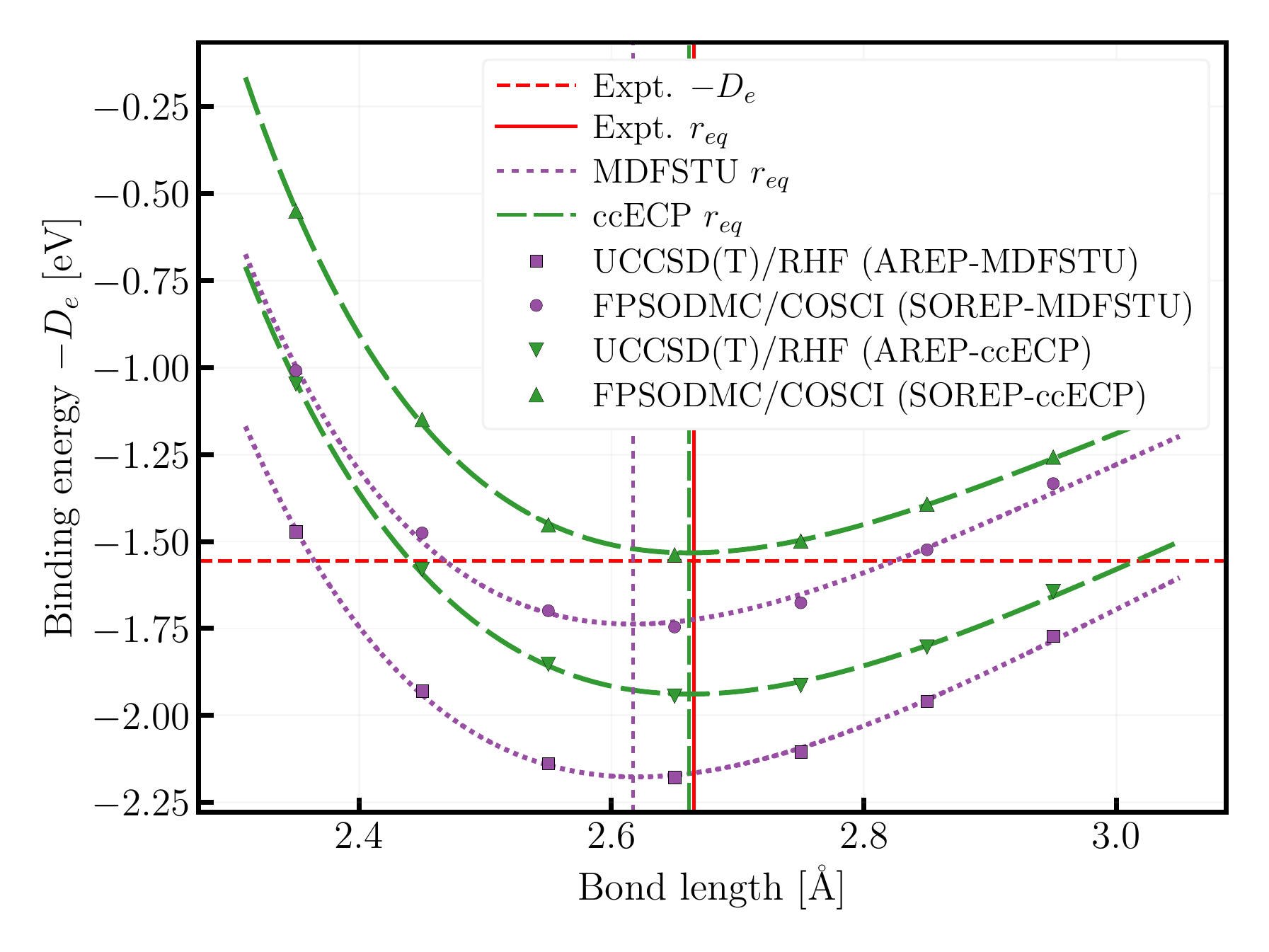}
\caption{I$_2$ dimer ($^1\Sigma_g$) binding curve with various methods.}
\label{fig:I2_mols}
\end{figure}

The overall accuracy is further corroborated 
in calculation of the iodine dimer bonding curve that shows 
significant improvements
already at the AREP level (Figure \ref{fig:I2_mols}).
We see that the calculation reproduces correctly the experimental bond length
unlike the previous construction which shows a bias towards smaller values.
Furthermore, the two-component spinor calculations show an excellent 
agreement with experimental atomization energy that clearly demonstrates the 
quality of ccECP.
Notice that the inclusion of explicit spin-orbit effect alleviates the overbinding about 0.5 eV for both ECPs. The excellent agreement of consistent improvement 
between explicit spin-orbit FPSODMC and CCSD(T) results is encouraging and suggests
comparable quality of the correlation description.

\begin{table}[!htbp]
\setlength{\tabcolsep}{4pt} %% default is 6pt
\small
\centering
\caption{
Iodine atomic excitation errors for MDFSTU versus ccECP in SOREP forms.
One set of errors are shown for full-relativistic X2C AE gaps using COSCI.
Another set of errors are calculated using FPSODMC and compared to experiments.
All values are in eV.
}
\label{tab:I_sorep}
\begin{adjustbox}{width=1.0\columnwidth,center}
%\resizebox{0.97\columwidth}{!}{%
\begin{tabular}{ll|rcc|r|c|cccc}
\hline\hline
\multirow{2}{*}{State} & \multirow{2}{*}{Term} & \multicolumn{3}{c|}{COSCI} &  \multirow{2}{*}{Expt.} & \multicolumn{2}{c}{FPSODMC} \\
\cline{3-5}
\cline{7-8}
& & AE & STU & ccECP &  & STU & ccECP \\
\hline
$5s^25p^5$ & $^{2}P_{3/2}$ &    0.000 &     0.000 &       0.000 &  0.000 &      0.0000 &        0.0000 \\
$5s^25p^6$ & $^{1}S_{0}$   &   -2.185 &     0.022 &       0.035 & -3.060 &     0.05(1) &       0.06(1) \\
$5s^25p^4$ & $^{3}P_{2}$   &    9.413 &     0.009 &      -0.046 & 10.451 &     0.05(1) &       0.04(2) \\
$5s^25p^3$ & $^{4}S_{3/2}$ &   27.262 &     0.097 &      -0.058 & 29.582 &     0.33(1) &       0.08(3) \\
\hline
$5s^25p^5$ & $^{2}P_{3/2}$ &    0.000 &     0.000 &       0.000 &  0.000 &      0.0000 &        0.0000 \\
        {} & $^{2}P_{1/2}$ &    0.962 &     0.030 &       0.024 &  0.942 &     0.01(1) &       0.00(2) \\
\hline
$5s^25p^4$ & $^{3}P_{2}$   &    0.000 &     0.000 &       0.000 &  0.000 &      0.0000 &        0.0000 \\
        {} & $^{3}P_{0}$   &    0.863 &     0.046 &       0.047 &  0.799 &     0.03(1) &       0.02(2) \\
        {} & $^{3}P_{1}$   &    0.869 &     0.019 &       0.018 &  0.878 &     0.08(1) &       0.04(2) \\
        {} & $^{1}D_{2}$   &    1.999 &    -0.029 &      -0.040 &  1.702 &     0.04(1) &      -0.01(2) \\
        {} & $^{1}S_{0}$   &    4.258 &    -0.104 &      -0.132 &  3.657 &    -0.08(1) &      -0.13(2) \\
\hline
$5s^25p^3$ & $^{4}S_{3/2}$ &    0.000 &     0.000 &       0.000 &  0.000 &      0.0000 &        0.0000 \\
        {} & $^{2}D_{3/2}$ &    1.959 &    -0.136 &      -0.159 &  1.451 &    -0.24(1) &      -0.07(2) \\
        {} & $^{2}D_{5/2}$ &    2.348 &    -0.101 &      -0.119 &  1.847 &    -0.14(1) &       0.08(2) \\
        {} & $^{2}P_{1/2}$ &    3.765 &    -0.183 &      -0.215 &  3.012 &    -0.24(1) &      -0.12(3) \\
        {} & $^{2}P_{3/2}$ &    4.375 &    -0.116 &      -0.138 &  3.674 &    -0.16(1) &      -0.06(3) \\
\hline
MAD        &               &          &     0.074 &       0.085 &        &     0.12(1) &       0.06(2) \\

\hline\hline
\end{tabular}
\end{adjustbox}
\end{table}

\subsection{Tellurium (Te)}
\subsubsection{AREP: Te}

\begin{figure}[!htbp]
\centering
\includegraphics[width=1.00\columnwidth]{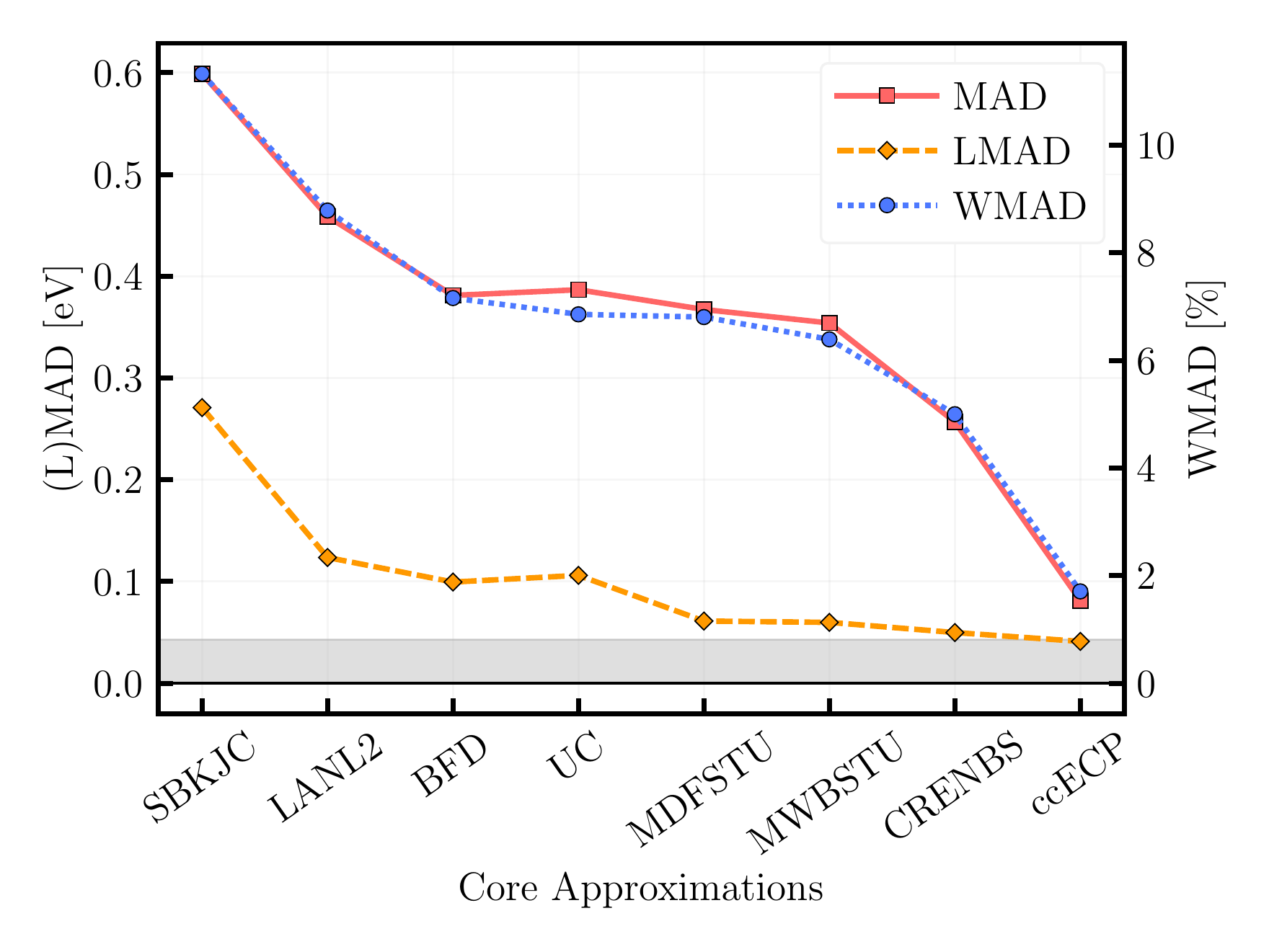}
\caption{
Te scalar relativistic (W/L)MADs for various core approximations. 
\mbox{[core] = [Kr]$4d^{10}$} (46 electrons).
RCCSD(T) method with unc-aug-cc-pwCVQZ(+diffuse terms) basis set was used.
}
\label{fig:Te_spectrum}
\end{figure}

\begin{figure*}[!htbp]
\centering
\begin{subfigure}{0.5\textwidth}
\includegraphics[width=\textwidth]{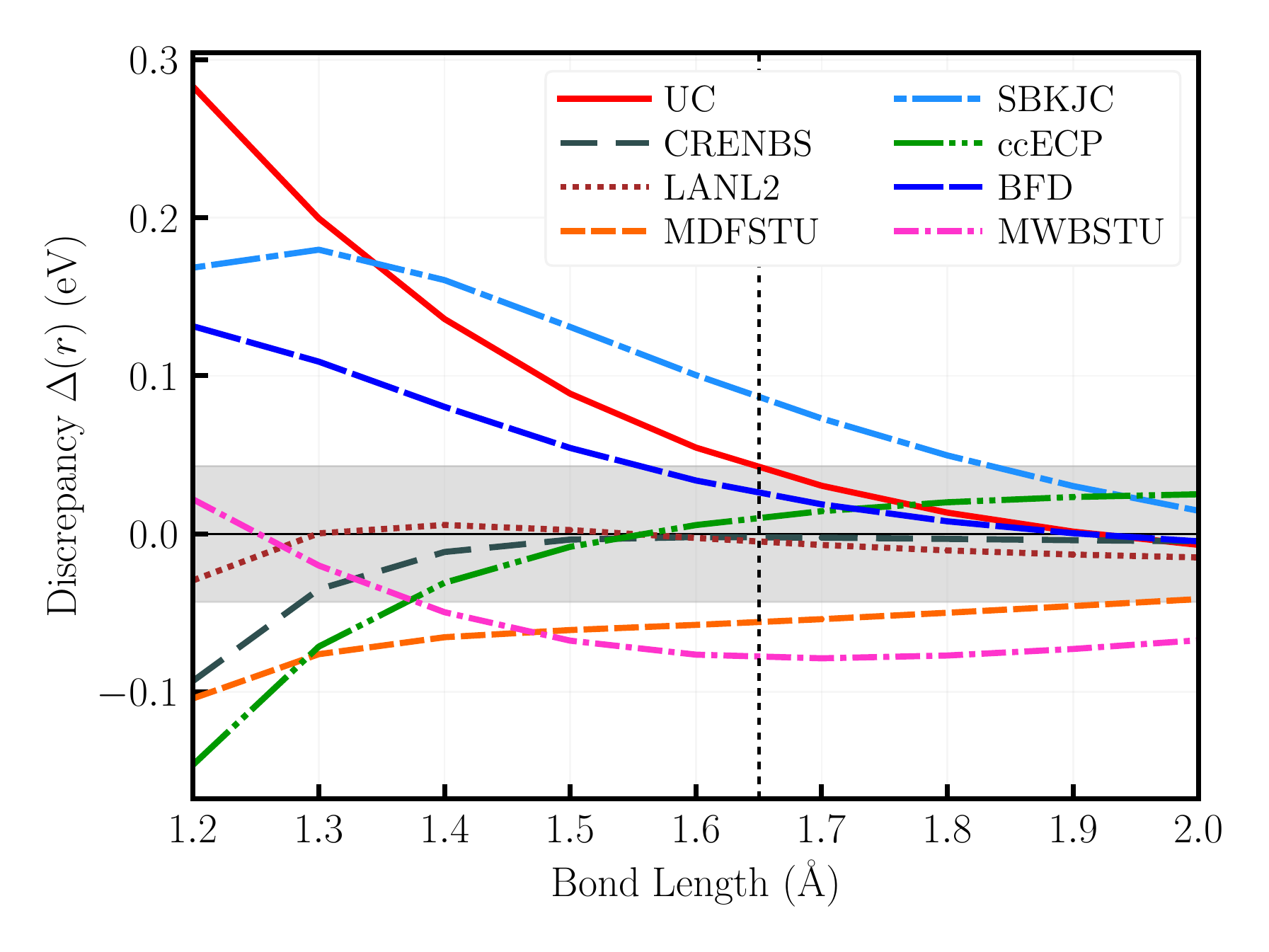}
\caption{TeH ($^2\Pi$) binding curve discrepancies}
%\label{fig:}
\end{subfigure}%
\begin{subfigure}{0.5\textwidth}
\includegraphics[width=\textwidth]{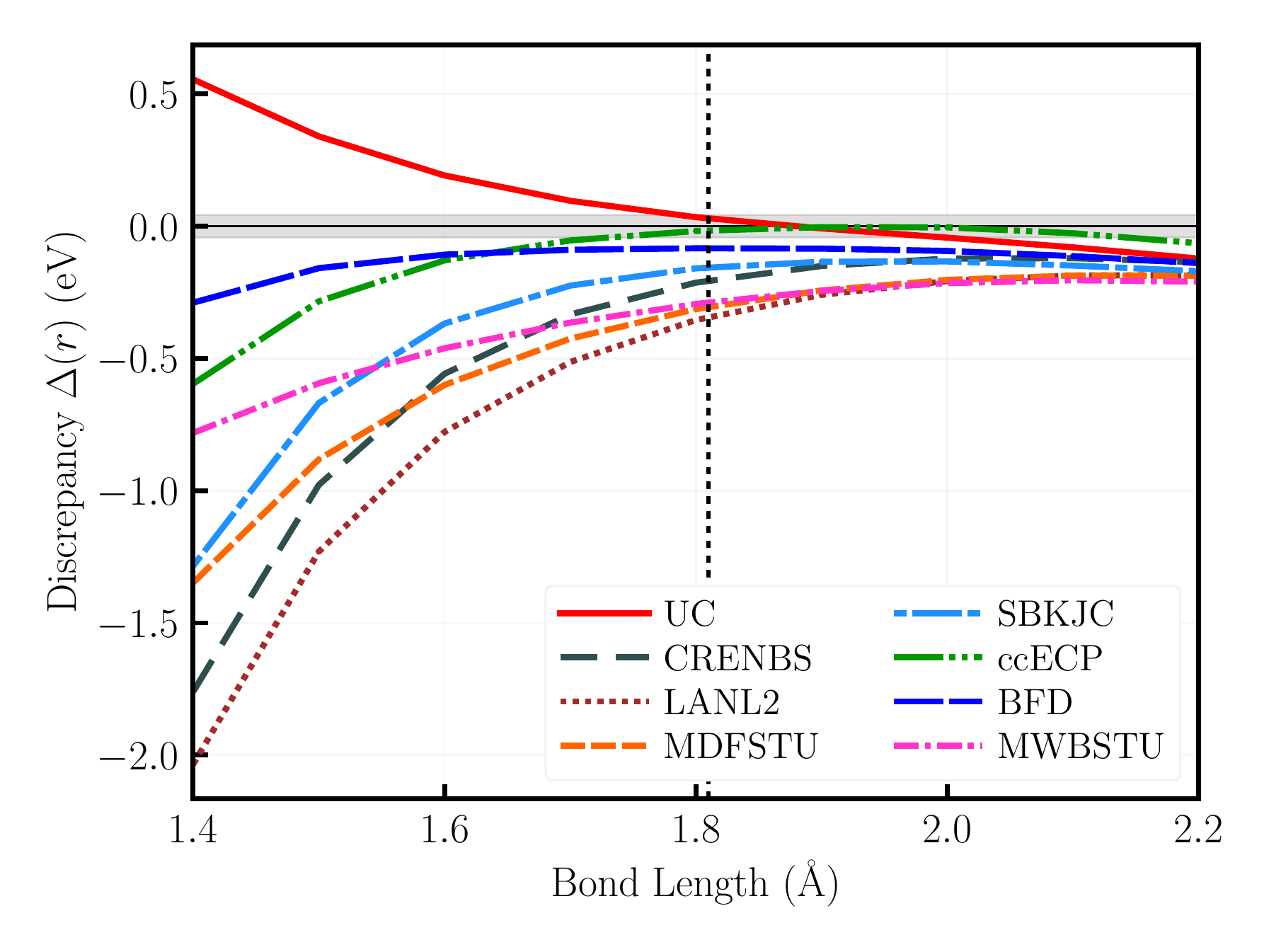}
\caption{TeO ($^3\Sigma$) binding curve discrepancies}
%\label{fig:}
\end{subfigure}
\caption{
Binding energy discrepancies for (a) TeH and (b) TeO molecules relative to scalar relativistic AE CCSD(T).
The shaded region indicates the band of chemical accuracy. The dashed vertical line represents the equilibrium geometry.
}
\label{fig:Te_mols}
\end{figure*}

The atomic spectrum and molecular binding discrepancy data for Te are provided in \fref{fig:Te_spectrum} and \fref{fig:Te_mols}. 
In the atomic spectrum, we observe a marginal improvement in ccECP LMAD compared with other ECPs.
Regarding MAD and WMAD, ccECP shows remarkable improvement with resulting much better description of higher atomic excitations.
The molecular data in \fref{fig:Te_mols} demonstrates the transferability test in TeH and TeO molecules.
In TeH, we see ccECP is well within chemical accuracy except very short bond lengths near the dissociation limit.
LANL2 and CRENBS appear to be better in TeH, however they overbind significantly in the oxide dimer TeO.
In fact, we see that all other ECPs noticeably overbind in TeO through all geometries, even at equilibrium bond length.
ccECP is very accurate near the equilibrium bond length and minor overbinding results 
only at the shortest bond lengths. 
Though BFD behaves slightly better at short bond lengths, it errs significantly at the equilibrium  which is vitally important for molecular and condensed matter properties.
Obviously, we achieve a better balance for these two types of bonds. Similar
compromise between hydride and oxide molecules has been seen from our previous ccECP constructions in 4p elements \cite{wang_new_2019}.

\subsubsection{SOREP: Te}

Tellurium SOREP atomic energy gap errors are provided in \tref{tab:Te_sorep}.
We notice that the MAD of COSCI gaps for ccECP is slightly larger than for MDFSTU and further improvements without compromising other ccECP aspects proved to be difficult.
(Exploiting the greater variational freedom of additional gaussians could provide some options in the future.)
The fixed-phase DMC calculations show similar results for ccECP and MDFSTU which compare the atomic gaps and spin-orbit splittings as referenced to experimental data. Some of the differences are larger than desired, however, this is to be expected due to small valence space, neglect of 
core relaxations and for FPSODMC using a single-reference trial function. The errors are larger for higher excitations and at this point it's not clear how
much would that affect accuracy in bonded settings.  Further research might be necessary if tests in bonded systems of interest would 
indicate that higher ccECP accuracy might be required.

\begin{table}[!htbp]
\setlength{\tabcolsep}{4pt} %% default is 6pt
\small
\centering
\caption{
Tellurium atomic excitation errors for MDFSTU versus ccECP in SOREP forms.
One set of errors are shown for full-relativistic X2C AE gaps using COSCI.
Another set of errors are calculated using FPSODMC and compared to experiments.
All values are in eV.
}
\label{tab:Te_sorep}
\begin{adjustbox}{width=1.0\columnwidth,center}
%\resizebox{0.97\columwidth}{!}{%
\begin{tabular}{ll|rcc|r|ccccc}
\hline\hline
\multirow{2}{*}{State} & \multirow{2}{*}{Term} & \multicolumn{3}{c|}{COSCI} &  \multirow{2}{*}{Expt.} & \multicolumn{2}{c}{FPSODMC} \\
\cline{3-5}
\cline{7-8}
& & AE & STU & ccECP &  & STU & ccECP \\
\hline
$5s^25p^4$ & $^{3}P_{2}$   &    0.0000 &     0.0000 &       0.0000 &   0.000 &     0.00  &         0.00    \\
$5s^25p^5$ & $^{2}P_{3/2}$ &   -0.9927 &    -0.0244 &       0.0254 &  -1.970 &   0.01(1) &      -0.02(1)   \\
$5s^25p^3$ & $^{4}S_{3/2}$ &    7.8285 &    -0.0680 &       0.0021 &   9.010 &   0.21(1) &       0.17(1)   \\
$5s^25p^2$ & $^{3}P_{2}$   &   25.4702 &     0.0505 &       0.2103 &  27.610 &   0.55(1) &       0.38(1)   \\
\hline                                                                             
$5s^25p^4$ & $^{3}P_{2}$   &  0.0000   &   0.0000   &  0.0000      &   0.000 &     0.00  &         0.00    \\
        {} & $^{3}P_{1}$   &  0.5773   &  -0.0651   &  0.0133      &   0.589 &   0.13(1) &       0.11(1)   \\
        {} & $^{3}P_{0}$   &  0.6304   &  -0.0454   &  0.0201      &   0.584 &   0.01(1) &       0.04(1)   \\
        {} & $^{1}D_{2}$   &  1.5787   &  -0.0045   &  0.0725      &   1.309 &   0.11(1) &       0.09(1)   \\
        {} & $^{1}S_{0}$   &  3.4514   &   0.0441   &  0.1714      &   2.876 &  -0.04(1) &      -0.09(1)   \\
\hline                                                                             
$5s^25p^5$ & $^{2}P_{3/2}$ &    0.0000 &     0.0000 &       0.0000 &   0.000 &     0.00  &         0.00    \\
        {} & $^{2}P_{1/2}$ &    0.6346 &    -0.0660 &       0.0139 &         &           &                 \\
\hline                                                                             
$5s^25p^3$ & $^{4}S_{3/2}$ &  0.0000   &  0.0000    &  0.0000      &   0.000 &     0.00  &         0.00    \\
        {} & $^{2}D_{3/2}$ &  1.7573   &  0.1126    &  0.1150      &   1.267 &  -0.18(1) &      -0.18(1)   \\
        {} & $^{2}D_{5/2}$ &  2.0117   &  0.0729    &  0.1211      &   1.540 &  -0.10(1) &      -0.13(1)   \\
        {} & $^{2}P_{1/2}$ &  3.2603   &  0.1372    &  0.1986      &   2.547 &  -0.21(1) &      -0.25(1)   \\
        {} & $^{2}P_{3/2}$ &  3.6511   &  0.0729    &  0.2097      &   2.980 &  -0.12(1) &      -0.19(1)   \\
\hline                                                                             
$5s^25p^2$ & $^{3}P_{0}$   &    0.0000 &     0.0000 &       0.0000 &   0.000 &     0.00  &         0.00    \\
        {} & $^{3}P_{1}$   &    0.5445 &    -0.0649 &       0.0191 &   0.589 &   0.09(1) &       0.11(1)   \\
        {} & $^{3}P_{2}$   &    1.0371 &    -0.0812 &       0.0469 &   1.012 &   0.07(1) &       0.12(1)   \\
        {} & $^{1}D_{2}$   &    2.4298 &    -0.0527 &       0.1352 &   2.152 &   0.12(1) &       0.10(1)   \\
        {} & $^{1}S_{0}$   &    4.4771 &     0.0656 &       0.2803 &         &           &                 \\
\hline
MAD        &               &           &    0.049   &       0.079  &         &   0.14(1) &       0.14(3)   \\    
\hline\hline
\end{tabular}
\end{adjustbox}
\end{table}

\subsection{Bismuth (Bi)}

\subsubsection{AREP: Bi}

\begin{figure}[!htbp]
\centering
\includegraphics[width=1.00\columnwidth]{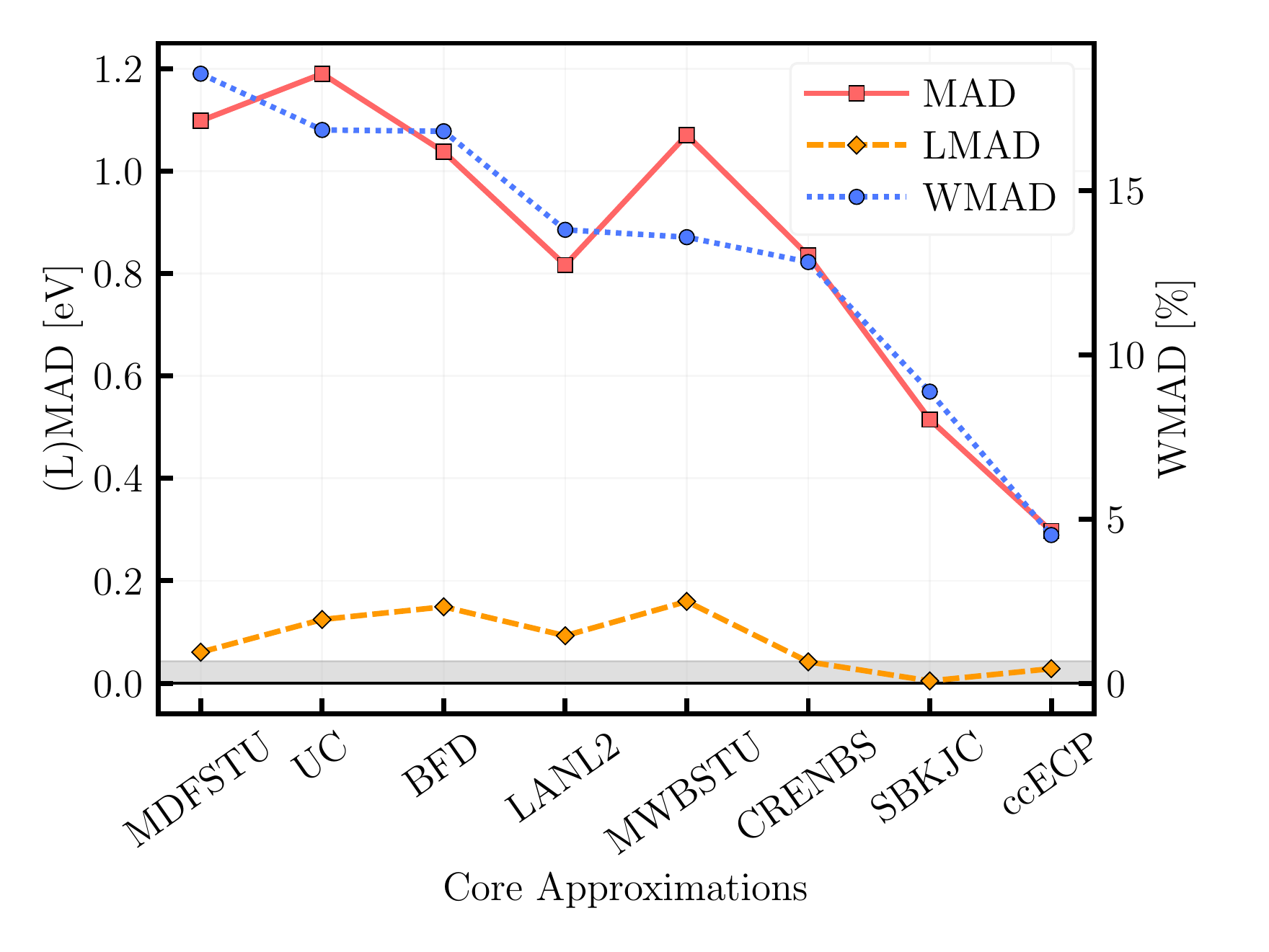}
\caption{
Bi scalar relativistic AE gap (W/L)MADs for various core approximations. 
\mbox{[core] = [Xe]$4f^{14}5d^{10}$} (78 electrons).
RCCSD(T) method with unc-aug-cc-pwCVTZ basis set was used.
}
\label{fig:Bi_spectrum}
\end{figure}

\begin{figure*}[!htbp]
\centering
\begin{subfigure}{0.5\textwidth}
\includegraphics[width=\textwidth]{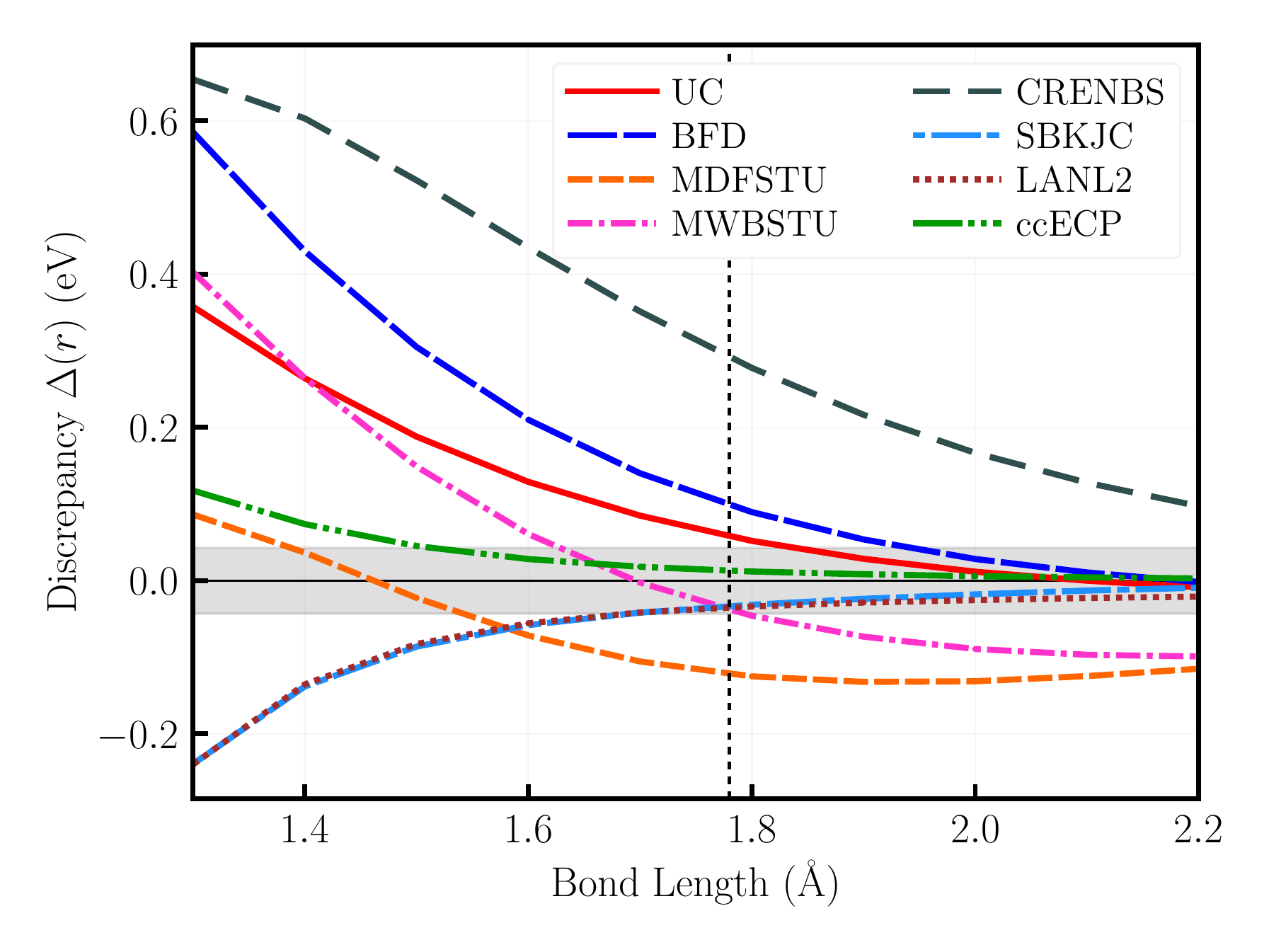}
\caption{BiH ($^3\Sigma$) binding curve discrepancies}
%\label{fig:}
\end{subfigure}%
\begin{subfigure}{0.5\textwidth}
\includegraphics[width=\textwidth]{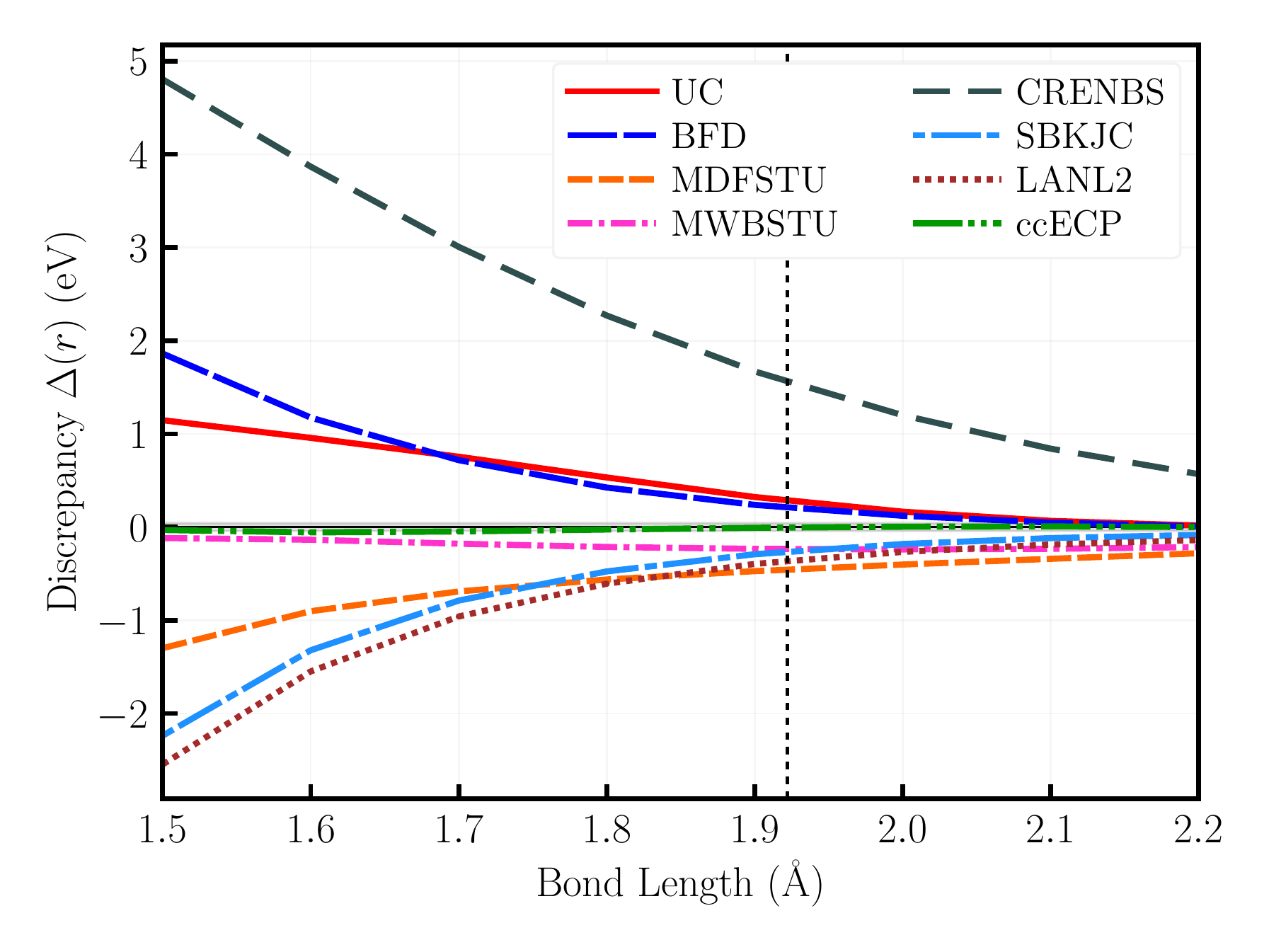}
\caption{BiO ($^2\Pi$) binding curve discrepancies}
%\label{fig:}
\end{subfigure}
\caption{
Binding energy discrepancies for (a) BiH and (b) BiO molecules relative to scalar relativistic AE CCSD(T).
The shaded region indicates the band of chemical accuracy. The dashed vertical line represents the equilibrium geometry.
}
\label{fig:Bi_mols}
\end{figure*}

%For Bi, we choose the core electrons to be [Xe]$4f^{14}5d^{10}$, i.e, the neutral valence occupation is $6s^{2}6p^{3}$.
\fref{fig:Bi_spectrum} shows the Bi atomic spectral errors of all considered core approximations
models considered.
Our ccECP displays the smallest MAD and WMAD of all the approximations, while LMAD is also within chemical accuracy.
Similarly, molecular errors are shown in \fref{fig:Bi_mols} for varying bond lengths. 
In BiH, ccECP results in the smallest errors which are mostly within the chemical accuracy with a slight underbinding exists near the dissociation limit.
For BiH, SBKJC and LANL2 ECPs show competitive errors, however they significantly overbind in BiO molecule with up to 2~eV errors.
On the other hand, ccECP errors in BiO are mostly within the chemical accuracy throughout the whole binding energy curve.
Note that UC severely underbinds in both molecules and shows larger errors in the atom compared to ccECP.
Overall, the data suggest that it might be possible to achieve better accuracy with proper form and optimizations even compared to AE systems with the same active space.

Another point of interest is that in this case the core/valence partitioning is not as clear-cut as for $5d$ elements that include the semi-core $5s,5p$ subshells into the valence space.
Specifically, this requires partitioning of $n=5$ principal quantum number, where $5d^{10}$ is in the core while $5f^{14}$ could be considered semi-core or valence space. Note that this type of
partitioning (i.e., the lowest one-particle eigenvalue does not correspond to $\ell=0$ channel) 
could result in significant errors in transition metals \cite{dolg_relativistic_2012}.
However, the errors seen in this case are similar to what was observed in isovalent elements such as N, P, As \cite{bennett_new_2017, bennett_new_2018, wang_new_2019} with analogous 
core/valence definitions.

\subsubsection{SOREP: Bi}

The SOREP spectral data for Bi is given in \tref{tab:Bi_sorep}.
Our chosen subset of valence states are composed of EA, IP, and single $d$/$f$ excitations. 
The inclusion of single $d$/$f$ excitation states are essential to constrain the spin-orbit splitting bias for corresponding channels. 
%The discrepancies of atomic energy gap for MDFSTU versus ccECP are directly compared to Expt. for FPSODMC/COSCI calculations. 
Overall, we see marginal reduction in MAD and more noticeable improvement in the ground state and in the first IP multiplet splitting.

\begin{figure}[!htbp]
\centering
\includegraphics[width=0.925\columnwidth]{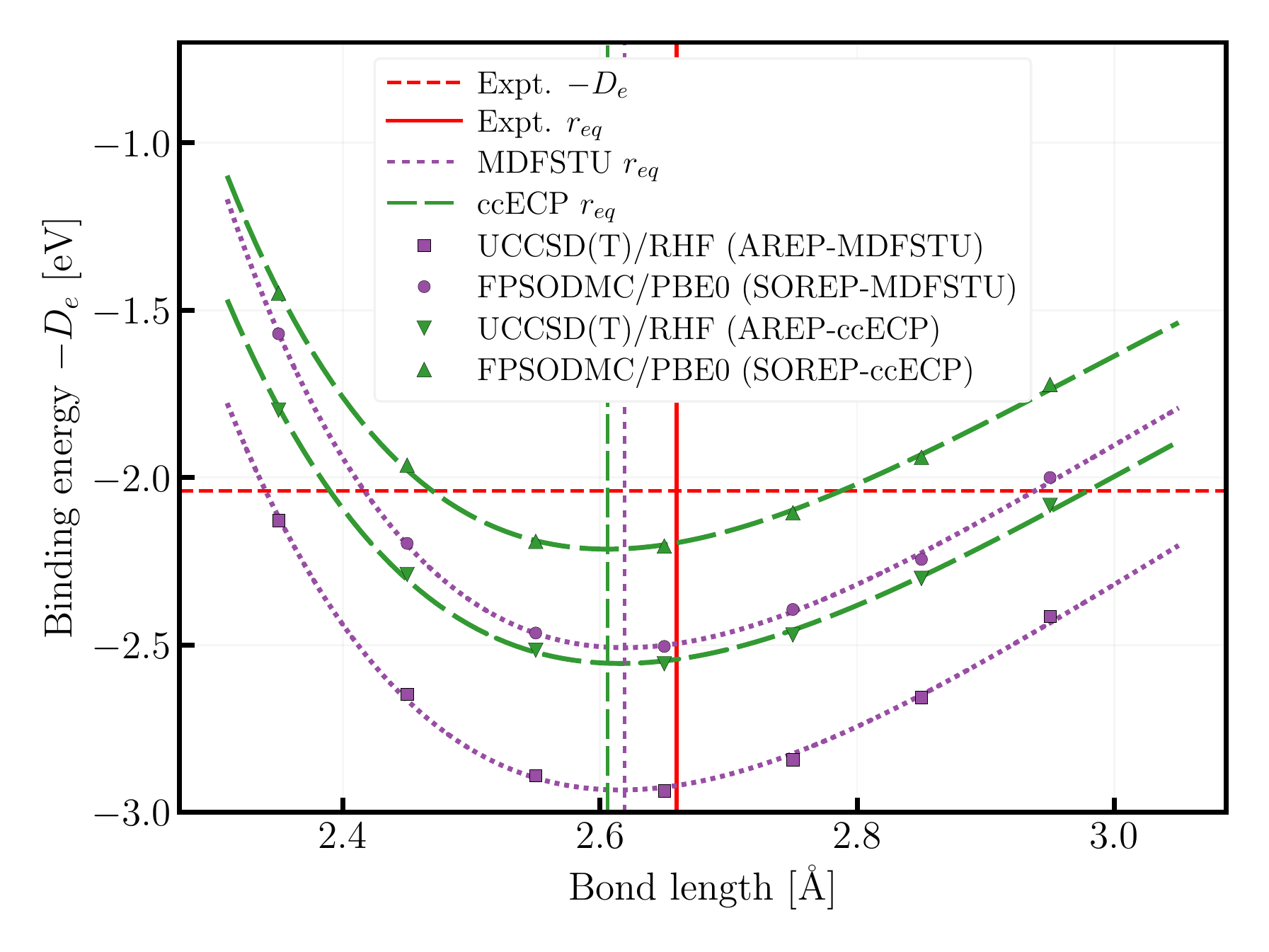}
\label{fig:Bi2}
\caption{Bi$_2$ dimer ($^1\Sigma_g$) binding curve with various methods. }
\label{fig:Bi2_mols}
\end{figure}

\fref{fig:Bi2_mols} shows the bismuth dimer binding curve.
The molecular calculations include AREP CCSD(T) and SOREP FPSODMC using PBE0 trial wave functions for both MDFSTU and ccECP. 
For both ECPs, the two-component spinor FPSODMC calculations show significant alleviation of overbinding that is present in AREP UCCSD(T). 
Our ccECP outperforms MDFSTU in both AREP and SOREP calculations being very close to experiments.

\begin{table}[htbp!]
\setlength{\tabcolsep}{4pt} %% default is 6pt
\small
\centering
\caption{
Bismuth atomic excitation errors for MDFSTU versus ccECP in SOREP forms.
One set of errors are shown for full-relativistic X2C AE gaps using COSCI.
Another set of errors are calculated using FPSODMC and compared to experiments.
All values are in eV.
}
\label{tab:Bi_sorep}
\begin{adjustbox}{width=1.0\columnwidth,center}
%\resizebox{0.97\columwidth}{!}{%
\begin{tabular}{ll|rcc|r|cccc}
\hline\hline
\multirow{2}{*}{State} & \multirow{2}{*}{Term} & \multicolumn{3}{c|}{COSCI} &  \multirow{2}{*}{Expt.} & \multicolumn{2}{c}{FPSODMC} \\
\cline{3-5}
\cline{7-8}
& & AE & STU & ccECP &  & STU & ccECP \\
\hline
$6s^26p^3$ & $^{4}S_{3/2}$ &    0.000 &     0.000 &       0.000 &  0.000 &      0.0000 &        0.0000 \\  
$6s^26p^4$ & $^{3}P_{2}$   &    0.034 &    -0.062 &       0.008 &  0.942 &    -0.04(1) &       0.08(1) \\  
$6s^26p^2$ & $^{3}P_{0}$   &    6.663 &    -0.079 &      -0.113 &  7.285 &     0.03(1) &       0.10(1) \\  
$6s^26p^1$ & $^{2}P_{1/2}$ &   22.621 &    -0.052 &      -0.166 & 23.988 &     0.33(1) &       0.26(1) \\  
\hline
$6s^26p^3$ & $^{4}S_{3/2}$ &    0.000 &     0.000 &       0.000 &  0.000 &      0.0000 &        0.0000 \\  
        {} & $^{2}D_{3/2}$ &    1.549 &     0.007 &      -0.036 &  1.415 &     0.07(1) &       0.03(2) \\  
        {} & $^{2}D_{5/2}$ &    2.140 &     0.011 &      -0.047 &  1.914 &     0.07(1) &       0.04(2) \\  
        {} & $^{2}P_{1/2}$ &    3.110 &     0.002 &      -0.093 &  2.685 &     0.08(1) &      -0.02(2) \\  
        {} & $^{2}P_{3/2}$ &    4.488 &     0.060 &      -0.050 &  4.111 &     0.11(1) &       0.06(2) \\  
\hline
$6s^26p^2$ & $^{3}P_{0}$   &    0.000 &     0.000 &       0.000 &  0.000 &      0.0000 &        0.0000 \\  
        {} & $^{3}P_{1}$   &    1.533 &     0.064 &       0.050 &  1.652 &     0.14(1) &       0.01(2) \\  
        {} & $^{3}P_{2}$   &    2.146 &     0.059 &       0.019 &  2.111 &     0.10(1) &      -0.03(2) \\  
        {} & $^{1}D_{2}$   &    4.307 &     0.126 &       0.041 &  4.207 &     0.21(1) &       0.01(2) \\  
        {} & $^{1}S_{0}$   &    5.944 &     0.095 &      -0.067 &  5.476 &     0.18(1) &       0.05(2) \\  
\hline
$6s^26p^1$ & $^{2}P_{1/2}$ &    0.000 &     0.000 &       0.000 &  0.000 &      0.0000 &        0.0000 \\  
        {} & $^{2}P_{3/2}$ &    2.597 &     0.103 &       0.057 &  2.577 &     0.12(1) &       0.10(1) \\  
\hline
$6s^25d^1$ & $^{2}D_{3/2}$ &    0.000 &     0.000 &       0.000 &  0.000 &      0.0000 &        0.0000 \\  
        {} & $^{2}D_{5/2}$ &    0.178 &     0.020 &      -0.000 &  0.780 &     0.61(1) &       0.53(2) \\  
\hline
$6s^25f^1$ & $^{2}F_{7/2}$ &    0.000 &     0.000 &       0.000 &  0.000 &      0.0000 &        0.0000 \\  
        {} & $^{2}F_{5/2}$ &    0.011 &     0.004 &      -0.002 &  0.012 &     0.03(1) &      -0.03(1) \\  
\hline
MAD        &               &          &     0.053 &       0.053 &         &    0.15(1) &      0.09(2) \\  
\hline\hline
\end{tabular}
\end{adjustbox}
\end{table}

\subsection{Silver (Ag)}

\subsubsection{AREP: Ag}

\begin{figure}[!htbp]
\centering
\includegraphics[width=1.00\columnwidth]{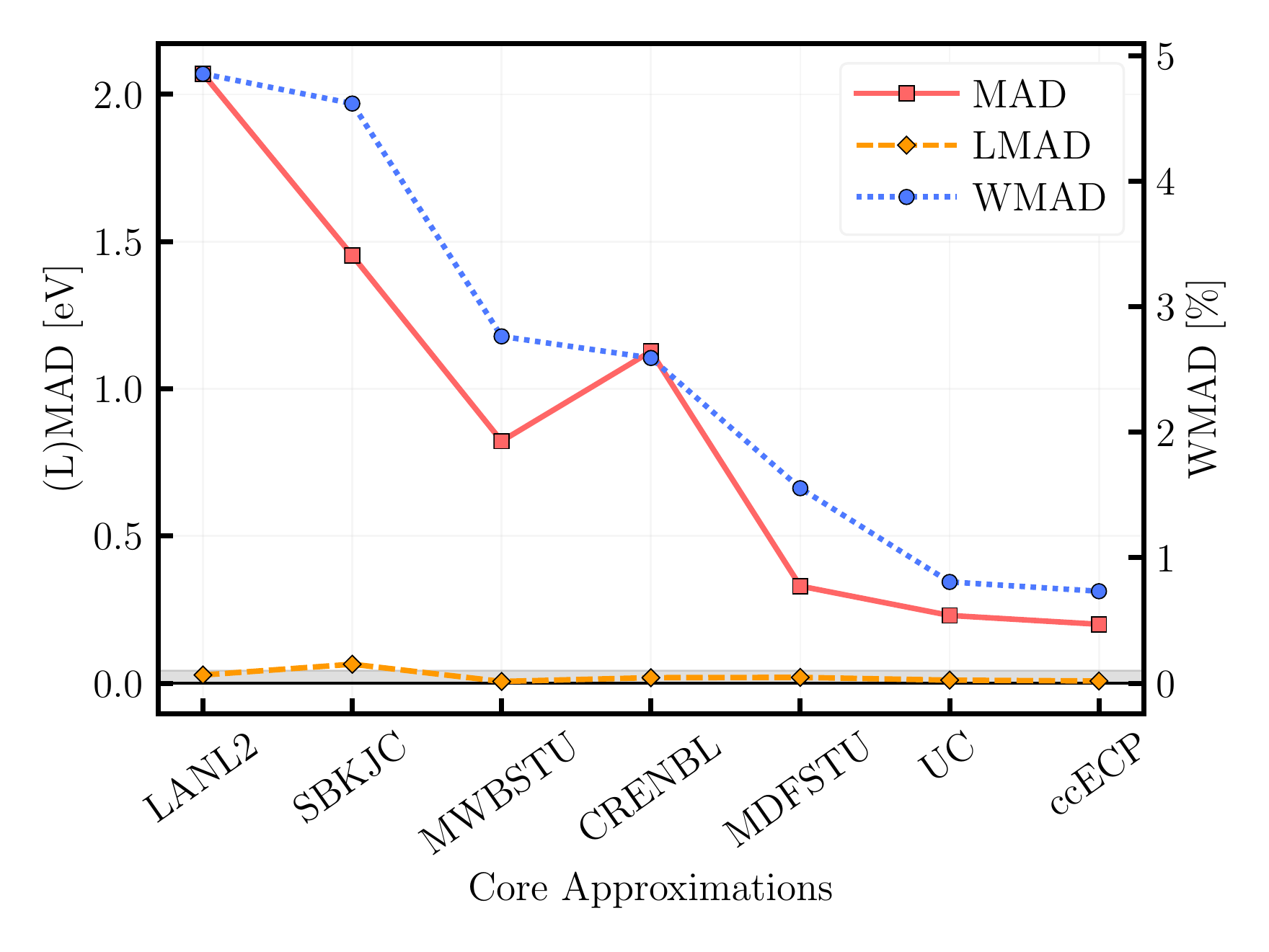}
\caption{
Ag scalar relativistic AE gap (W/L)MADs for various core approximations. 
\mbox{[core] = [Ar]$3d^{10}$} (28 electrons).
RCCSD(T) method with unc-aug-cc-pwCV5Z basis set was used.
}
\label{fig:Ag_spectrum}
\end{figure}

\begin{figure*}[!htbp]
\centering
\begin{subfigure}{0.5\textwidth}
\includegraphics[width=\textwidth]{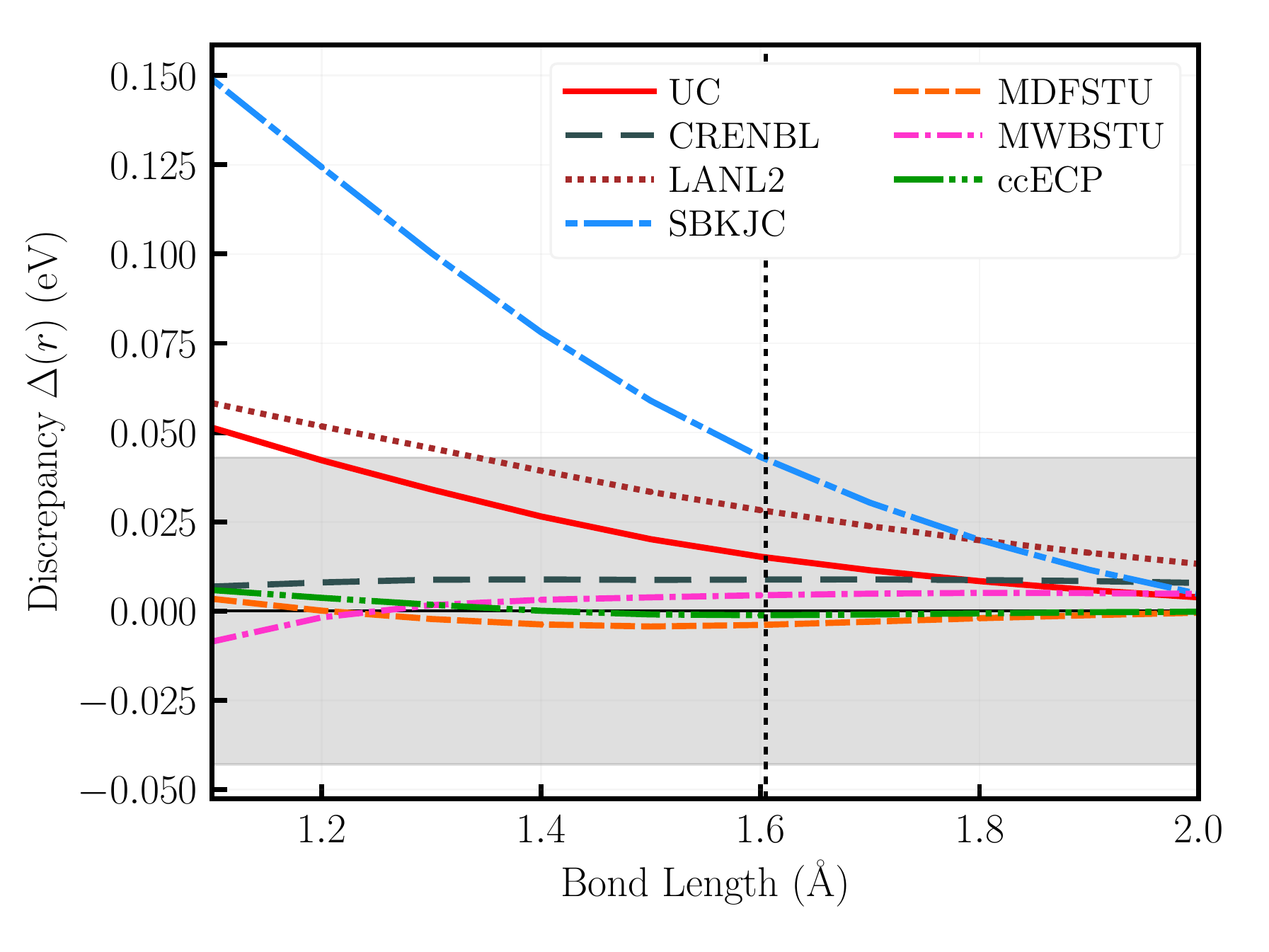}
\caption{AgH ($^1\Sigma$) binding curve discrepancies}
%\label{fig:}
\end{subfigure}%
\begin{subfigure}{0.5\textwidth}
\includegraphics[width=\textwidth]{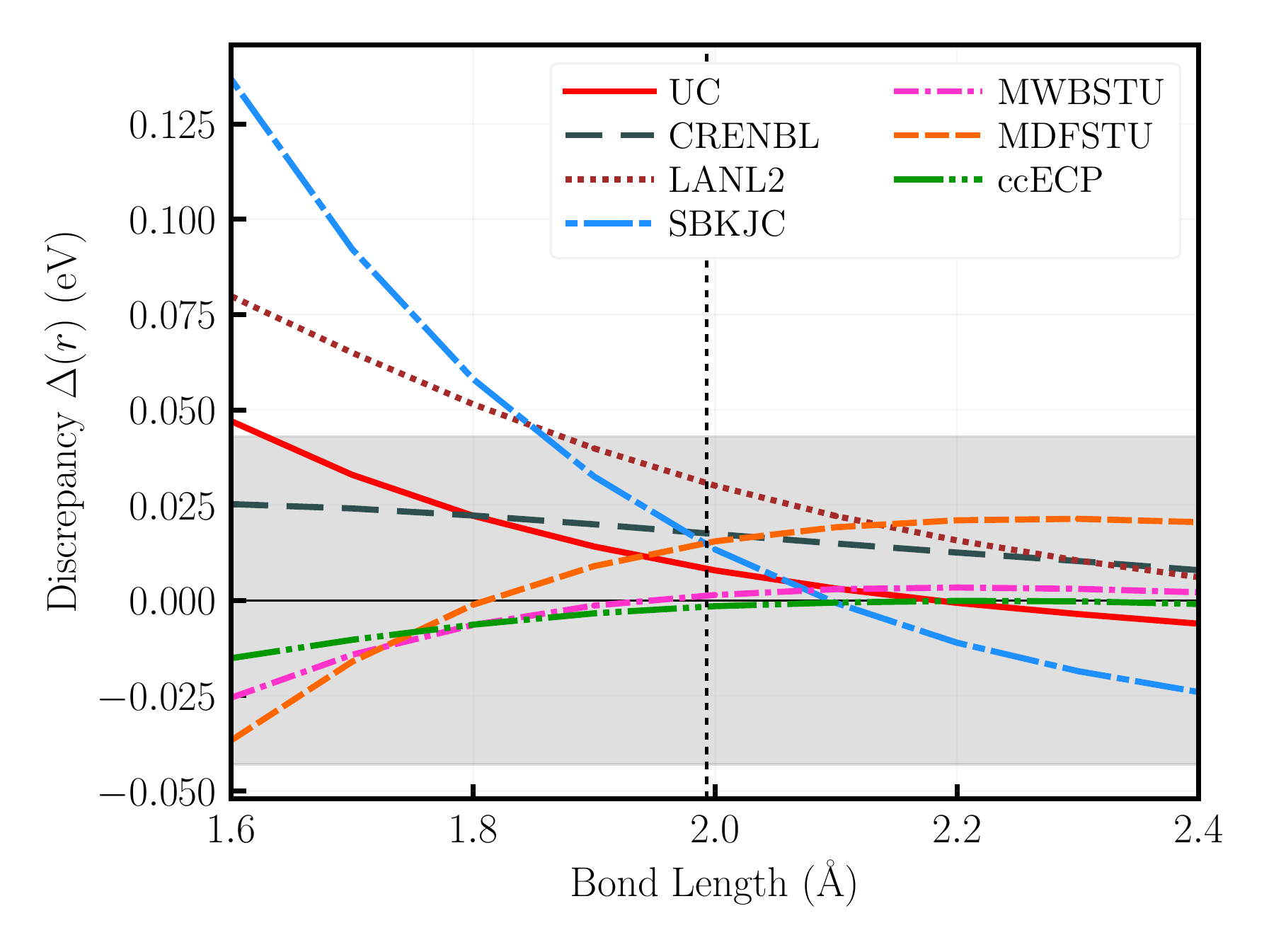}
\caption{AgO ($^2\Pi$) binding curve discrepancies}
%\label{fig:}
\end{subfigure}
\caption{
Binding energy discrepancies for (a) AgH and (b) AgO molecules relative to scalar relativistic AE CCSD(T).
The shaded region indicates the band of chemical accuracy. The dashed vertical line represents the equilibrium geometry.
}
\label{fig:Ag_mols}
\end{figure*}

For silver, the averaged relativistic atomic and molecular results are shown in \fref{fig:Ag_spectrum} and \fref{fig:Ag_mols}. 
%The ECP is designated as [Kr]$3d^{10}$ core with electrons in quantum number $n \leq 3$ shells replaced and the ground state valence space is  $4s^{2}4p^{6}4d^{10}5s^{1}$.
%Our ccECP shows comparable LMAD with other core approximations due to the simple closed $d$ shell.
All ECPs show quantitatively good accuracy for LMAD due to the simple closed $d$ shell.
If we consider broader states in spectrum, MAD and WMAD reveal significant improvement achieved comparing to other ECPs.
In \fref{fig:Ag_mols}, AgH is quite well described by most ECPs and ccECP is among the best ones. 
Improvement is more noticeable in AgO, though most core approximations maintain the accuracy for all geometries, ccECP and MWBSTU show the best performance with almost perfect agreement with AE binding energies.
Note that in both molecules, ccECP is closest to AE results while keeping the discrepancies flat throughout, therefore, it provide the highest accuracy of all core approximations.

\subsubsection{SOREP: Ag}

The SOREP atomic and molecular results are shown in \tref{tab:Ag_sorep}, \tref{tab:Ag_ccsd}. 
The COSCI agreement of MDFSTU and ccECP with AE gaps is remarkably good, with MAD less than 0.01 eV. 
The closed-shell electronic configurations lead to Ag being a special case resulting in a single determinant COSCI wavefunction even in spin-orbit relativistic REL-CCSD(T) calculations (Au is a similar case, too).
This enables us to perform explicit spin-orbit relativistic REL-CCSD(T) calculations for the lowest charged states with results collected in \tref{tab:Ag_ccsd}.
We observe the gaps errors smaller than chemical accuracy when compared to experimental data for both ECPs.

For the same states we carry out FPSODMC calculations with single-reference
trial functions in order to probe for the corresponding fixed-phase biases (Table \ref{tab:Ag_sorep}). Here we see discrepancies  
 from 0.1 eV up to 0.34 eV for the highest state. These types of errors are not unexpected 
 for single-reference due to increased mixing of higher excitations which 
 results from lowering the 
 symmetry from $LS$-coupling to $J$-coupling. We verified this argument in the part 
 devoted to tungsten where we constructed trial functions based on medium size Configuration Interaction (CI) expansions and we observed corresponding diminishing of fixed-phase biases.
On the other hand, the FPSODMC method shows more favorable results of meV bias in multiplet splittings where REL-CCSD(T) method proved to be problematic in DIRAC code.
This demonstrates the quality level in dealing with spin-orbit splittings with the developed fixed-phase method \cite{melton_spin-orbit_2016}.
Therefore we employed J-MAD that includes only bias from spin-orbit splitting states for FPSODMC accuracy assessment from Expt. values.
Obviously, the J-MAD does not represent the ultimate accuracy but gives reasonable approximation for accuracy of SO terms.

\begin{table}[!htbp]
\setlength{\tabcolsep}{4pt} %% default is 6pt
\small
\centering
\caption{
Silver atomic excitation errors for MDFSTU versus ccECP in SOREP forms.
One set of errors are shown for full-relativistic X2C AE gaps using COSCI.
Another set of errors are calculated using FPSODMC and compared to experiments.
All values are in eV.
}
\label{tab:Ag_sorep}
\begin{adjustbox}{width=1.0\columnwidth,center}
%\resizebox{0.97\textwidth}{!}{%
\begin{tabular}{ll|rcc|r|cccc}
\hline\hline
\multirow{2}{*}{State} & \multirow{2}{*}{Term} & \multicolumn{3}{c|}{COSCI} &  \multirow{2}{*}{Expt.} & \multicolumn{2}{c}{FPSODMC}  \\
\cline{3-5}
\cline{7-8}
& & AE & STU & ccECP &  & STU & ccECP \\
%State & Term & COSCI/AE & COSCI/STU & COSCI/ccECP &  Expt. & FPSODMC/STU & FPSODMC/ccECP & REP-CCSD(T)/STU & REP-CCSD(T)/ccECP \\
\hline
$4d^{10}5s^1$  &  $^{2}S_{1/2}$  &     0.000 &     0.000 &       0.000 &  0.000 &       0.000 &         0.000  \\
$4d^{10}5s^2$  &  $^{1}S_{0}$    &    -0.117 &     0.000 &       0.001 & -1.304 &    -0.35(1) &      -0.27(3)  \\
$4d^{10}5p^1$  &  $^{2}P_{1/2}$  &     3.013 &    -0.001 &       0.002 &  3.664 &     0.06(1) &       0.12(3)  \\
$4d^{10}$      &  $^{1}S_{0}$    &     6.340 &    -0.002 &      -0.002 &  7.576 &     0.19(1) &       0.28(3)  \\
$4d^{9}$       &  $^{2}D_{5/2}$  &    25.999 &    -0.006 &       0.020 & 29.061 &    -0.28(1) &       0.34(3)  \\
\hline                                                                               
$4d^{10}5p^1$  &  $^{2}P_{1/2}$  &     0.000 &     0.000 &       0.000 &  0.000 &       0.000 &         0.000  \\
           {}  &  $^{2}P_{3/2}$  &     0.074 &     0.001 &       0.002 &  0.114 &     0.01(1) &      -0.01(3)  \\
\hline                                                                               
$4d^{9}$       &  $^{2}D_{5/2}$  &     0.000 &     0.000 &       0.000 &  0.000 &       0.000 &         0.000  \\
           {}  &  $^{2}D_{3/2}$  &     0.571 &     0.030 &       0.011 &  0.571 &     0.01(1) &       0.03(3)  \\
\hline                                                                               
\multicolumn{2}{l|}{J-MAD}    &           &           &             &        &   0.01(1)   &       0.02(3)  \\
\hline                                                                                        
MAD            &                 &           &     0.007 &       0.006 &        &             &                \\
\hline\hline
\end{tabular}
%}
\end{adjustbox}
\end{table}

\begin{table}
\setlength{\tabcolsep}{4pt} %% default is 6pt
\small
\centering
\caption{
Silver atomic excitation errors for MDFSTU versus ccECP in SOREP forms.
The set of errors are calculated using REL-CCSD(T) from DIRAC code and compared to experiments.
All values are in eV.
}
\label{tab:Ag_ccsd}
\begin{tabular}{ll|r|rrrrr}
\hline\hline
\multirow{2}{*}{State} & \multirow{2}{*}{Term} & \multirow{2}{*}{Expt.} & \multicolumn{2}{r}{REP-CCSD(T)} \\
\cline{4-5}
& & & STU & ccECP \\
\hline
$4d^{10}5s^1$  &  $^{2}S_{1/2}$  &     0.000 &         0.000 &             0.000 \\
$4d^{10}5s^2$  &  $^{1}S_{0}$    &    -1.304 &         0.005 &             0.002 \\
$4d^{10}5p^1$  &  $^{2}P_{1/2}$  &     3.664 &        -0.029 &            -0.024 \\
$4d^{10}$      &  $^{1}S_{0}$    &     7.576 &        -0.034 &            -0.027 \\
$4d^{9}$       &  $^{2}D_{5/2}$  &    29.061 &         0.007 &             0.031 \\
\hline                                          
\multicolumn{2}{l|}{CC-MAD}        &           &         0.015 &             0.017 \\
\hline\hline
\end{tabular}
\end{table}

\subsection{Gold (Au)}
\subsubsection{AREP: Au}

\begin{figure}[!htbp]
\centering
\includegraphics[width=1.00\columnwidth]{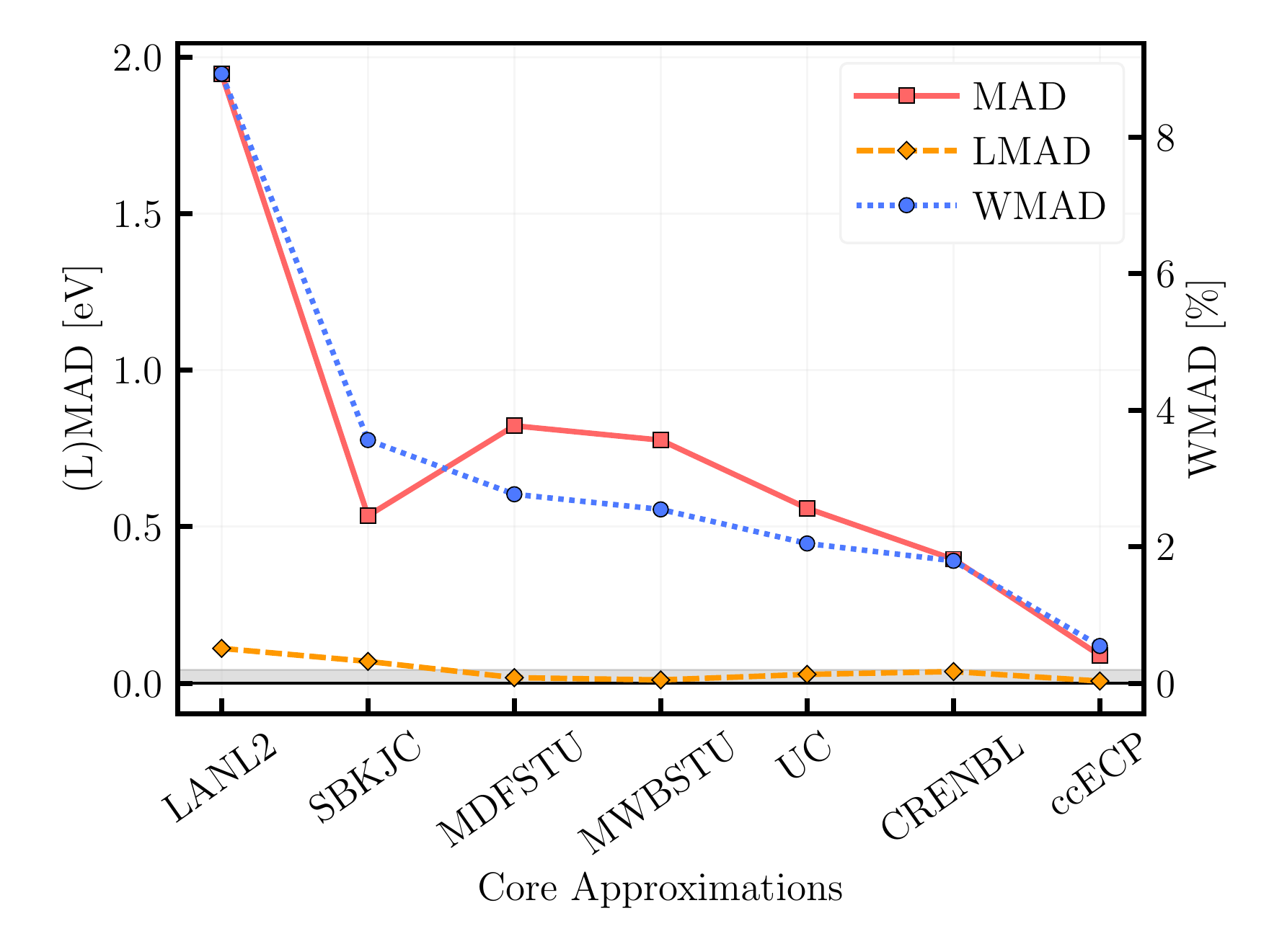}
\caption{
Au scalar relativistic AE gap (W/L)MADs for various core approximations. 
\mbox{[core] = [Kr]$4d^{10}4f^{14}$} (60 electrons).
RCCSD(T) method with unc-aug-cc-pwCVTZ basis set was used.
}
\label{fig:Au_spectrum}
\end{figure}

\begin{figure*}[!htbp]
\centering
\begin{subfigure}{0.5\textwidth}
\includegraphics[width=\textwidth]{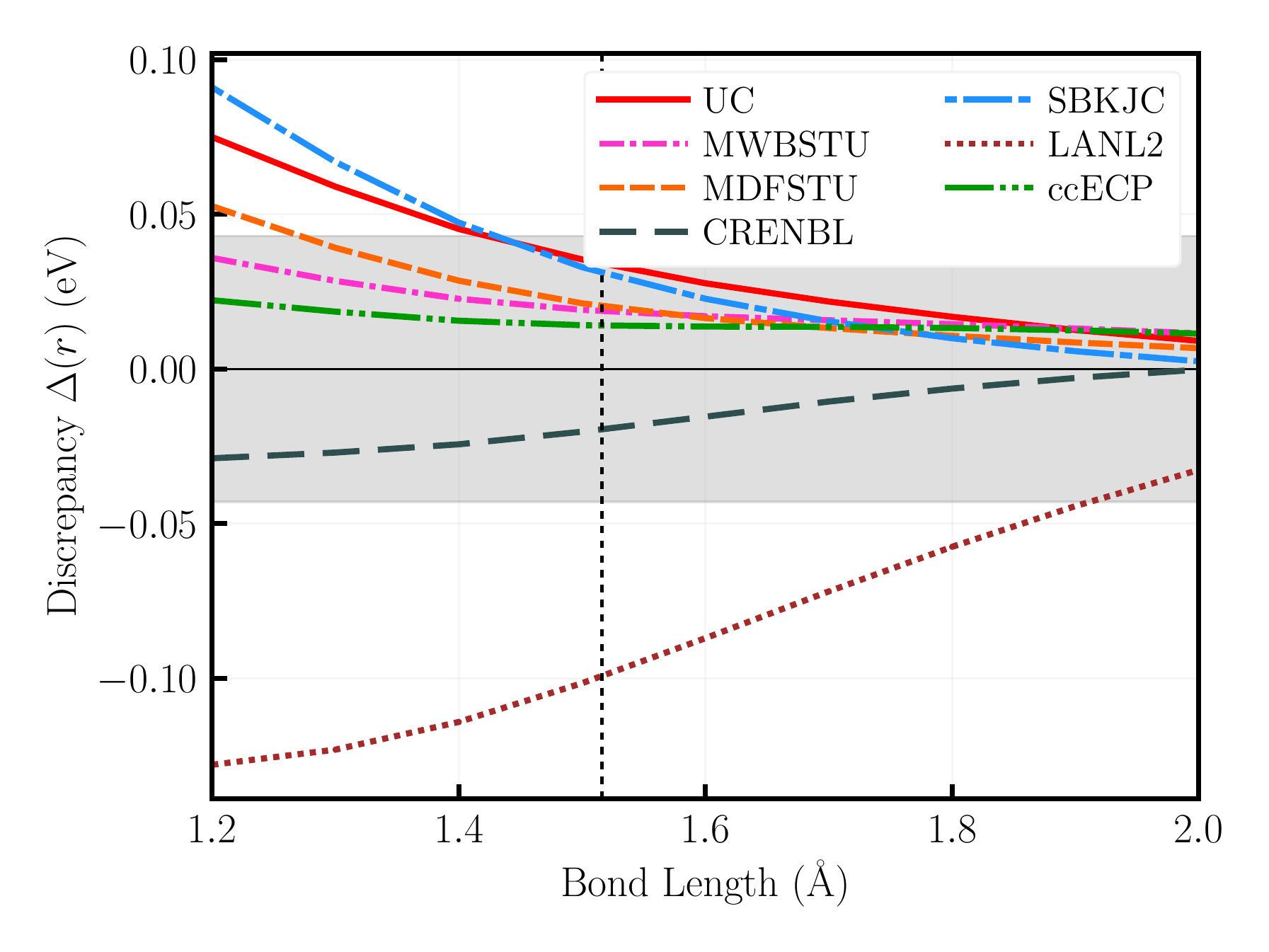}
\caption{AuH ($^1\Sigma$) binding curve discrepancies}
%\label{fig:}
\end{subfigure}%
\begin{subfigure}{0.5\textwidth}
\includegraphics[width=\textwidth]{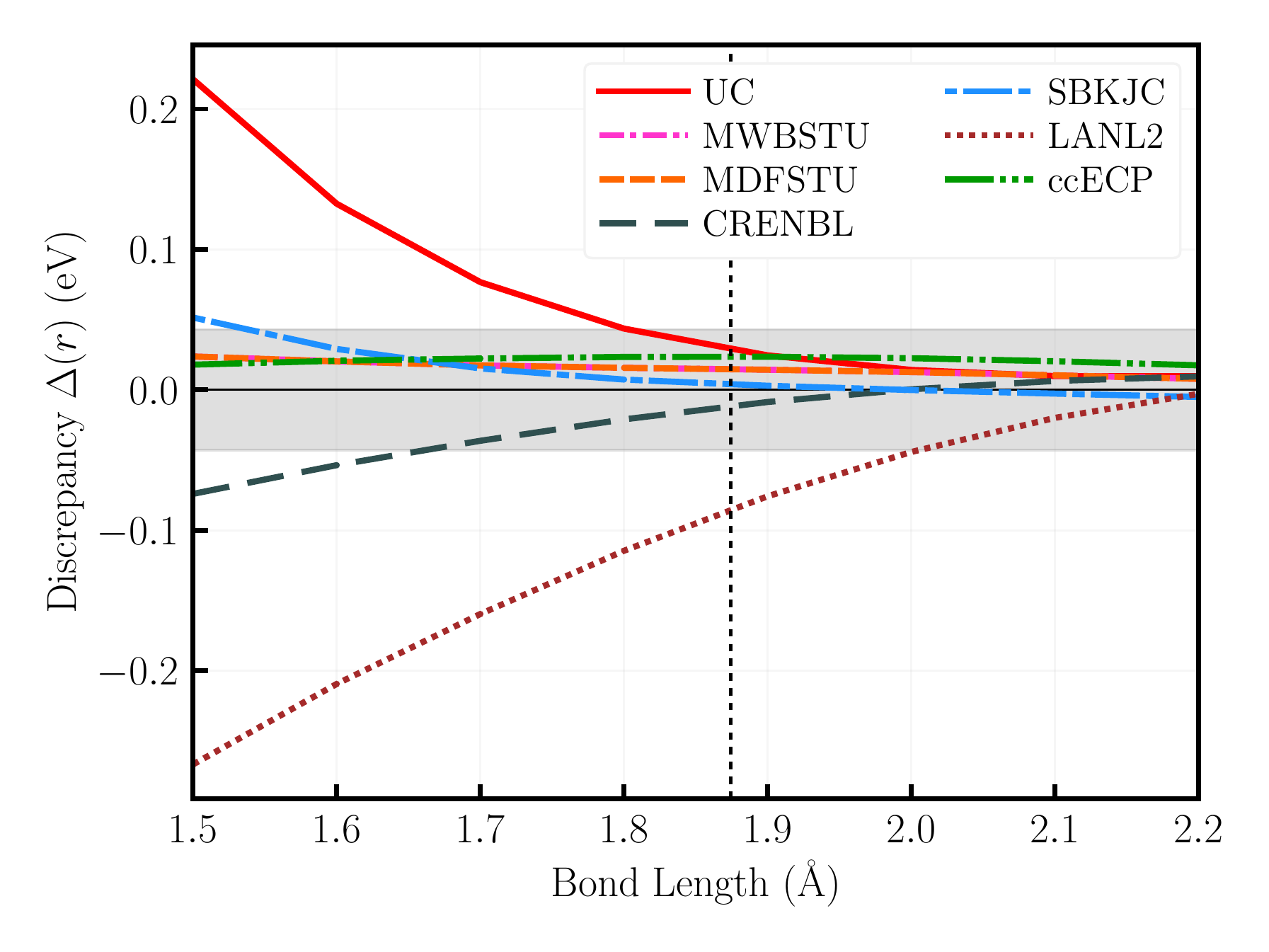}
\caption{AuO ($^2\Pi$) binding curve discrepancies}
%\label{fig:}
\end{subfigure}
\caption{
Binding energy discrepancies for (a) AuH and (b) AuO molecules relative to scalar relativistic AE CCSD(T).
The shaded region indicates the band of chemical accuracy. The dashed vertical line represents the equilibrium geometry.
}
\label{fig:Au_mols}
\end{figure*}

\fref{fig:Au_spectrum} shows the spectral errors for each Au ECP investigated in this work. 
Our ccECP far outperforms the others in MAD and WMAD, with the LMAD remaining well within the chemical accuracy. 
For the molecular binding energy curves in \fref{fig:Au_mols}, our  ccECP remained well within chemical accuracy over the range of geometries tested. 
Specifically, for AuH our ccECP has the lowest discrepancy from the equilibrium bond length to the most compressed geometry we tested, whereas most other ECPs show pronounced underbinding.
For AuO, the ccECP performs consistently with the discrepancy remaining very small at all bond lengths.
In both molecules and in the atomic spectrum, the UC approximation is significantly outperformed by our ccECP.

\subsubsection{SOREP: Au}

The atomic data for Au MDFSTU and ccECP is listed in Tables \ref{tab:Au_sorep}, \ref{tab:Au_sorep_CC}. 
In Table \ref{tab:Au_sorep} we observe that for multiplet splittings both ECPs show close performance using FPSODMC approach.
For the atomic calculations, we provide additional REP-CCSD(T) charged states for both ECPs (Table \ref{tab:Au_sorep_CC}).
Though FPSODMC results for some charged states show biases of $\approx$ 0.1 - 0.3 eV  
due to limits of single-reference trial functions, REL-CCSD(T) calculations exhibit uniform consistency almost fully within the chemical accuracy bounds.
The remarkable agreement of REL-CCSD(T) calculations with experimental excitations demonstrate the quality for both ECPs.  The scattered FPSODMC biases for charged states that result from varying mixing of higher excitations of the same symmetry clearly require more thorough study with trial functions that include sufficiently large active spaces.
Overall, ccECP and MDFSTU show similar excellent accuracy for both charged excitatons and for multiplet splittings.

\begin{figure}[!htbp]
\centering
\includegraphics[width=0.925\columnwidth]{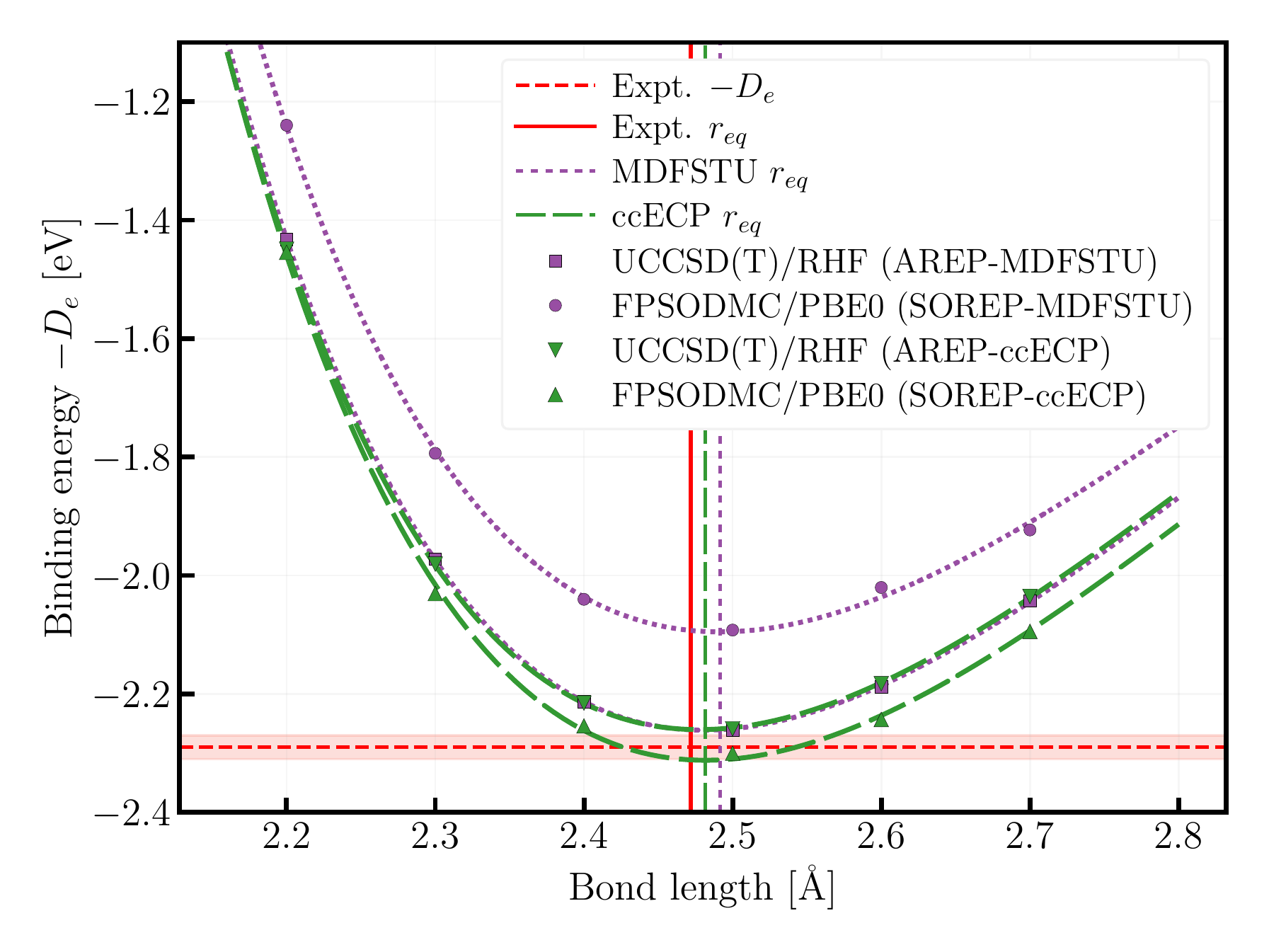}
\caption{Au$_2$ dimer ($^1\Sigma_g$) binding curve with various methods. The band shows the available experimental uncertainty which is 0.02 eV.}
\label{fig:Au2_mols}
\end{figure}

Further tests were carried out in Au$_2$ dimer, Figure \ref{fig:Au2_mols}.
Here, we find similar and satisfying agreement with experiment for equilibrium bond length and also binding energy using AREP-CCSD(T) method.
However, for ccECP the ultimate accuracy manifests in full SOREP setting. Note that
STU underbinds the dimer by about 0.25 eV while ccECP improves slightly the bond length equlibrium and the binding energy with very negligible constant overbinding.

\begin{table}
\setlength{\tabcolsep}{4pt} %% default is 6pt
\small
\centering
\caption{
Gold atomic excitation errors for MDFSTU versus ccECP in SOREP forms.
One set of errors are shown for full-relativistic X2C AE gaps using COSCI.
Another set of errors are calculated using FPSODMC and compared to experiments.
All values are in eV.
}
\label{tab:Au_sorep}
\begin{adjustbox}{width=1.0\columnwidth,center}
%\resizebox{0.97\textwidth}{!}{%
\begin{tabular}{ll|rcc|r|cccc}
\hline\hline
\multirow{2}{*}{State} & \multirow{2}{*}{Term} & \multicolumn{3}{c|}{COSCI} &  \multirow{2}{*}{Expt.} & \multicolumn{2}{c}{FPSODMC}  \\
\cline{3-5}
\cline{7-8}
& & AE & STU & ccECP &  & STU & ccECP \\
\hline
$5d^{10}6s^1$  &  $^{2}S_{1/2}$  &    0.000 &     0.000 &       0.000 &  0.000 &       0.000 &         0.000  \\
$5d^{10}6s^2$  &  $^{1}S_{0}$    &   -0.648 &     0.016 &       0.022 & -2.310 &    -0.35(4) &      -0.22(3)  \\
$5d^{9}6s^2$   &  $^{2}D_{5/2}$  &    1.284 &    -0.040 &      -0.038 &  1.140 &    -0.25(4) &      -0.14(3)  \\
$5d^{10}6p^1$  &  $^{2}P_{1/2}$  &    4.024 &     0.008 &       0.003 &  4.630 &    -0.10(3) &      -0.10(3)  \\
$5d^{10}$      &  $^{1}S_{0}$    &    7.704 &     0.026 &       0.005 &  9.230 &     0.26(4) &       0.25(3)  \\
$5d^{9}$       &  $^{2}D_{5/2}$  &   26.266 &     0.026 &      -0.033 & 29.430 &     0.15(3) &       0.16(3)  \\
\hline
$5d^{9}6s^2$   &  $^{2}D_{5/2}$  &    0.000 &     0.000 &       0.000 &  0.000 &       0.000 &         0.000  \\
           {}  &  $^{2}D_{3/2}$  &    1.480 &     0.061 &      -0.020 &  1.520 &     0.02(3) &      -0.06(3)  \\
\hline
$5d^{10}6p^1$  &  $^{2}P_{1/2}$  &    0.000 &     0.000 &       0.000 &  0.000 &       0.000 &         0.000  \\
           {}  &  $^{2}P_{3/2}$  &    0.349 &     0.011 &      -0.005 &  0.470 &     0.02(3) &       0.00(3)  \\
\hline
$5d^{9}$       &  $^{2}D_{5/2}$  &    0.000 &     0.000 &       0.000 &  0.000 &       0.000 &         0.000  \\
           {}  &  $^{2}D_{3/2}$  &    1.558 &     0.071 &      -0.010 &        &             &                \\
\hline
$5d^{10}5f^1$  &  $^{2}F_{7/2}$  &    0.000 &     0.000 &       0.000 &  0.000 &       0.000 &         0.000  \\
           {}  &  $^{2}F_{5/2}$  &    0.103 &     0.018 &       0.008 &  0.000 &    -0.09(4) &      -0.05(4)  \\
\hline
\multicolumn{2}{l|}{J-MAD}   &          &           &             &        &     0.05(3) &       0.03(3)  \\
\hline
MAD            &                &          &     0.031 &       0.016 &        &             &                \\
\hline\hline
\end{tabular}
\end{adjustbox}
\end{table}

\begin{table}[!htbp]
\setlength{\tabcolsep}{4pt} %% default is 6pt
\small
\centering
\caption{
Gold atomic excitation errors for MDFSTU versus ccECP in SOREP forms.
The set of errors are calculated using REL-CCSD(T) from DIRAC code and compared to experiments.
All values are in eV.
}
\label{tab:Au_sorep_CC}
\begin{tabular}{ll|r|rrrrr}
\hline\hline
\multirow{2}{*}{State} & \multirow{2}{*}{Term} & \multirow{2}{*}{Expt.} & \multicolumn{2}{r}{REP-CCSD(T)} \\
\cline{4-5}
& & & STU & ccECP \\
\hline
$5d^{10}6s^1$  &  $^{2}S_{1/2}$ &   0.000   &    0.000   &    0.000   \\
$5d^{10}6s^2$  &  $^{1}S_{0}$   &  -2.310   &    0.086   &    0.025   \\
$5d^{9}6s^2$   &  $^{2}D_{5/2}$ &   1.140   &    0.010   &    0.035   \\
$5d^{10}6p^1$  &  $^{2}P_{1/2}$ &   4.630   &   -0.044   &   -0.046   \\
$5d^{10}$      &  $^{1}S_{0}$   &   9.230   &   -0.004   &   -0.025   \\
$5d^{9}$       &  $^{2}D_{5/2}$ &  29.430   &   -0.010   &   -0.052   \\
\hline                                          
\multicolumn{2}{l|}{CC-MAD}        &           &         0.026 &             0.031 \\
\hline\hline
\end{tabular}
\end{table}

\subsection{Tungsten (W)}
\subsubsection{AREP: W}

\begin{figure}[!htbp]
\centering
\includegraphics[width=1.00\columnwidth]{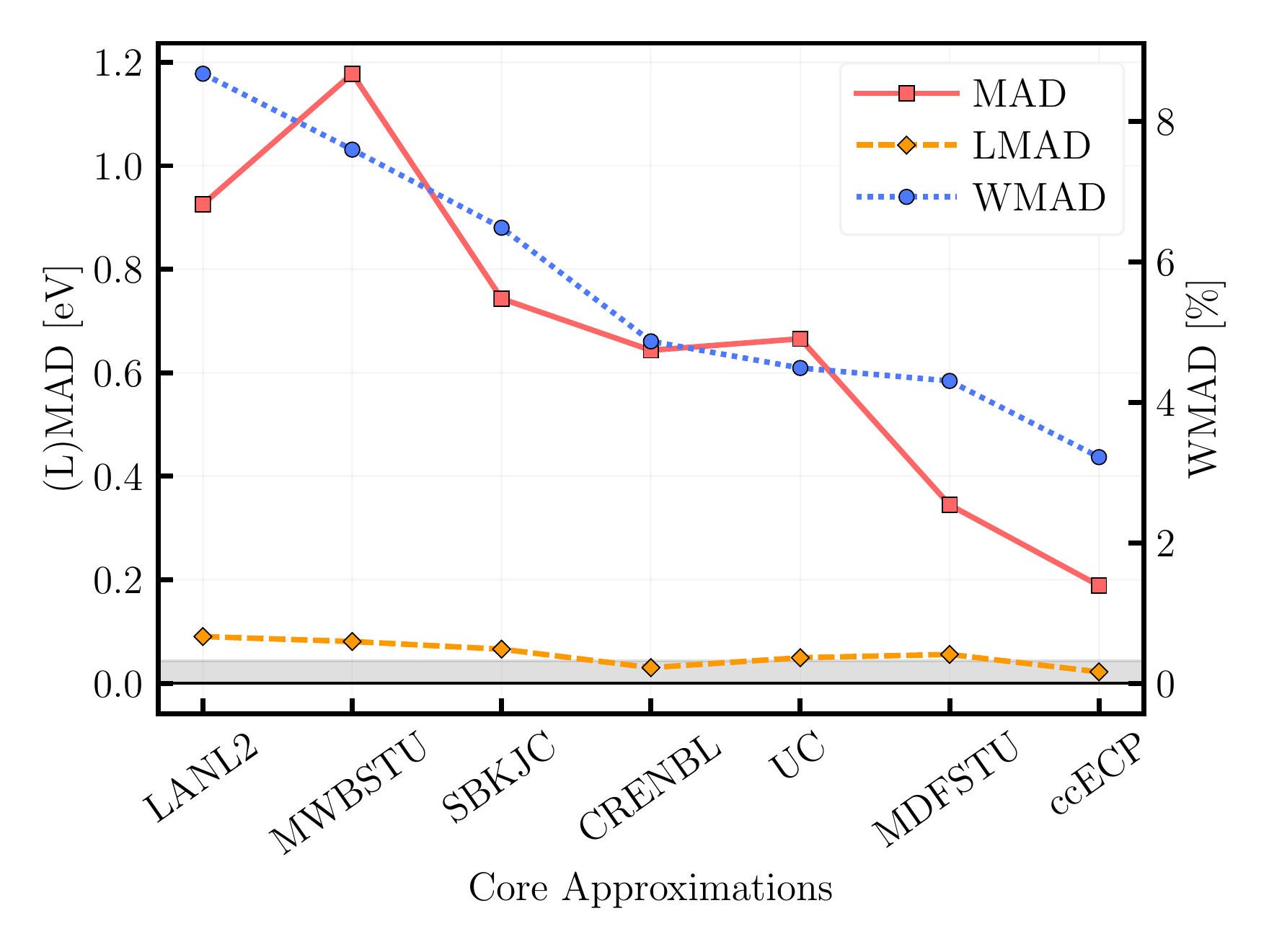}
\caption{
W scalar relativistic AE gap (W/L)MADs for various core approximations. 
\mbox{[core] = [Kr]$4d^{10}4f^{14}$} (60 electrons).
RCCSD(T) method with unc-aug-cc-pwCVTZ basis set was used.
}
\label{fig:W_spectrum}
\end{figure}

\begin{figure*}[!htbp]
\centering
\begin{subfigure}{0.5\textwidth}
\includegraphics[width=\textwidth]{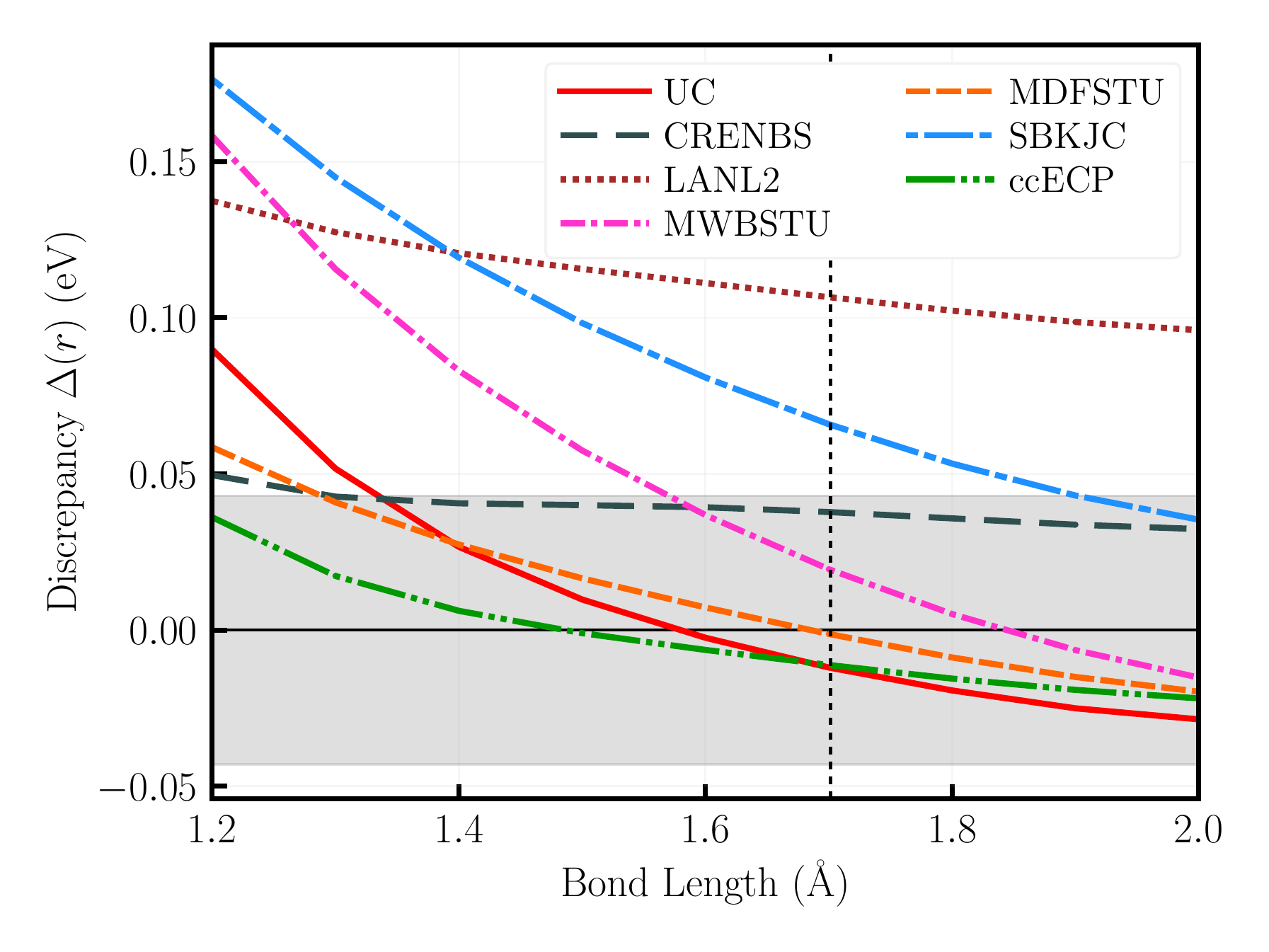}
\caption{WH ($^6\Sigma$) binding curve discrepancies}
%\label{fig:}
\end{subfigure}%
\begin{subfigure}{0.5\textwidth}
\includegraphics[width=\textwidth]{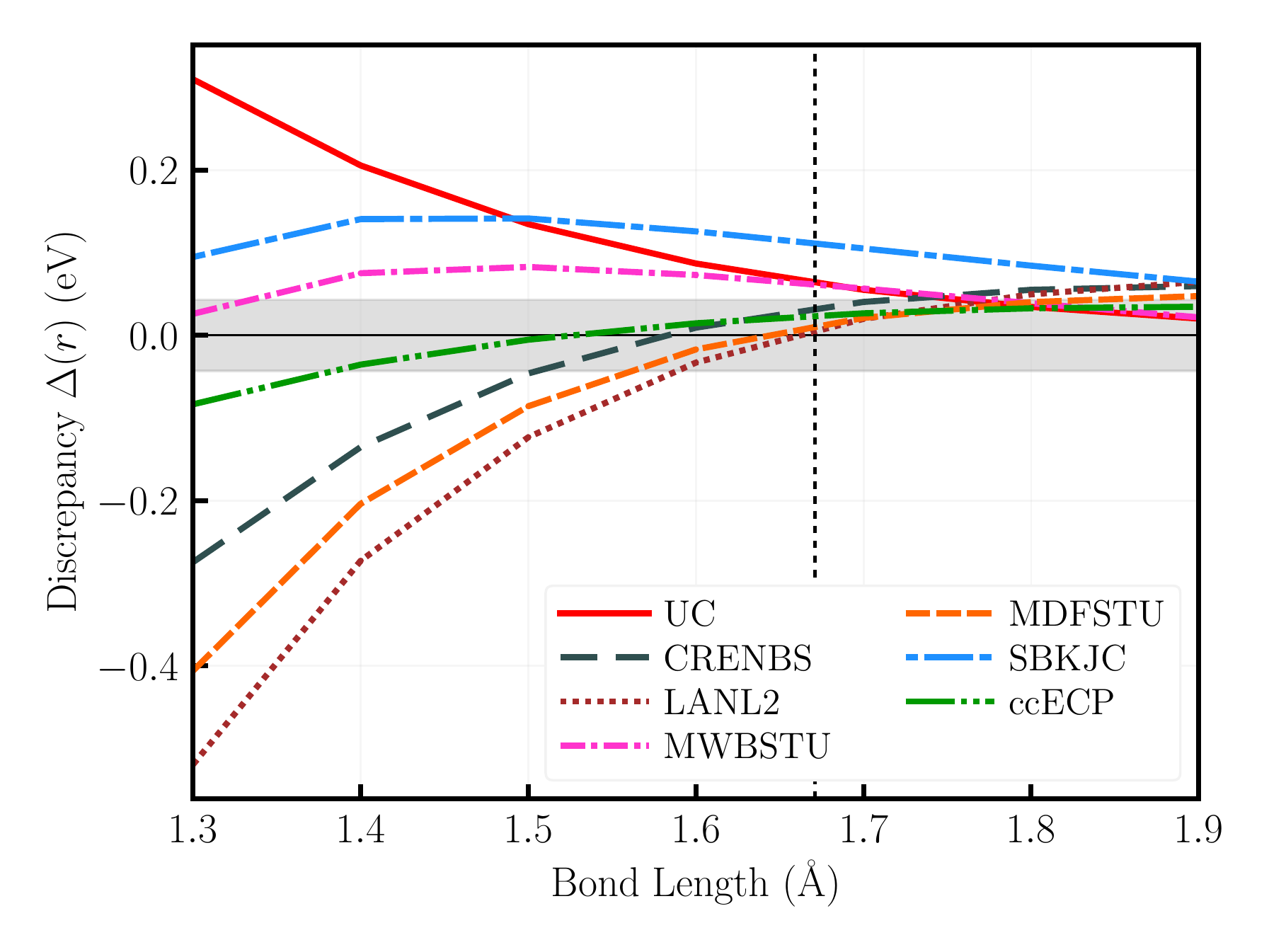}
\caption{WO ($^3\Sigma$) binding curve discrepancies}
%\label{fig:}
\end{subfigure}
\caption{
Binding energy discrepancies for (a) WH and (b) WO molecules relative to scalar relativistic AE CCSD(T).
The shaded region indicates the band of chemical accuracy. The dashed vertical line represents the equilibrium geometry.
}
\label{fig:W_mols}
\end{figure*}

%For the case of tungsten, we adopt the core same as Au and Ir with the electrons within quantum number n=4 reside as the core electrons. 
\fref{fig:W_spectrum} and \ref{fig:W_mols} present the bias of W atomic spectra and discrepancies for W molecular dimers, respectively. 
This is again a case where we show the accuracy of our constructed ccECP to be higher consistently
for atomic and molecular properties than other core approximations included.
%either in atomic or molecular tests.
In \fref{fig:W_spectrum}, we show that the MAD and WMAD are significantly reduced for ccECP.
Also as the metrics of evaluating the low-lying states errors, ccECP LMAD is contained within the chemical accuracy.
The other ECPs do not maintain the errors within the chemical accuracy.
We have also achieved substantial improvement in molecular properties.
In \fref{fig:W_mols}, there is a clear tendency of underbinding of the hydride dimer and overbinding the oxide dimer for all ECPs. 
In WH molecule, our constructed ccECP retains the bias inside the chemical accuracy band for the entire binding curve.
For WO, ccECP also stands out when compared with previously tabulated ECPs and overal provides 
the best balance of accuracy in all tested systems.

\subsubsection{SOREP: W}

Using the SOREP Hamiltonian the tungsten atomic energy gaps are shared in \fref{tab:W_sorep} for MDFSTU versus ccECP.
Our optimization of ccECP reduces MAD to approximately one third of MDFSTU when referenced to COSCI/AE gaps.
Not surprisingly, fixed-phase calculations show encouraging agreement of multiplet splitting gaps with experimental values but significant errors appear in for charged states, which were scrutinized also in constructions for gold and silver.
To further investigate the origin of these errors, we extend our calculations 
to CI expansions and to related DMC calculations of the ground state ($5d^46s^2$, $^{5}D_{0}$) and excited state ($5d^56s^1$, $^{7}S_{3}$) for W ccECP, Table \ref{tab:W_sorep_CC}.
Clearly, we show that in AREP with substantial multi-reference wave functions the gap approaches AREP estimated experimental value.
Similarly, SOREP calculations reveal that applying CI expansions boost the accuracy further towards the experimental value. As it has been observed also previously, both
FPSODMC and CI with restricted COSCI trial function is inadequate and exhibits the incorrect ground state occupancy  $5d^56s^1$.
This shows that both explicit treatment of spin-orbit and accurate correlation is crucial for the atomic spectrum calculations and this in particular is true for $5d$ mid series elements. 
It is reassuring that both
MR-CISD and subsequent FPSODMC/MR-CISD calculations correctly predict the order of these two states.
Therefore, our estimation of FPSODMC gap errors exclude the charged configuration gaps and only keeps the J-splitting gaps in J-MAD.

\begin{figure}[!htbp]
\centering
\includegraphics[width=0.925\columnwidth]{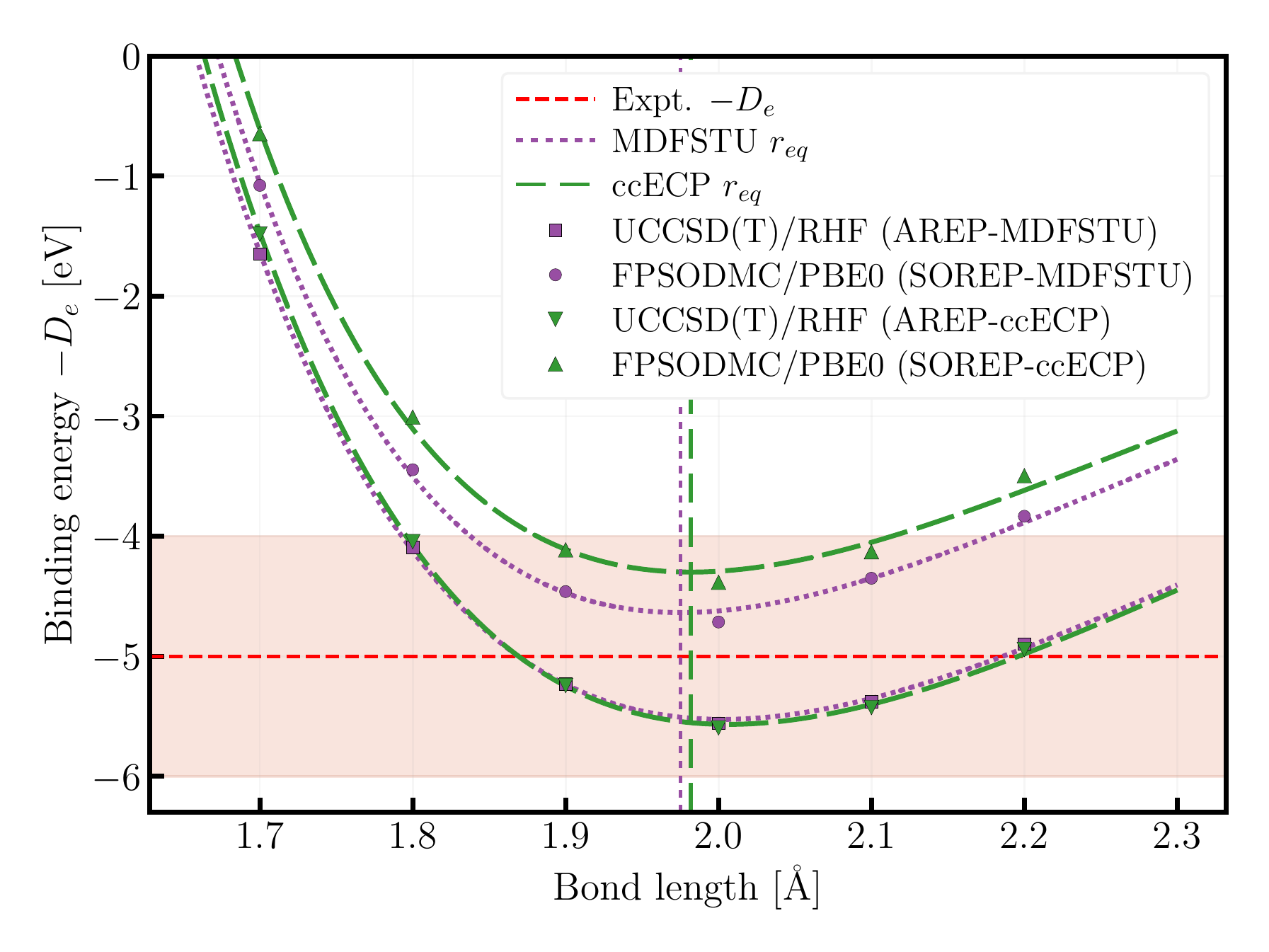}
\label{fig:W2}
\caption{W$_2$ dimer ($^1\Sigma_g$) binding curve with various methods. The band here shows the available experimental binding uncertainty which is 1 eV \cite{morse_clusters_1986}.
}
\label{fig:W2_mols}
\end{figure}

\fref{fig:W2_mols} shows the W dimers in AREP CCSD(T) and two-component FPSODMC calculations.
To the best of our knowledge, an experimental data is lacking here since the most accurate experimental 
value was estimated to be 5(1) eV \cite{morse_clusters_1986}.
Interestingly, we find MDFSTU and ccECP cross validate each other in both AREP and REP calculations with very similar binding curves. 
All of the binding energies for both ECPs are inside the estimations from the previous work.
Both ECPs predict the equilibrium both length near 1.95 \r{A}.

\begin{table}[htbp!]
\setlength{\tabcolsep}{4pt} %% default is 6pt
\small
\centering
\caption{
Tungsten atomic excitation errors for MDFSTU versus ccECP in SOREP forms.
One set of errors are shown for full-relativistic X2C AE gaps using COSCI.
Another set of errors are calculated using FPSODMC and compared to experiments.
All values are in eV.
}
\label{tab:W_sorep}
\begin{adjustbox}{width=1.0\columnwidth,center}
%\resizebox{0.97\columwidth}{!}{%
\begin{tabular}{ll|rcc|r|ccccc}
\hline\hline
\multirow{2}{*}{State} & \multirow{2}{*}{Term} & \multicolumn{3}{c|}{COSCI} &  \multirow{2}{*}{Expt.} & \multicolumn{2}{c}{FPSODMC} \\
\cline{3-5}
\cline{7-8}
& & AE & STU & ccECP &  & STU & ccECP \\
\hline
$5s^25p^65d^{4}6s^2$   &  $^{5}D_{0}$    &    0.000 &     0.000 &       0.000 &    0.000 &    0.00      &   0.00       \\
$5s^25p^65d^{5}6s^2$   &  $^{6}S_{5/2}$  &    0.016 &     0.078 &       0.021 &   -0.815 &    0.14(3)   &   0.01(2)    \\
$5s^25p^65d^{5}6s^1$   &  $^{7}S_{3}$    &   -0.733 &     0.113 &       0.057 &    0.366 &    0.64(4)   &   0.48(2)    \\
$5s^25p^65d^{4}6s^1$   &  $^{6}D_{1/2}$  &    5.850 &     0.031 &      -0.005 &    7.864 &    0.52(3)   &   0.44(2)    \\
$5s^25p^65d^{4}$       &  $^{5}D_{0}$    &   20.685 &     0.020 &      -0.053 &  24.2(2) &     1.1(2)   &   1.0(2)     \\
\hline                                                                                              
$5s^25p^65d^{4}6s^2$   &  $^{5}D_{0}$    &    0.000 &     0.000 &       0.000 &    0.000 &    0.00      &   0.00       \\
          {}           &  $^{5}D_{1}$    &    0.123 &     0.026 &      -0.003 &    0.207 &    0.14(3)   &   0.04(2)    \\
          {}           &  $^{5}D_{2}$    &    0.296 &     0.052 &      -0.008 &    0.412 &    0.11(3)   &   0.03(2)    \\
          {}           &  $^{5}D_{3}$    &    0.486 &     0.072 &      -0.014 &    0.599 &    0.14(3)   &  -0.00(2)    \\
          {}           &  $^{5}D_{4}$    &    0.683 &     0.086 &      -0.021 &    0.771 &    0.16(3)   &  -0.02(2)    \\
\hline                                                                                              
$5s^25p^65d^{4}6s^1$   &  $^{6}D_{1/2}$  &    0.000 &     0.000 &       0.000 &    0.000 &    0.00      &   0.00       \\
          {}           &  $^{6}D_{3/2}$  &    0.116 &     0.020 &      -0.003 &    0.188 &    0.02(3)   &   0.05(2)    \\
          {}           &  $^{6}D_{5/2}$  &    0.273 &     0.041 &      -0.009 &    0.393 &    0.12(3)   &   0.08(2)    \\
          {}           &  $^{6}D_{7/2}$  &    0.454 &     0.061 &      -0.015 &    0.585 &    0.14(3)   &   0.02(2)    \\
          {}           &  $^{6}D_{9/2}$  &    0.648 &     0.076 &      -0.022 &    0.762 &    0.13(3)   &   0.00(2)    \\
\hline                                                                                              
$5s^25p^65d^{4}6s^1$   &  $^{5}D_{0}$    &    0.000 &     0.000 &       0.000 &    0.000 &    0.00      &   0.00       \\
          {}           &  $^{5}D_{1}$    &    0.148 &     0.031 &      -0.004 &    0.280 &    0.18(3)   &   0.10(1)    \\
          {}           &  $^{5}D_{2}$    &    0.348 &     0.060 &      -0.010 &    0.553 &    0.22(3)   &   0.09(2)    \\
          {}           &  $^{5}D_{3}$    &    0.561 &     0.081 &      -0.019 &    0.778 &    0.22(3)   &   0.10(2)    \\
          {}           &  $^{5}D_{4}$    &    0.777 &     0.094 &      -0.027 &    0.953 &    0.19(3)   &   0.05(2)    \\
\hline                                                                                             
$5s^25p^66s^26p^1$     &  $^{2}P_{1/2}$  &    0.000 &     0.000 &       0.000 &          &              &              \\
          {}           &  $^{2}P_{3/2}$  &    1.697 &     0.023 &      -0.054 &          &              &              \\
\hline                                                                                                      
$5s^25p^66s^26f^1$     &  $^{2}F_{5/2}$  &    0.000 &     0.000 &       0.000 &          &              &              \\
          {}           &  $^{2}F_{7/2}$  &    0.008 &     0.003 &      -0.028 &          &              &              \\
\hline                                                                                                      
$5s^25p^5$             &  $^{2}P_{3/2}$  &    0.000 &     0.000 &       0.000 &          &              &              \\
          {}           &  $^{2}P_{1/2}$  &   11.059 &     0.144 &       0.020 &          &              &              \\
\hline         
J-MAD               &                 &          &           &             &          &    0.15(3)   &   0.05(2)    \\
\hline
MAD                    &                 &          &     0.059 &       0.021 &          &              &              \\
\hline\hline
\end{tabular}
\end{adjustbox}
\end{table}

\begin{table}
\centering
\caption{Atomic excitations of W ground state $5d^46s^2$ and excited state $5d^56s^1$ of ccECP pseudoatom using various methods.
Note the improvement of gaps as more accurate methods are used.
``AREP Est. Expt.'' represents the estimated experimental value for AREP gap by weighted $J-$averaging.
}
\label{tab:W_sorep_CC}
\begin{tabular}{lr}
\hline\hline
Method                              & Gap [eV]  \\
\hline
\multicolumn{2}{c}{AREP} \\
\hline
FNDMC/HF        & -0.65(2)  \\   
FNDMC/PBE0      & -0.54(2)  \\
FNDMC/CASCI\footnote{Active space = (30o, 6e$^{-}$) $\approx 10$k dets.} 
                & -0.42(3) \\
CCSD(T)         & -0.31     \\
AREP Est. Expt. & -0.18     \\

\hline
\multicolumn{2}{c}{SOREP} \\
\hline

COSCI           & -0.790    \\
FPSODMC/COSCI   & -0.11(3)  \\
FPSODMC/CISD\footnote{Frozen-core, truncated CISD $\approx 10$k dets.}
                &  0.169    \\
Expt.           &  0.366    \\

\hline\hline
\end{tabular}
\end{table}

\subsection{Palladium (Pd)}
\subsubsection{AREP: Pd}

\begin{figure}[!htbp]
\centering
\includegraphics[width=1.00\columnwidth]{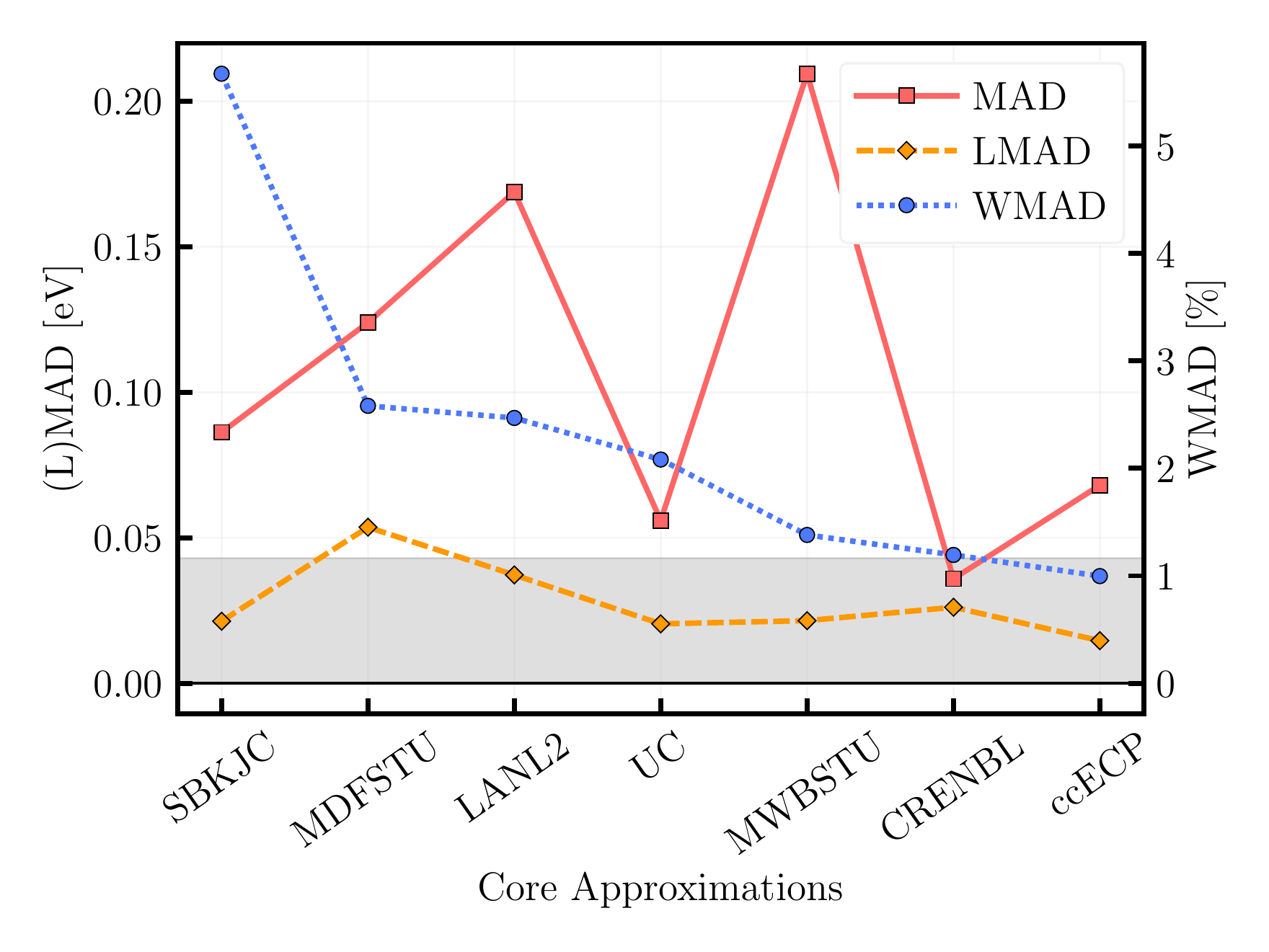}
\caption{
Pd scalar relativistic AE gap (W/L)MADs for various core approximations. 
\mbox{[core] = [Ar]$3d^{10}$} (28 electrons).
RCCSD(T) method with unc-aug-cc-pwCVTZ basis set was used.
}
\label{fig:Pd_spectrum}
\end{figure}

\fref{fig:Pd_spectrum} shows  atomic excitation errors for Pd ccECP.
The Pd ccECP outperforms most of the other ECPs in all metrics with a couple notable exceptions.
The CRENBL ECP had a slight advantage with the raw MAD from all of the states chosen, but the ccECP performed better at replicating the energies at low-lying states and this is demonstrated by the WMAD statistic which weights the errors of smaller gaps more heavily.

For the molecular binding Pd was one of the elements where we used the SEFIT/MEFIT method.
Compared to our initial optimizations using our more conventional spectral fitting method, the SEFIT/MEFIT method led to much better molecular binding curve discrepancies with comparable or unchanged atomic spectrum performance and provided energy curves with greater accuracy than
most contending ECPs.

\begin{figure*}[!htbp]
\centering
\begin{subfigure}{0.5\textwidth}
\includegraphics[width=\textwidth]{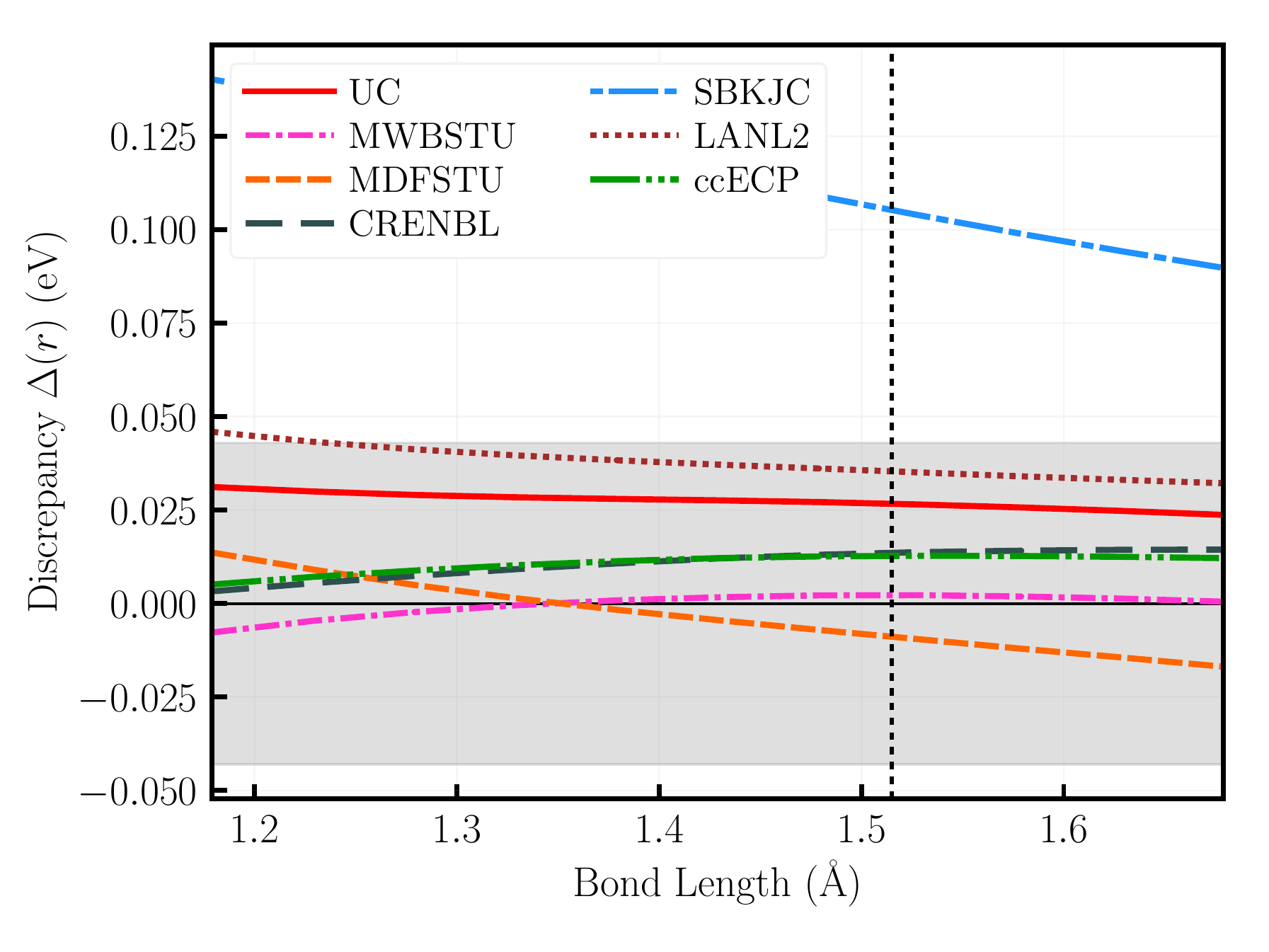}
\caption{PdH ($^2\Sigma$) binding curve discrepancies}
%\label{fig:}
\end{subfigure}%
\begin{subfigure}{0.5\textwidth}
\includegraphics[width=\textwidth]{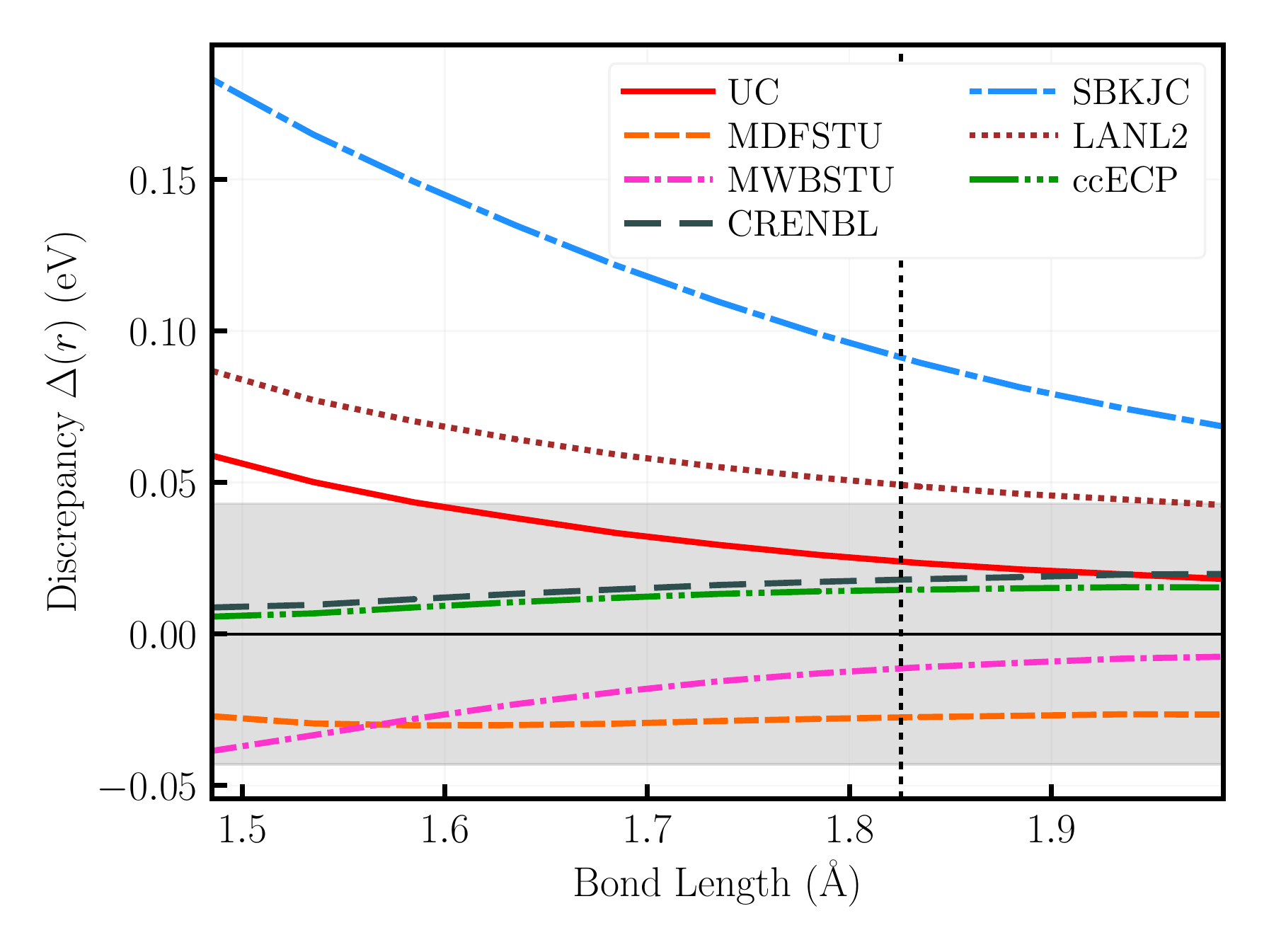}
\caption{PdO ($^3\Pi$) binding curve discrepancies}
%\label{fig:}
\end{subfigure}
\caption{
Binding energy discrepancies for (a) PdH and (b) PdO molecules relative to scalar relativistic AE CCSD(T).
The shaded region indicates the band of chemical accuracy. The dashed vertical line represents the equilibrium geometry.
}
\label{fig:Pd_mols}
\end{figure*}

Figure \fref{fig:Pd_mols} provides the PdH and PdO molecular binding curve plots.
Here most ECPs remain within the bounds of chemical accuracy over the entire range of geometries tested with the exception of SBKJC and LANL2.
Overall CRENBL and ccECP show the smallest errors for these molecules.

\subsubsection{SOREP: Pd}

Table \ref{tab:Pd_sorep} provides the SOREP atomic gap errors using COSCI and FPSODMC.
We again see only minor improvements from MDFSTU in COSCI method while FPSODMC errors are comparable.
Overall, both SOREP ECPs show high quality in terms of multiplet splittings.
We expect to see even smaller errors for all gaps as the trial wave function quality is increased.

\begin{table}[!htbp]
\setlength{\tabcolsep}{4pt} %% default is 6pt
\centering
\small
\caption{
Palladium atomic excitation errors for MDFSTU versus ccECP in SOREP forms.
One set of errors are shown for full-relativistic X2C AE gaps using COSCI.
Another set of errors is calculated using FPSODMC with reference to experiments.
All values are in eV.
}
\label{tab:Pd_sorep}
\begin{adjustbox}{width=1.0\columnwidth,center}
%\resizebox{0.97\columwidth}{!}{%
\begin{tabular}{ll|rcc|r|ccccc}
\hline\hline
\multirow{2}{*}{State} & \multirow{2}{*}{Term} & \multicolumn{3}{c|}{COSCI} &  \multirow{2}{*}{Expt.} & \multicolumn{2}{c}{FPSODMC} \\
\cline{3-5}
\cline{7-8}
& & AE & STU & ccECP &  & STU & ccECP \\
\hline
$4s^24p^64d^{10}$      &  $^{1}S_{3}$    &      0.000 &     0.000 &       0.000 &  0.000 &       0.000 &         0.000 \\
$4s^24p^64d^{10}5s^1$  &  $^{2}S_{1/2}$  &      0.236 &     0.009 &       0.008 & -0.562 &    -0.33(3) &      -0.36(2) \\
$4s^24p^64d^{9}5s^1$   &  $^{2}[5/2]_3$  &     -0.067 &    -0.012 &      -0.036 &  0.814 &    -0.22(2) &      -0.23(2) \\
$4s^24p^64d^{9}$       &  $^{2}S_{5/2}$  &      6.273 &    -0.009 &      -0.044 &  8.337 &     0.05(2) &       0.03(2) \\
$4s^24p^64d^{8}$       &  $^{3}F_{4}$    &     23.662 &     0.034 &      -0.078 & 27.770 &     0.26(2) &       0.12(2) \\
\hline
$4s^24p^64d^{9}5s^1$   &  $^{2}[5/2]_3$  &      0.000 &     0.000 &       0.000 &  0.000 &       0.000 &         0.000 \\
          {}           &  $^{2}[5/2]_2$  &      0.175 &     0.010 &      -0.008 &  0.148 &    -0.02(2) &       0.01(2) \\
\hline
$4s^24p^64d^{9}$       &  $^{2}S_{5/2}$  &      0.000 &     0.000 &       0.000 &  0.000 &       0.000 &         0.000 \\
          {}           &  $^{2}S_{3/2}$  &      0.436 &     0.042 &      -0.038 &  0.439 &    -0.02(3) &      -0.03(2) \\
\hline
$4s^24p^64d^{8}$       &  $^{3}F_{4}$    &      0.000 &     0.000 &       0.000 &  0.000 &       0.000 &         0.000 \\
          {}           &  $^{3}F_{3}$    &      0.394 &     0.041 &      -0.015 &  0.400 &     0.10(2) &       0.05(2) \\
          {}           &  $^{3}F_{2}$    &      0.604 &     0.050 &      -0.024 &  0.581 &     0.02(2) &      -0.06(2) \\
\hline
$4s^24p^1$             &  $^{2}P_{1/2}$  &      0.000 &     0.000 &       0.000 &        &             &               \\
          {}           &  $^{2}P_{3/2}$  &      6.640 &     0.093 &       0.000 &        &             &               \\
\hline
J-MAD                  &                 &            &           &             &        &     0.05(2) &       0.05(2) \\
\hline
MAD                    &                 &            &     0.033 &       0.028 &        &             &               \\
\hline\hline
\end{tabular}
\end{adjustbox}
\end{table}

\subsection{Iridium (Ir)}
\subsubsection{AREP: Ir}

\begin{figure}[!htbp]
\centering
\includegraphics[width=1.00\columnwidth]{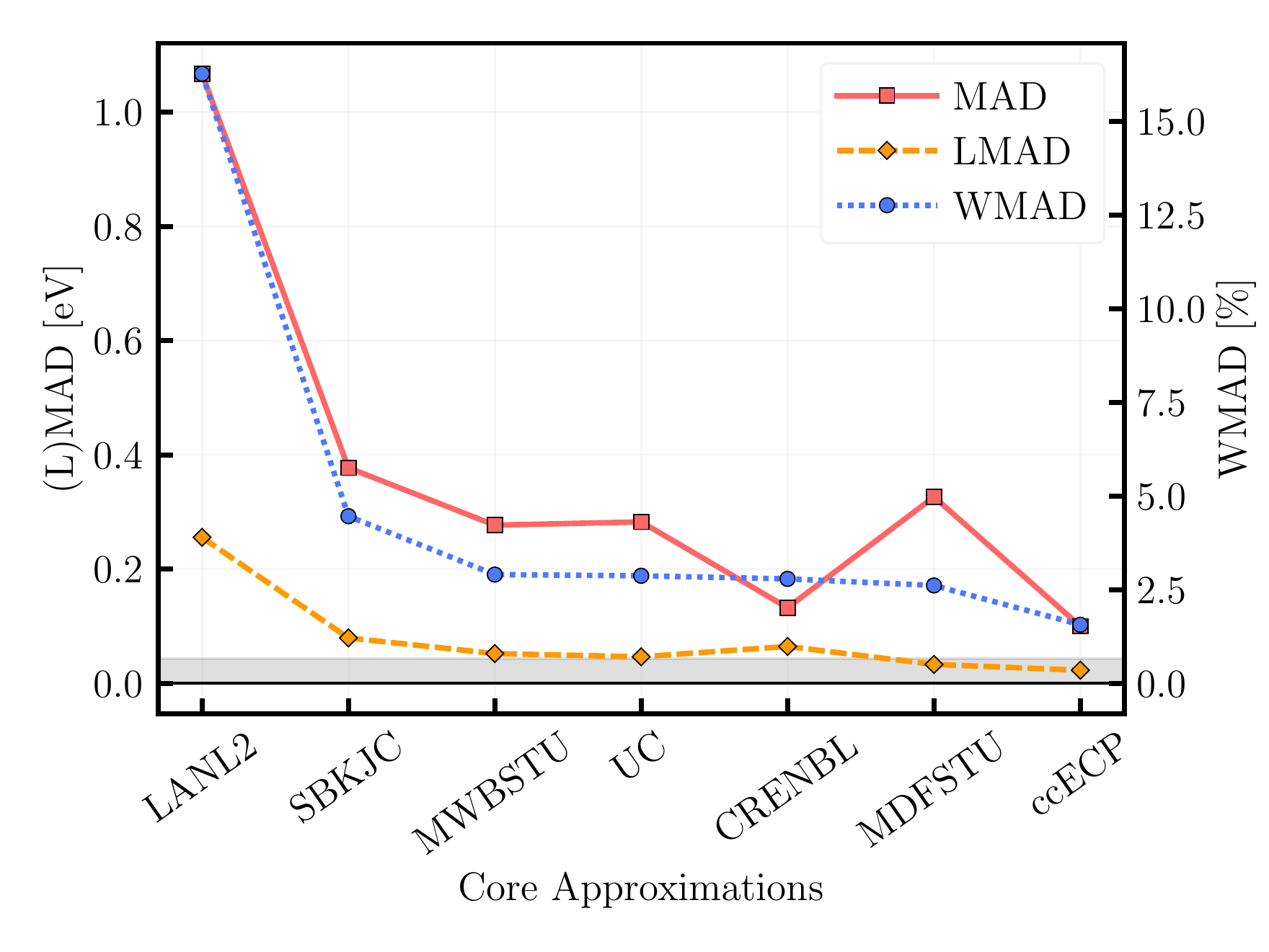}
\caption{
Ir scalar relativistic AE gap (W/L)MADs for various core approximations. 
\mbox{[core] = [Kr]$4d^{10}4f^{14}$} (60 electrons).
RCCSD(T) method with unc-aug-cc-pwCVTZ basis set was used.
}
\label{fig:Ir_spectrum}
\end{figure}

\begin{figure*}[!htbp]
\centering
\begin{subfigure}{0.5\textwidth}
\includegraphics[width=\textwidth]{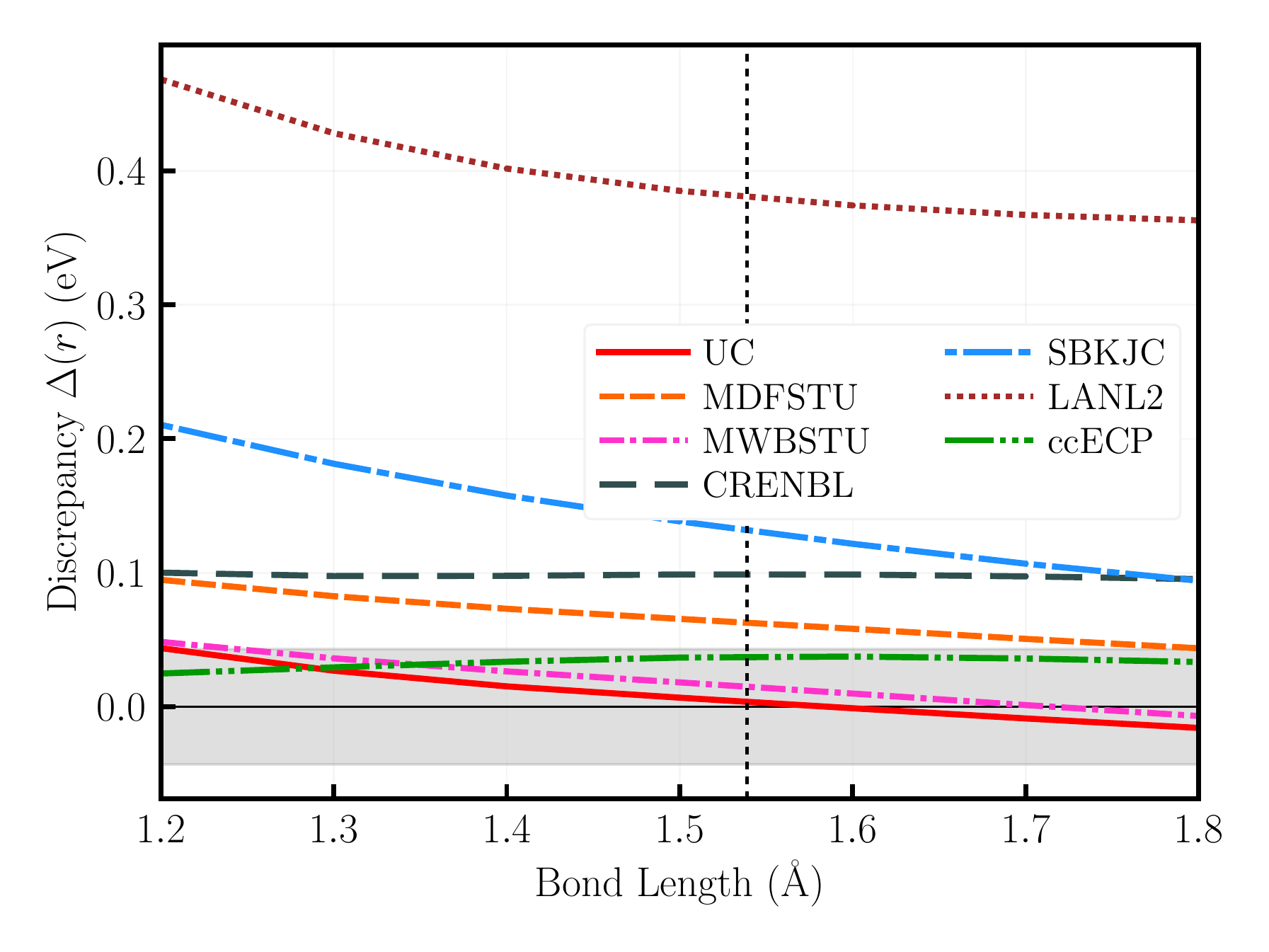}
\caption{IrH ($^3\Delta$) binding curve discrepancies}
%\label{fig:}
\end{subfigure}%
\begin{subfigure}{0.5\textwidth}
\includegraphics[width=\textwidth]{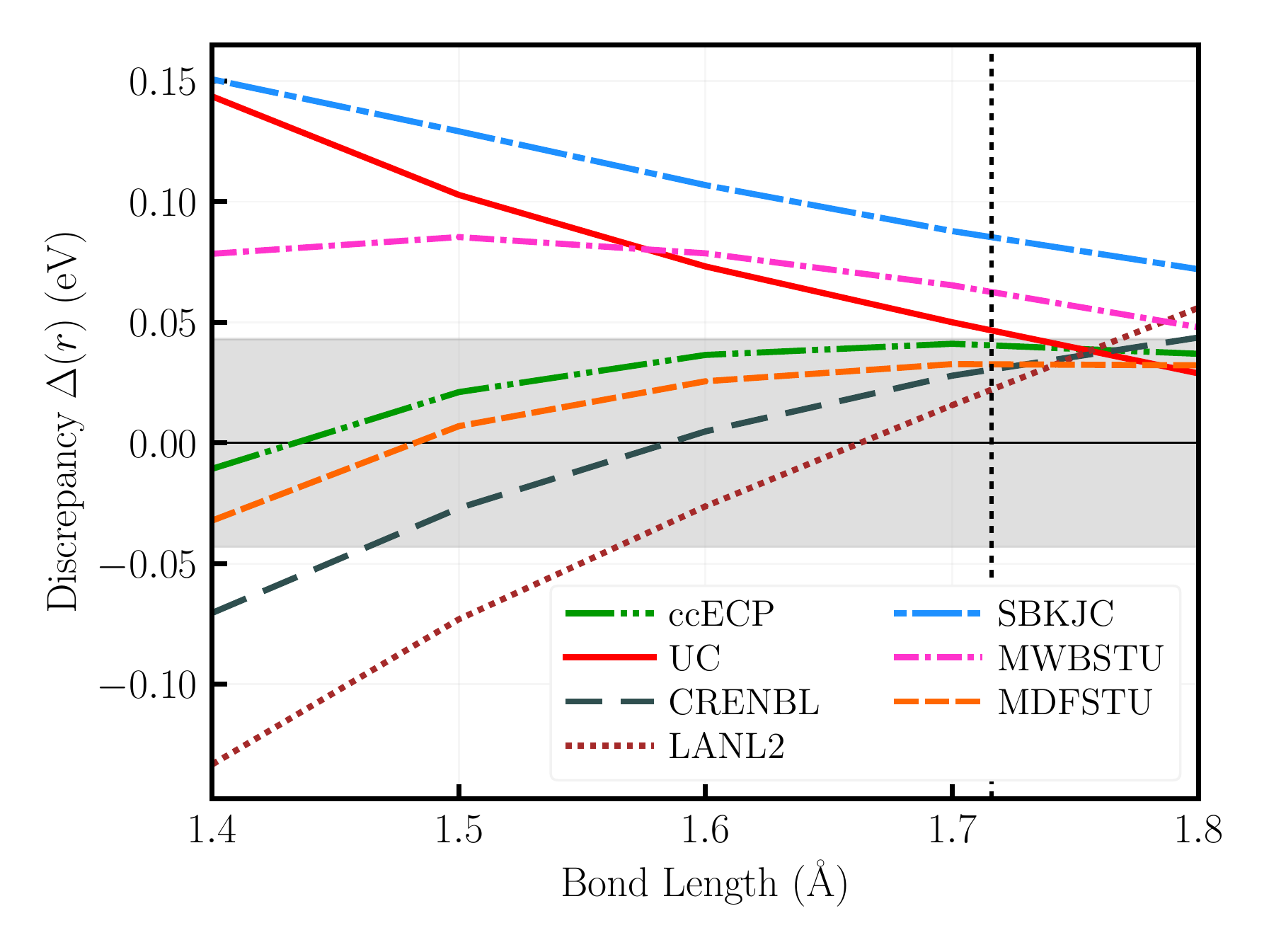}
\caption{IrO ($^2\Sigma$) binding curve discrepancies}
%\label{fig:}
\end{subfigure}
\caption{
Binding energy discrepancies for (a) IrH and (b) IrO molecules relative to scalar relativistic AE CCSD(T).
The shaded region indicates the band of chemical accuracy. The dashed vertical line represents the equilibrium geometry.
}
\label{fig:Ir_mols}
\end{figure*}

Looking at \fref{fig:Ir_spectrum} one can see the Ir spectral errors of various ECPs tested within this work.
The Ir ccECP exceeded the accuracy of all the other ECPs for this element in all metrics. 
The LMAD was comfortably within the chemical accuracy, and the MAD was much smaller than most other contenders apart from the MDFSTU ECP which achieved a similar MAD.
\fref{fig:Ir_mols} shows the molecular binding energy discrepancy for both IrH and IrO. 
Half of the ECPs tested were outside of the chemical accuracy for IrH over the entire range of geometries.
The Ir ccECP does not have the smallest discrepancy at all points for either molecule, but when considering both systems, it has the most balanced biases that are always remaining within
the chemical accuracy.

\subsubsection{SOREP: Ir}

We provide Ir SOREP atomic excitation errors in \tref{tab:Ir_sorep}.
Optimization of ccECP provides an accurate spectrum for COSCI atomic gaps for the entire set of states with the MAD of 0.018 eV, which is less than one third of MDFSTU MAD.
However, considering the higher accuracy FPSODMC calculations, we see that ccECP and STU give comparable MADs for spin-orbit splitting states referenced to experimental data, about 0.2 eV.
Although we see an improvement in MAD of $J-$splitting in the first IP state, the large discrepancies in ground state multiplet gaps overshadow this improvement.
Further inspection shows that the ground state multiplet splittings change order when going from COSCI to experimental values.
This is another indication that electron correlations must be accurately accounted to properly describe these low-lying states. Clearly, a more extensive study of correlation for these
states is required and we plan to address it in future.
A similar case was observed in W atom where incorrect ordering of states were obtained unless higher order CI expansions were used (Table \ref{tab:W_sorep_CC}).

\begin{table}[htbp!]
\setlength{\tabcolsep}{4pt} %% default is 6pt
\centering
\small
\caption{
\footnotesize{Iridium atomic excitation errors for MDFSTU versus ccECP in SOREP forms.
One set of errors are shown for full-relativistic X2C AE gaps using COSCI.
Another set of errors are calculated using FPSODMC and compared to experiments.
All values are in eV.}
}
\label{tab:Ir_sorep}
\begin{adjustbox}{width=1.0\columnwidth,center}
%\resizebox{0.97\columwidth}{!}{%
\begin{tabular}{ll|rcc|r|c|cccc}
\hline\hline
\multirow{2}{*}{State} & \multirow{2}{*}{Term} & \multicolumn{3}{c|}{COSCI} &  \multirow{2}{*}{Expt.} & \multicolumn{2}{c}{FPSODMC} \\
\cline{3-5}
\cline{7-8}
& & AE & STU & ccECP &  & STU & ccECP \\
\hline
$5s^25p^65d^{7}6s^2$   &  $^{4}F_{9/2}$  &     0.000 &     0.000 &       0.000 &    0.000 &     0.00    &       0.00    \\
$5s^25p^65d^{8}6s^2$   &  $^{6}S_{5/2}$  &     0.120 &     0.000 &       0.039 &   -1.565 &    -0.33(3) &      -0.28(3) \\
$5s^25p^65d^{8}6s^1$   &  $^{4}F_{9/2}$  &     0.284 &    -0.001 &       0.045 &    0.351 &     0.08(3) &       0.11(3) \\
$5s^25p^65d^{7}6s^1$   &  $^{5}F_{5}$    &     6.997 &     0.012 &       0.004 &    8.967 &     0.34(4) &       0.35(3) \\
$5s^25p^65d^{7}$       &  $^{4}F_{9/2}$  &    23.290 &     0.020 &      -0.010 &  26.0(3) &     -0.1(3) &      -0.1(3)  \\
\hline                                        
$5s^25p^65d^{7}6s^2$   &  $^{4}F_{9/2}$  &     0.000 &     0.000 &       0.000 &    0.000 &     0.00    &       0.00    \\
          {}           &  $^{4}F_{7/2}$  &     0.782 &     0.088 &       0.009 &    0.784 &     0.01(3) &      -0.05(3) \\
          {}           &  $^{4}F_{5/2}$  &     1.075 &     0.080 &      -0.004 &    0.717 &    -0.35(3) &      -0.36(3) \\
          {}           &  $^{4}F_{3/2}$  &     1.202 &     0.052 &      -0.021 &    0.506 &    -0.66(3) &      -0.71(3) \\
\hline                                        
$5s^25p^65d^{8}6s^1$   &  $^{4}F_{9/2}$  &     0.000 &     0.000 &       0.000 &    0.000 &     0.00    &       0.00    \\
          {}           &  $^{4}F_{7/2}$  &     0.512 &     0.038 &       0.002 &    0.530 &     0.07(3) &       0.05(3) \\
          {}           &  $^{4}F_{5/2}$  &     0.940 &     0.068 &      -0.003 &    0.873 &    -0.06(3) &      -0.02(3) \\
          {}           &  $^{4}F_{3/2}$  &     1.084 &     0.056 &      -0.010 &    1.115 &     0.06(3) &       0.09(3) \\
\hline                                        
$5s^25p^65d^{7}6s^1$   &  $^{5}F_{5}$    &     0.000 &     0.000 &       0.000 &    0.000 &     0.00    &       0.00    \\
          {}           &  $^{5}F_{4}$    &     0.596 &     0.055 &       0.003 &    0.594 &     0.05(4) &      -0.07(3) \\
          {}           &  $^{5}F_{3}$    &     1.015 &     0.092 &       0.003 &    1.015 &     0.09(4) &      -0.03(3) \\
          {}           &  $^{5}F_{2}$    &     1.241 &     0.096 &      -0.003 &    1.402 &     0.26(4) &       0.14(3) \\
          {}           &  $^{5}F_{1}$    &     1.376 &     0.099 &      -0.006 &    1.483 &     0.21(4) &       0.05(3) \\
\hline                                        
$5s^25p^65d^{7}$       &  $^{4}F_{9/2}$  &     0.000 &     0.000 &       0.000 &          &             &               \\
          {}           &  $^{4}F_{7/2}$  &     0.848 &     0.094 &       0.008 &          &             &               \\
          {}           &  $^{4}F_{5/2}$  &     1.159 &     0.082 &      -0.007 &          &             &               \\
          {}           &  $^{4}F_{3/2}$  &     1.289 &     0.048 &      -0.027 &          &             &               \\
\hline                                        
$5s^25p^66s^26p^1$     &  $^{2}P_{1/2}$  &     0.000 &     0.000 &       0.000 &          &             &               \\
          {}           &  $^{2}P_{3/2}$  &     3.601 &     0.016 &       0.031 &          &             &               \\
\hline                                        
$5s^25p^66s^26f^1$     &  $^{2}F_{5/2}$  &     0.000 &     0.000 &       0.000 &          &             &               \\
          {}           &  $^{2}F_{7/2}$  &     0.106 &     0.040 &      -0.001 &          &             &               \\
\hline                                        
$5s^25p^5$             &  $^{2}P_{3/2}$  &     0.000 &     0.000 &       0.000 &          &             &               \\
          {}           &  $^{2}P_{1/2}$  &    16.084 &     0.169 &      -0.114 &          &             &               \\
\hline                                                                        
J-MAD               &                 &           &           &             &          &     0.20(4) &       0.17(3) \\
\hline                                                                        
MAD                    &                 &           &     0.060 &       0.018 &          &             &               \\
\hline\hline
\end{tabular}
%}
\end{adjustbox}
\end{table}

\subsection{Molybdenum (Mo)}
\subsubsection{AREP: Mo}

\begin{figure}[!htbp]
\centering
\includegraphics[width=1.00\columnwidth]{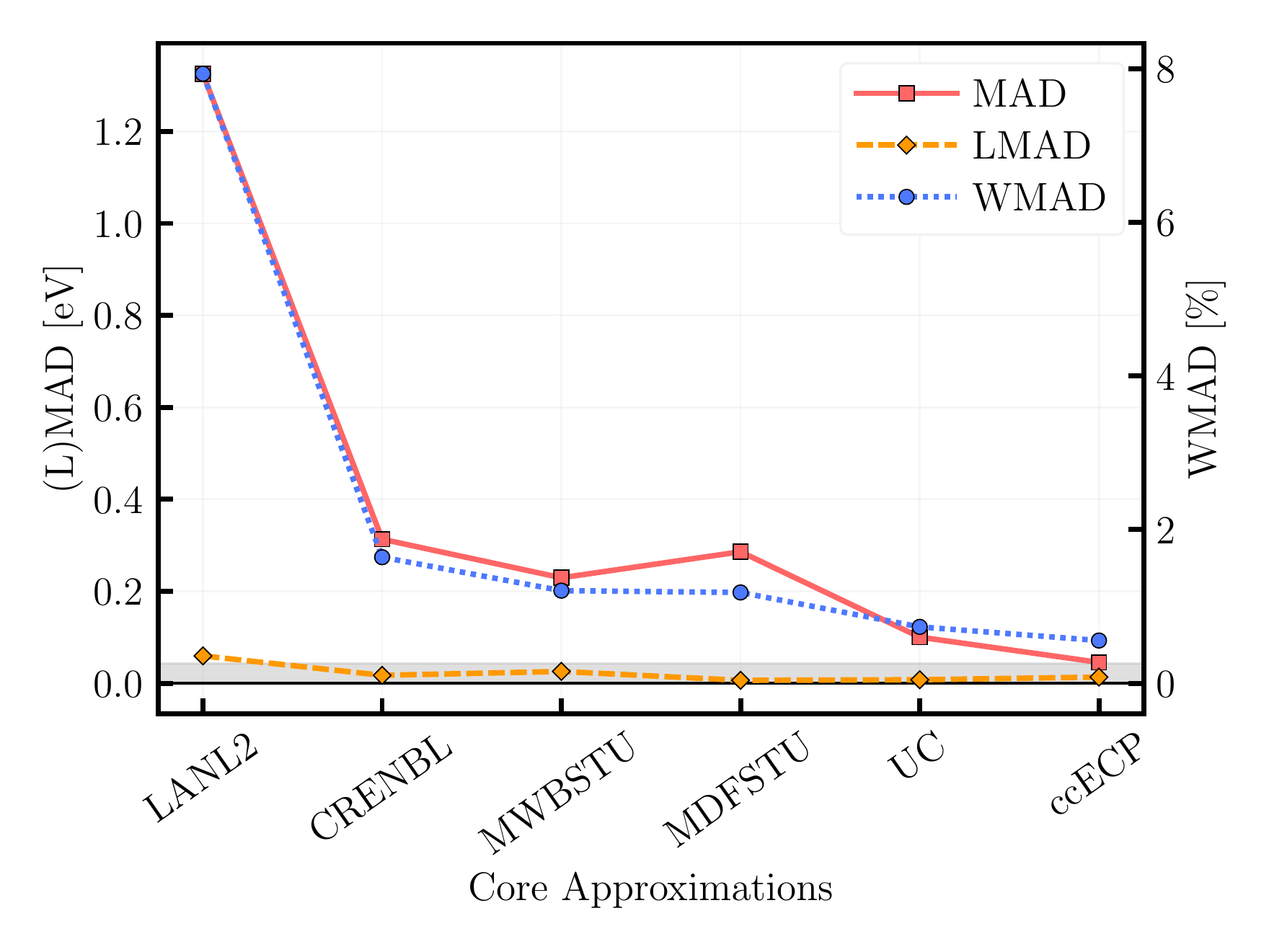}
\caption{
Mo scalar relativistic AE gap (W/L)MADs for various core approximations. 
\mbox{[core] = [Ar]$3d^{10}$} (28 electrons).
RCCSD(T) method with unc-aug-cc-pwCVTZ basis set was used.
}
\label{fig:Mo_spectrum}
\end{figure}

\begin{figure*}[!htbp]
\centering
\begin{subfigure}{0.5\textwidth}
\includegraphics[width=\textwidth]{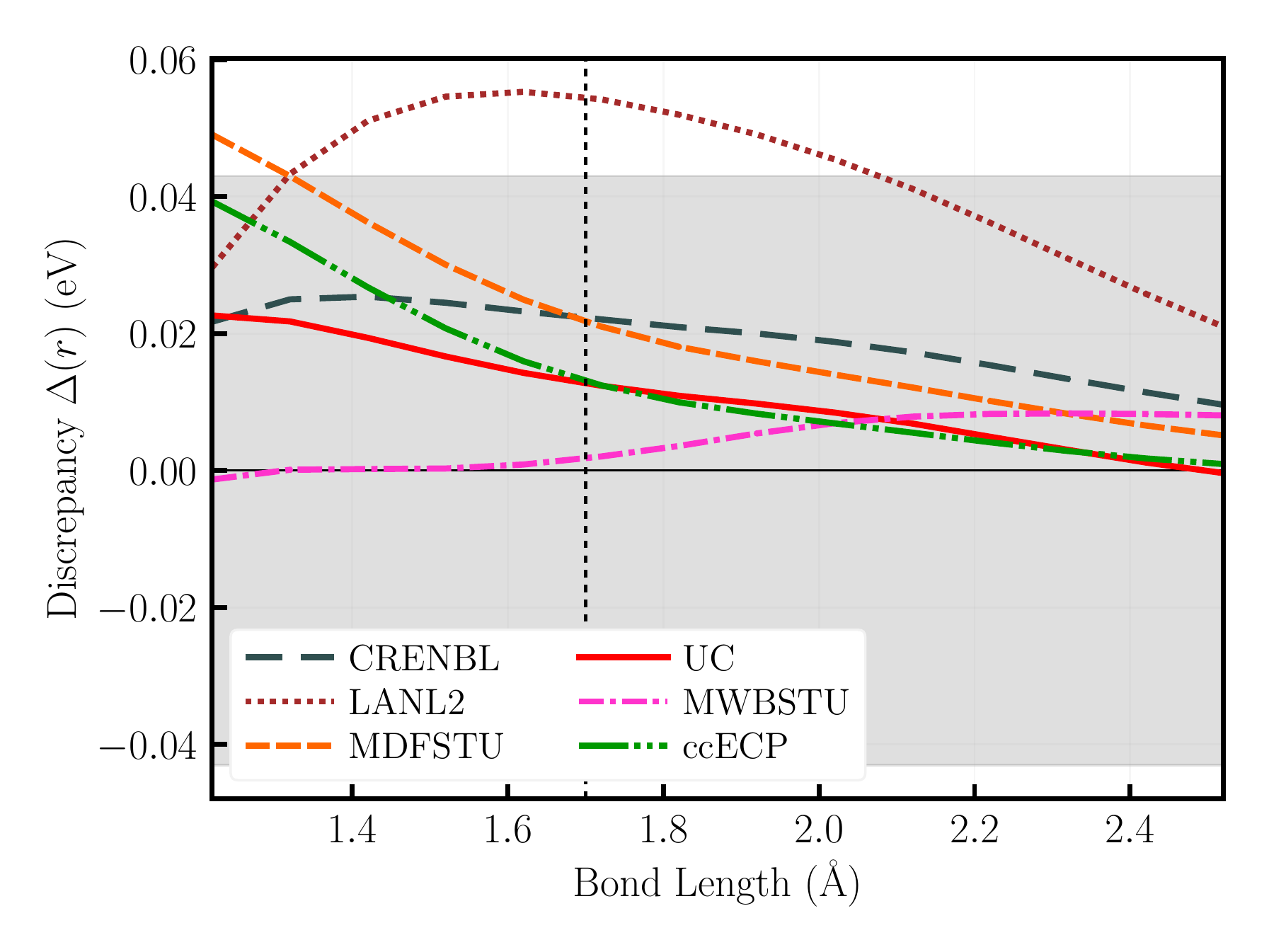}
\caption{MoH ($^6\Sigma$) binding curve discrepancies}
%\label{fig:}
\end{subfigure}%
\begin{subfigure}{0.5\textwidth}
\includegraphics[width=\textwidth]{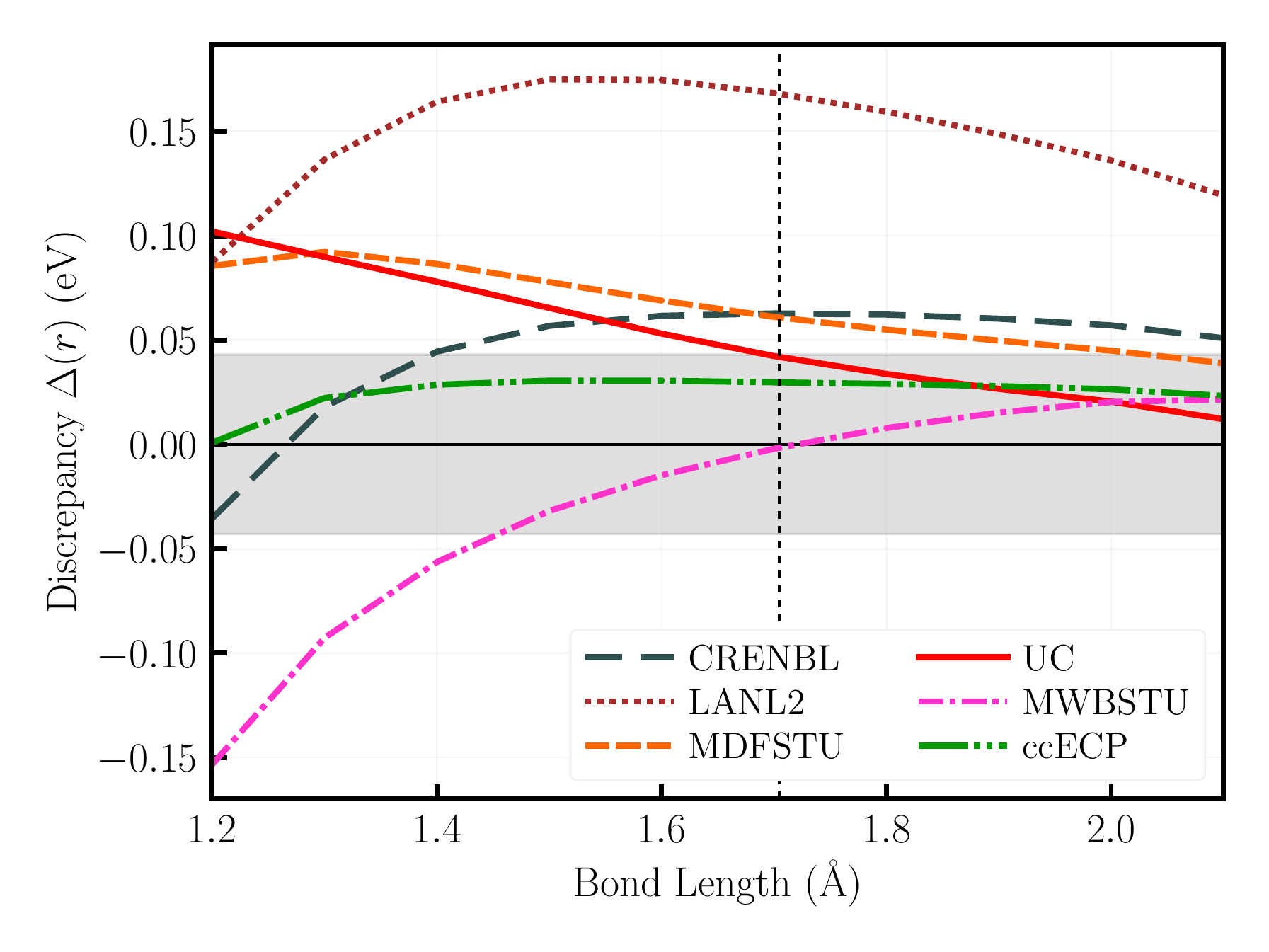}
\caption{MoO ($^5\Pi$) binding curve discrepancies}
%\label{fig:}
\end{subfigure}
\caption{
Binding energy discrepancies for (a) MoH and (b) MoO molecules relative to scalar relativistic AE CCSD(T).
The shaded region indicates the band of chemical accuracy. The dashed vertical line represents the equilibrium geometry.
}
\label{fig:Mo_mols}
\end{figure*}

In \fref{fig:Mo_spectrum} and \fref{fig:Mo_mols}, we share the Mo atomic spectral errors and molecular binding discrepancies for various ECPs. 
Figure \ref{fig:Mo_spectrum} clearly shows that the developed Mo ccECP outperforms all the other ECPs in MAD and WMAD. Note that the MAD of our ccECP is refined to chemical accuracy which indicates the high accuracy is achieved for the full span of deep ionizations. 
Moreover, the LMAD, representing the precision of the low-lying states energies, is within the chemical accuracy at a notably low level. 
MDFSTU ECP has a slightly better LMAD, yet the MAD and WMAD is not comparable with ccECP, suggesting that ccECP is a more comprehensive solution as a robust effective core potential. 
In Figure \ref{fig:Mo_mols}, it is shown that the ccECP binding energy discrepancies are within chemical accuracy for all geometries in both MoH and MoO molecules. 
In the hydride, MWBSTU shows flatter and much smaller errors in shorter bond lengths. 
However in the oxide, ccECP is accurate for all bond lengths, while all other core approximations (including MWBSTU) deviate outside of chemical accuracy in some parts of the curve.

\subsubsection{SOREP: Mo}

\tref{tab:Mo_sorep} provide the Mo atomic excitations errors for MDFSTU and ccECP.
Obviously, the optimization of SO ccECP results in a more accurate spectrum.
The MAD for COSCI calculations of ccECP is decreased by almost a magnitude when compared with MDFSTU results.
Although the accuracy is not fully achieved in fixed-phase calculations with restricted quality trial function, the MAD of the low-lying spin-orbit splitting spectrum referenced to experiments is about 0.08 eV which is significantly smaller than for MDFSTU.

\begin{table}[!htbp]
\setlength{\tabcolsep}{4pt} %% default is 6pt
\centering
\small
\caption{
Molybdenum atomic excitation errors for MDFSTU versus ccECP in SOREP forms.
One set of errors are shown for full-relativistic X2C AE gaps using COSCI.
Another set of errors is calculated using FPSODMC with reference to experiments.
All values are in eV.
}
\label{tab:Mo_sorep}
\begin{adjustbox}{width=1.0\columnwidth,center}
%\resizebox{0.97\columwidth}{!}{%
\begin{tabular}{ll|rcc|r|ccccc}
\hline\hline
\multirow{2}{*}{State} & \multirow{2}{*}{Term} & \multicolumn{3}{c|}{COSCI} &  \multirow{2}{*}{Expt.} & \multicolumn{2}{c}{FPSODMC} \\
\cline{3-5}
\cline{7-8}
& & AE & STU & ccECP &  & STU & ccECP \\
\hline
$4s^24p^64d^{5}5s^1$   &  $^{7}S_{3}$    &    0.000 &     0.000 &     0.000 &    0.000 &       0.000 &         0.000 \\
$4s^24p^64d^{5}5s^2$   &  $^{6}S_{5/2}$  &    0.693 &    -0.007 &    -0.007 &   -0.747 &      -1.21(3) &        -0.67(5) \\
$4s^24p^64d^{4}5s^2$   &  $^{5}D_{0}$    &    2.189 &    -0.068 &    -0.039 &    1.360 &      -1.18(3) &        -0.74(4) \\
$4s^24p^64d^{5}$       &  $^{6}S_{5/2}$  &    8.274 &     0.629 &    -0.078 &    7.092 &      -0.62(5) &        -0.16(5) \\
$4s^24p^64d^{4}$       &  $^{5}D_{0}$    &   21.200 &    -0.057 &    -0.024 &   23.252 &      -0.46(4) &         0.25(5) \\
\hline                                                                        
$4s^24p^64d^{5}5s^1$   &  $^{7}S_{3}$    &    0.000 &     0.000 &     0.000 &    0.000 &       0.000 &         0.00 \\
          {}           &  $^{5}S_{2}$    &    1.870 &    -0.009 &    -0.009 &    1.335 &      -0.4(1) &        -0.24(4) \\
          {}           &  $^{5}G_{2}$    &    2.725 &    -0.066 &    -0.063 &    2.063 &      -0.66(6) &        -0.34(4) \\
\hline                                                                        
$4s^24p^64d^{4}5s^2$   &  $^{5}D_{0}$    &    0.000 &     0.000 &     0.000 &    0.000 &       0.000 &         0.00 \\
          {}           &  $^{5}D_{1}$    &    0.024 &     0.005 &     0.005 &    0.022 &      -0.14(3) &         0.03(5) \\
          {}           &  $^{5}D_{2}$    &    0.070 &     0.015 &     0.013 &    0.061 &      -0.05(4) &         0.09(4) \\
          {}           &  $^{5}D_{3}$    &    0.133 &     0.027 &     0.024 &    0.111 &      -0.28(4) &         0.16(4)\\
          {}           &  $^{5}D_{4}$    &    0.209 &     0.040 &     0.036 &    0.171 &      -0.08(3) &         0.07(3) \\
\hline                                                                        
$4s^24p^64d^{5}$       &  $^{6}S_{5/2}$  &    0.000 &     0.000 &     0.000 &    0.000 &       0.000 &         0.000 \\
          {}           &  $^{4}G_{5/2}$  &    0.063 &    -0.507 &     0.009 &    1.884 &      -0.26(3) &         0.00(4) \\
          {}           &  $^{4}G_{7/2}$  &    0.195 &    -0.966 &     0.030 &    1.901 &      -0.20(4) &        -0.05(6) \\
          {}           &  $^{4}G_{9/2}$  &    0.431 &    -0.773 &     0.082 &    1.913 &      -0.26(4) &         0.00(4) \\
          {}           &  $^{4}G_{11/2}$ &    0.726 &    -0.570 &    -0.020 &    1.915 &      -0.34(2) &        -0.16(4) \\
\hline                                                                        
$4s^24p^64d^{4}$       &  $^{5}D_{0}$    &    0.000 &     0.000 &     0.000 &    0.000 &       0.000 &         0.000 \\
          {}           &  $^{5}D_{1}$    &    0.026 &     0.006 &     0.005 &    0.030 &      -0.04(3) &        -0.016(4) \\
          {}           &  $^{5}D_{2}$    &    0.076 &     0.016 &     0.014 &    0.083 &       0.04(6) &         0.01(4) \\
          {}           &  $^{5}D_{3}$    &    0.143 &     0.028 &     0.025 &    0.152 &       0.14(5) &         0.01(3) \\
          {}           &  $^{5}D_{4}$    &    0.226 &     0.042 &     0.037 &    0.232 &       0.11(5) &        -0.07(4) \\
\hline                                                                        
$4s^24p^65s^26p^1$     &  $^{2}P_{1/2}$  &    0.000 &     0.000 &     0.000 &          &             &               \\
          {}           &  $^{2}P_{3/2}$  &    0.564 &     0.097 &    -0.006 &          &             &               \\
\hline                                                                        
$4s^24p^5$             &  $^{2}P_{3/2}$  &    0.000 &     0.000 &     0.000 &          &             &               \\
          {}           &  $^{2}P_{1/2}$  &    2.948 &     0.050 &    -0.013 &          &             &               \\
\hline                                                                        
J-MAD               &                 &          &           &           &          &       0.200 &         0.080 \\
\hline                                                                        
MAD                    &                 &          &     0.199 &     0.027 &          &             &               \\
\hline\hline
\end{tabular}
\end{adjustbox}
\end{table}

\subsection{Summary of results}

\subsubsection{Averages: AREP}

\begin{figure}[!htbp]
\centering
\includegraphics[width=1.00\columnwidth]{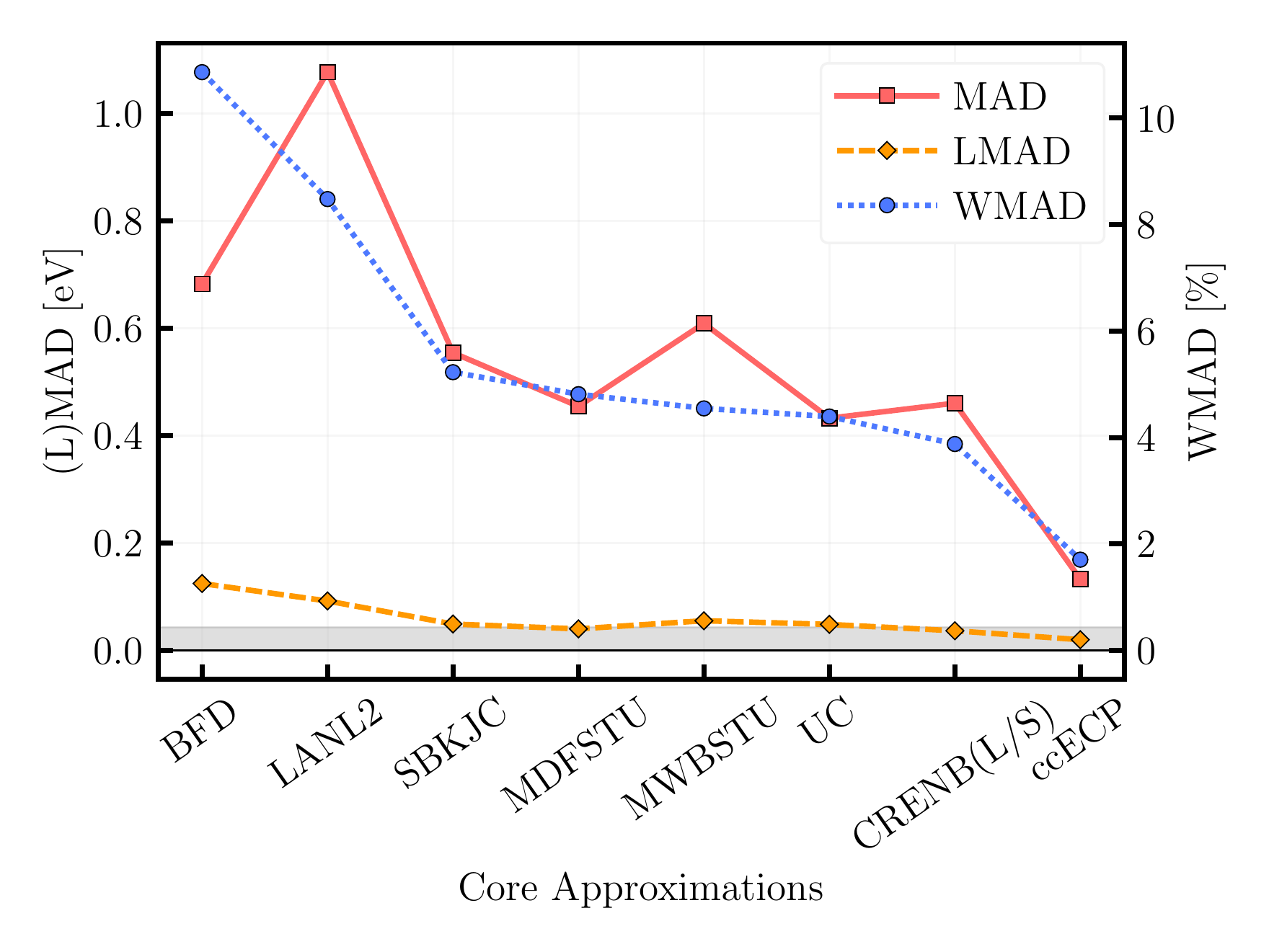}
\caption{
Scalar relativistic AE gap (W/L)MADs for various core approximations.
The values correspond to the averages of all elements considered.
}
\label{fig:all_spectrum}
\end{figure}

\fref{fig:all_spectrum} and \tref{morse:all_ecps} give the summary of AREP atomic spectrum errors and AREP molecular property discrepancies for all elements considered in this work.
In general, we have achieved substantial improvements for both atomic and molecular results.
For the atomic spectrum, our ccECP show significant improvement in all metrics, LMAD, MAD, and WMAD.
The LMAD metric includes only low-lying states composed of EA, IP, and IP2 and 
it is well within chemical accuracy and slightly better than the rest of core approximations.
The MAD and WMAD show similar picture to previous elements with the lowest errors for ccECP, with an overall dramatic improvements in a wide range of excitation energies.
\tref{morse:all_ecps} provide the collected results of various binding parameter errors for all elements as obtained by fits in Eqn. \eqref{morse_pot}.
Clearly, our ccECPs give the highest accuracy with the lowest errors for all parameters characterizing the molecular bonding. 
In addition, considering both atomic or molecular errors, ccECPs show higher accuracy than the existing ECPs, but also than UC results as discussed earlier.

\begin{table}[!htbp]
\centering
\caption{Mean absolute deviations of binding parameters for various core approximations with respect to AE correlated data for I, Te, Ag, Pd, Mo, Bi, Au, Ir, W related molecules. All parameters were obtained using Morse potential fit. The parameters shown are dissociation energy $D_e$, equilibrium bond length $r_e$, vibrational frequency $\omega_e$ and binding energy discrepancy at dissociation bond length $D_{diss}$.
}
\label{morse:all_ecps}
\begin{tabular}{l|rrrrrrrrrr}
\hline\hline
{} & $D_e$(eV) & $r_e$(\AA) & $\omega_e$(cm$^{-1}$) & $D_{diss}$(eV) \\
\hline
BFD            &    0.078(6) &   0.018(1) &                 22(3) &        0.41(5) \\
CRENBL(S)      &    0.115(5) &  0.0183(9) &                 26(3) &        0.40(4) \\
LANL2          &    0.118(5) &  0.0122(9) &                 16(3) &        0.36(4) \\
MDFSTU         &    0.096(5) &  0.0094(9) &                 10(3) &        0.24(4) \\
MWBSTU         &    0.050(5) &  0.0056(9) &                 10(3) &        0.10(4) \\
SBKJC          &    0.089(4) &  0.0120(8) &                 19(2) &        0.30(4) \\
UC             &    0.040(5) &  0.0104(9) &                 13(3) &        0.19(4) \\
ccECP          &    0.018(5) &  0.0022(9) &                  6(3) &        0.07(4) \\
\hline\hline
\end{tabular}
\end{table}

\subsubsection{Averages: SOREP}

\begin{figure*}[!htbp]
\centering
\begin{subfigure}{0.5\textwidth}
\includegraphics[width=\textwidth]{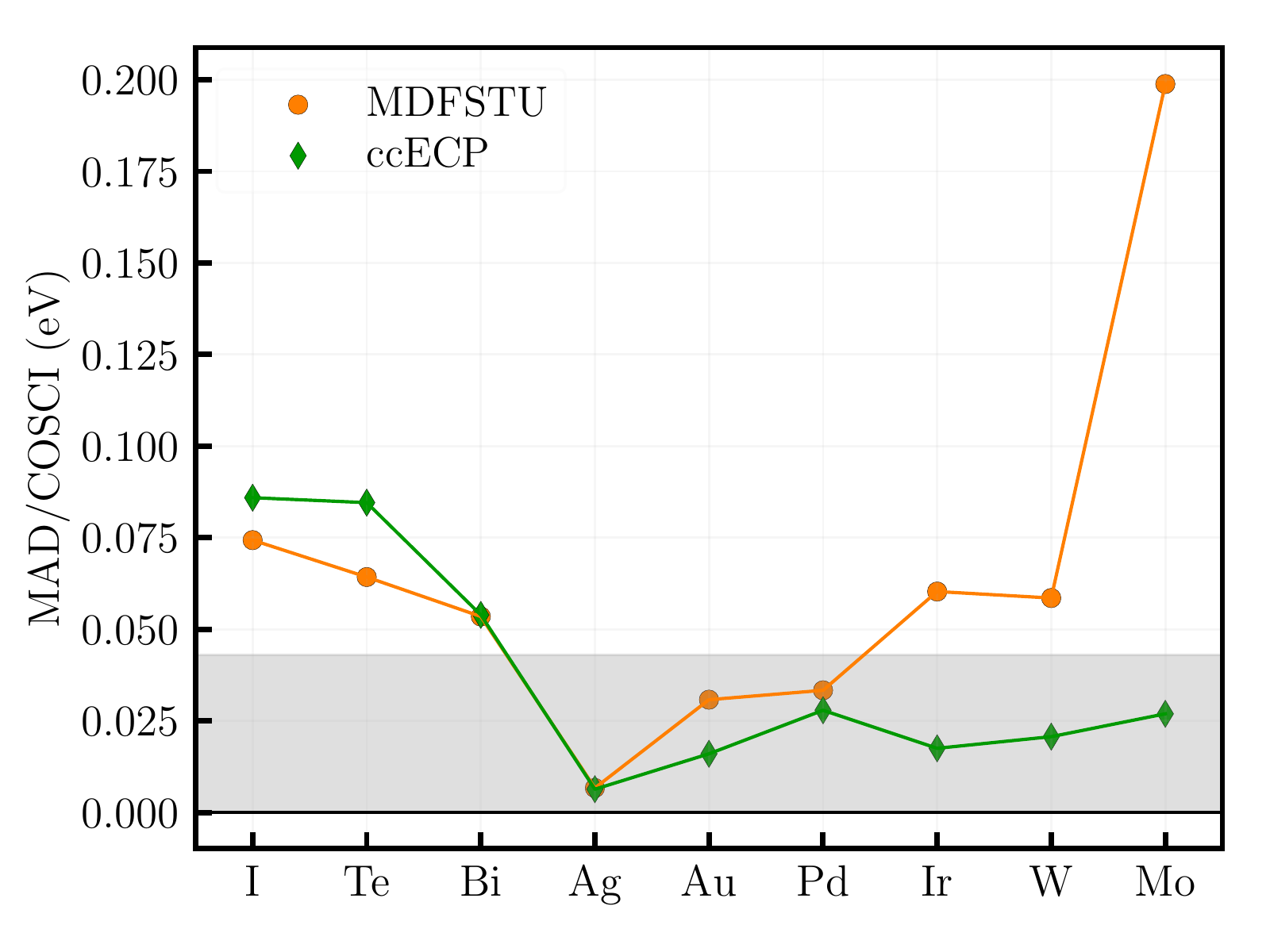}
\caption{Atomic COSCI gap MADs}
\label{fig:sorep_mad_avg_a}
\end{subfigure}%
\begin{subfigure}{0.5\textwidth}
\includegraphics[width=\textwidth]{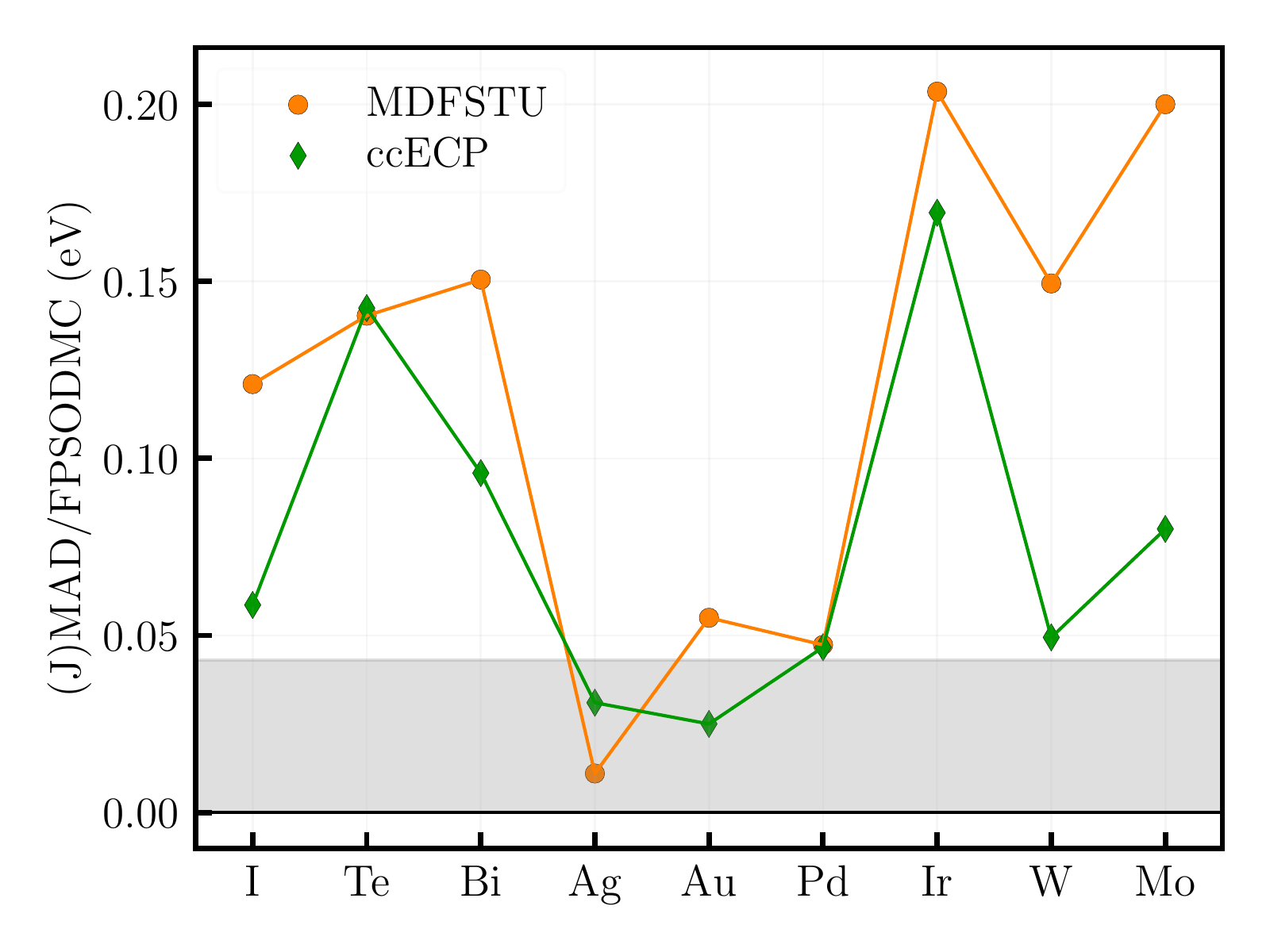}
\caption{Atomic FPSODMC gap (J)MADs}
\label{fig:sorep_mad_avg_b}
\end{subfigure}
\caption{
Comparison of spin-orbit atomic gap MADs for MDFSTU and ccECP, a) using COSCI level of theory and referenced to corresponding COSCI AE gaps and (b) in single-reference FPSODMC calculations that are referenced to experiments for all states with main group elements, I, Te, Bi and spin-orbit splittings for the remaining transition metal elements.
The shaded region indicates the band of chemical accuracy.
}
\label{fig:sorep_mad_avg}
\end{figure*}

We provide the summary of SOREP atomic excitations MADs for MDFSTU and ccECP in \fref{fig:sorep_mad_avg}.
The atomic COSCI gaps are used for the optimization of spin-orbit terms and they are plotted in \fref{fig:sorep_mad_avg_a}.
As commented above ccECPs shows mildly higher or on par MADs for main group elements and consistent significant improvements for transition elements.
Not surprisingly, the slightly higher MADs in main group elements are due to the fixed AREP part in spin-orbit terms optimization since that restricts the total variational freedom, however, it helps
the optimization efficiency. Although further minor refinements might be possible, the overall
expected gains are deemed as marginal due to the dominant source of bias from large cores.
In transition metals, the larger valence space alleviates this deficiency and provides the COSCI MADs with chemical accuracy for all cases.
Especially for Mo atom, we see a dramatic improvement of MADs from previously observed
$\approx$ 0.2 eV which is reduced by almost an order of magnitude.
In \fref{fig:sorep_mad_avg_b}, we present the assessments of MDFSTU and ccECPs in FPSODMC calculations referenced to experimental data.
Note that for the transition metals, we include only spin-orbit splitting states since we find the accuracy of FPSODMC as being somewhat limited for the charged states in single-reference trial setting. Clearly, this calls for more elaborated QMC study with multi-reference trial functions.
We see that ccECP shows overall reduced MAD compared to MDFSTU except Te with similar large MADs and Ag with MADs maintained within desirable chemical accuracy.
Interestingly, though \fref{fig:sorep_mad_avg_a} shows slightly larger COSCI MAD for the main group elements while the FPSODMC MAD is slightly lower.
We anticipate that overall our AREP ccECPs will boost the accuracy of spin-orbit calculations, especially for the charged states where both charge relaxation and correlation play significant roles.
For transition metals, noticeable gains in accuracy and consistency are obtained for Au, Ag, W, and Mo while Ir shows modest improvements.
This has been discussed in Ir section where the ground state spin-orbit splitting appears to be significantly different from other cases and perhaps it might be further refined in future.
Generally speaking, ccECP shows consistent improvements in the accuracy compared to MDFSTU.

We show the SOREP transferability tests for I, Bi, Au, and W in \fref{fig:I2_mols}, \fref{fig:Bi2_mols}, \fref{fig:Au2_mols}, \fref{fig:W2_mols} which are monoatomic dimer binding curves, respectively.
For I$_2$ and Bi$_2$, we are pleased to see consistent improvements from AREP CCSD(T) to SOREP FPSODMC calculations as well as going from MDFSTU to ccECP.
Especially in I$_2$, the constructed ccECP in FPSODMC shows near exact equilibrium bond length and binding energy compared to Expt.
In certain aspects, the full $d$ shell in Au leads to simpler binding picture for Au$_2$ and we indeed see excellent accuracy of both MDFSTU  and ccECP in the CCSD(T)/AREP binding parameters agreeing with experimental data.
When we further consider explicit spin-orbit effect, the exceptional accuracy is not maintained for MDFSTU while ccECP shows very consistent performance with desired properties in both AREP and SOREP level.
For W$_2$, we have provided our calculations though experimental data are lacking to our best knowledge.
The estimated binding energy from previous work is given 5(1) eV, which all calculations are inside the range.
In this case, MDFSTU and ccECP behave almost the same and cross validate each other.
We believe the calculations provide a new reliable reference for further studies of this system.

{\em Basis sets and K-B formats.}
The derived ccECPs are accompanied by basis sets up to 6Z level for main group elements and 5Z level for transition metal elements \cite{pseudopotential_library}. The cited library includes also Kleinmann-Bylander transformed 
forms and corresponding files for use with plane wave codes. In general, very good convergence 
is achieved for cut-offs below $\approx$ 200~Ry for main group elements and $\approx$ 400~Ry for transition metal elements which enable routine calculations of 
solids and 2D materials. Further details can be found in Supplementary Information.
All ccECP and corresponding basis sets in various code formats can be found at 
\url{https://pseudopotentiallibrary.org}

\section{Conclusions}
\label{sec:conclusions}

In this work, we present newly constructed correlation consistent effective core potentials for heavy elements  I, Te, Bi, Ag, Au, Pd, Ir, Mo, and W. 
Following the same convention of our previous constructions for first three row elements, the valence spaces are the most generally used for main groups elements, I, Bi, Te, including only $n^{th}$ $s$ and $p$ electrons, where $n = \{ 5, 6\}$ is the largest main quantum number.
For the 4d and 5d transition metal elements, we chose a larger valence space by incorporating the semi-core $s$ and $p$ electrons with the outer-layer $s$ and $d$ electrons.

Our primary goal was to generate highly accurate ccECPs for the mentioned elements incorporating many-body theories and explicit spin-orbit effect.
Intuitively, the construction is partitioned into the AREP (spin-averaged) part and subsequent SO (spin-orbit) part.
Such methodology relies on the quality of AREP since the SO part plays the role as extensive refinements.
To obtain highly accurate AREP ccECPs, we follow previous scheme of many-body construction method that involves an iterative process in corresponding calculations of all-electron atoms using Coupled Cluster methods and optimizations of objective functions that include weighted atomic spectra, norm-conservations, extensive quality/transferability tests in molecular binding curves for hydride and oxide dimers.
The spin-orbit optimization applies the iso-spectrality of low-lying states and corresponding spin-orbit splittings.
Further assessments of the constructed ccECPs from spin-orbit splittings and several molecular dimer binding calculations are carried out the two-component spinor FPSODMC calculations.

We find that the main source of biases is the AREP part. 
This appears to be similar to the observation for 
$3d$ transition metal elements where the Hartree-Fock levels produced the dominant source of inaccuracies.

The ccECPs enable finding spin-orbit splittings with errors of 0.05 - 0.1 eV when compared with experiments in most cases and show remarkable accuracy in the dimer molecular binding for the cases we tested.

The comparisons with previously constructed sets (CRENBL, STU, SBKJC, etc) show that ccECPs are overall much more accurate and consistent
in minimizing the biases. The main group elements show minor errors beyond chemical accuracy in dimers at very small bond lengths
which has been observed already for $3s3p$ main group elements although for those cases the errors are in general larger.

We also test the ECPs on FPSODMC calculations for particular for homonuclear dimers and 
{%\color{blue} 
with single-reference trial functions. 
 The agreement with CCSD(T) is reasonably good although biases of a few tenths of eV occur in particular cases. Similarly to some of the atomic calculations, these discrepancies are assigned to limited accuracy of the trial fixed-phase generated by single-reference as the tests on selected tungsten systems with CI trial states illustrate. More systematic study of the
 fixed-phase biases with more accurate CI trial functions is left for future work due to already significant length of this paper. We also mention that
 considering the lack of accurate experimental data, our calculations of the W$_2$ dimer provide an independent prediction of the binding parameters.}

We believe that our study paves the way for accurate many-body valence space calculations with heavy atoms by providing a new generation of effective core potentials for several elements that are present in technologically important materials. The spin-orbit terms are included in a two-component spinor formalism. The new ccECPs are systematically more accurate and show better consistency with all-electron settings for atoms as well as molecular oxides and hydrides with additional benchmarks and periodic system calculations left for future work.

%\bigskip
%\bigskip
%\textbf{Supplementary Material} 
\section{Supplementary Material} 

Additional information about ccECPs can be found in supplementary material.
Therein, calculated AE spectra are given for each element and also corresponding discrepancies of various core approximations.
AE, UC, and various ECP molecular fit parameters for hydrides, oxides, or dimers are provided.
The ccECPs in semi-local and Kleinman-Bylander projected forms as well as optimized Gaussian valence basis sets in various input formats (\textsc{Molpro}, \textsc{GAMESS}, \textsc{NWChem}) can be found at website \cite{pseudopotential_library}.

The input/output files and supporting data generated in this work are published in Materials Data Facility \cite{blaiszik_materials_2016, blaiszik_data_2019} and can be found in Ref \cite{data-facility}.

%\bigskip
%\bigskip
\section{Acknowledgments}

The authors thank Cody A. Melton for the kind help with spin-orbit QMC calculations.
We are also grateful to Paul R. C. Kent for the reading of the manuscript and helpful suggestions.

This work has been supported by the U.S. Department of Energy, Office of Science, Basic Energy Sciences, Materials Sciences and Engineering Division, as part of the Computational Materials Sciences Program and Center for Predictive Simulation of Functional Materials.

This research used resources of the National Energy Research Scientific Computing Center (NERSC), a U.S. Department of Energy Office of Science User Facility operated under Contract No. DE-AC02-05CH11231. 
This research used resources of the Argonne Leadership Computing Facility, which is a DOE Office of Science User Facility supported under contract DE-AC02-06CH11357. 
This research also used resources of the Oak Ridge Leadership Computing Facility, which is a DOE Office of Science User Facility supported under Contract DE-AC05-00OR22725.

This paper describes objective technical results and analysis. Any subjective views or opinions that might be expressed in the paper do not necessarily represent the views of the U.S. Department of Energy or the United States Government.

\bibliography{main.bib}

\end{document}